\documentclass[a4paper,12pt,oneside]{amsart}
\usepackage{graphicx}
\usepackage{epigraph}
\usepackage{amssymb}
\usepackage{paralist}

\usepackage{amscd}
\usepackage{dsfont}
\usepackage{bm}

\newtheorem{theorem}{Theorem}[section]
\newtheorem{lemma}[theorem]{Lemma}

\newtheorem{rem}[theorem]{Remark}   
\newtheorem{rems}[theorem]{Remarks}   
\newtheorem{col}[theorem]{Corollary }  

\newtheorem{definition}[theorem]{Definition}

\theoremstyle{remark}

\numberwithin{equation}{section}

\newcommand{\abs}[1]{\left\vert#1\right\vert}

\def \1{\mathds{1}}
\def \0{\mathsf{0}}

\def\A{\mathcal{A}}
\def\Aa{ \mathbb{A}}
\def\B{\mathfrak{B}}
\def\C{\mathfrak{C}}
\def\P{\mathcal{P}}

\def\Ee{ \mathbb{E}}
\def\Mm{ \mathbb{M}}
\def\Mmm{ \mathrm{M}}
\def\Sym{ \mathds{S}}
\def\Nn{\scriptscriptstyle{N}}
\def\Tt{\scriptscriptstyle{T}}

\def\N{ \mathcal{N}}

\def\X{ {\mathcal{X}}}
\def\V{ {\mathcal{V}}}
\def\M{ {\mathcal{M}}}

\def\mq{ m^q}
\def\mp{ m^p}
\def\m{ \mathrm{m}}
\def\q{ {\mathsf{q}}}
\def\p{ {\mathsf{p}}}

\def\X{ {\mathcal{X}}}
\def\F{ {\mathcal{F}}}

\def\Zz{ \Z^2_{\scriptscriptstyle\mathrm{NT}}}

\def\H{ {\mathcal{H}}}
\def\Z{ {\mathbb{Z}}}

\def\S{ {\mathcal{S}}}

\def\R{ {\mathcal{R}}}

\def\Re{ {\mathrm{Re}}}

\begin{document}
 
\title[Many-particle quantum chaos]{Classical foundations of   many-particle quantum chaos}

\author{Boris Gutkin, Vladimir Osipov}
\address{
${}^\dag$ Faculty of Physics, University Duisburg-Essen, Lotharstr. 1, 47048 Duisburg, Germany;
}
\email{boris.gutkin@uni-duisburg-essen.de, vladimir.al.osipov@gmail.com}
\begin{abstract}

In the framework of  semiclassical theory the universal properties of quantum systems with classically chaotic dynamics  can be accounted for  through   correlations between partner periodic orbits with small action differences. So far, however, the scope of this approach has been mainly limited to systems of  a few particles with low-dimensional phase spaces.  In the present work we consider  $N$-particle chaotic systems with local homogeneous interactions, where $N$ is not necessarily small. Based on a  model of coupled cat maps  we demonstrate emergence of a new mechanism for correlation between periodic orbit actions.   In particular, we show the existence of  partner orbits which are specific to many-particle systems. For a sufficiently large $N$  these new partners  dominate the spectrum of correlating periodic orbits and seem to be necessary for  construction of a consistent many-particle semiclassical theory. 

\end{abstract}

\maketitle
\section{Motivation and Goals}

\epigraph{``Sadly, searching for periodic orbits will never become as popular as a week on C\^{o}te d'Azur, or publishing yet another log-log plot in Phys. Rev. Letters.''}{--- \textit{P. Cvitanovi$\acute{c}$, et al., \cite{Predrag}}}

Already at the dawn of the quantum era  it was realized that  properties of quantum systems  crucially depend on their classical dynamics. While the eigenvalues of  integrable systems can be explicitly related to the set of integer numbers by the Bohr-Sommerfeld quantisation rules no such   regular structure  exists for  systems with complex dynamics. At first sight, the energy spectrum of complex systems like nuclei    resembles  a structureless set of  random numbers very much dependent  on particularities of the system. This, however, turns out to be not entirely true.
 The famous Wigner-Dyson-Mehta conjecture asserts  that the spectrum of complex quantum systems on the scales of the mean level spacing between eigenvalues  is universal  and can be described by Random Matrix Ensembles within the same symmetry class. In particular, all $n$-point correlation functions of eigenvalues can be derived analytically by using Random Matrix Theory (RMT). Indeed, such  spectral statistics have  been observed in  many real and numerical experiments with  various systems ranging from compound nuclei to complex molecules and atoms. 
Furthermore,  it was realized in the early 1980s that even  single-particle systems generically exhibit the same universality if their classical dynamics are chaotic.  
In the last three  decades a substantial progress has been achieved in  understanding of the origins of this universality through  the application  of the semiclassical theory  to quantum chaotic systems. In particular, the Gutzwiller's trace formula allows to express eigenvalue  density function through the sum of unstable classical periodic orbits. By using this representation  the  correlations between  system  eigenenergies can be straightforwardly related  to the correlations  between actions of periodic orbits. This approach was pioneered by M.~Berry in the seminal paper~\cite{berry1984}, where the diagonal correlations between  periodic orbits were taken into account.  It was  also  noted there and in subsequent works~\cite{UzyArgaman, UzyPrimack}  that in order to obtain the full RMT result one would need to include  non-trivial correlations between periodic orbits as well.

On the quantitative level the  non-trivial correlations between periodic orbits  were first taken into account  in the groundbreaking work of M.~Sieber and K.~Richter~\cite{sieber}.  On the basis of classical chaotic dynamics they demonstrated existence of periodic orbit pairs with close actions. 
Generally speaking, such orbits traverse approximately the same points of the configuration space but in a different time order. By  including the \textit{Sieber-Richter pairs}, and their natural generalizations  it turned out to be   possible to derive the  full RMT result for universal spectral correlations~\cite{haake}.

So far, however, this remarkable progress  has by and large been restricted  to the systems composed of just few particles.  Although, formally Hamiltonian system of $N$ particles in $d$ dimensions can be thought as just one particle in $Nd$ dimension, such  dynamical interpretation does not  allow to use automatically the ``single-particle'' semiclassical  theory when $N$  grows simultaneously with $\hbar^{-1}$. This is because the standard semiclassical limit assumes  a fixed  dimensionality of the system while the effective Planck's constant  $\hbar_{eff}$ tends to zero. 
Another, often considered ``semiclassical'' theory corresponds to the thermodynamic limit, where    $N\to\infty$ but $\hbar_{eff}$ is fixed. As can be clearly seen on the  example of    bosonic systems, this  limit  essentially differs from the one where $N$  is fixed and $\hbar_{eff}\to 0$~\cite{richter2013, richter2014}. Indeed for interacting  bosons  the two limits   correspond  to very different classical Hamiltonians associated with the first and the  second quantisation of the problem, respectively.   At the present, very little  is known   about  the limit  where  the number of particles $N$ and the effective dimension $\hbar^{-1}_{eff}$ of the one-particle Hilbert space  grow simultaneously. This is especially frustrating considering that  the original Wigner's conjecture relates to many particle systems. 
 One might wonder  whether the ``single-particle'' correlation mechanism between periodic orbits is still relevant for   $N$-particle chaotic systems. \textit{Do these correlations  lead to a spectral universality, when  $N$ is large?  At which energy scales? Does the answer  depend on the details of the Hamiltonian?} 

The central goal of the present paper is to show that $N$-particle chaotic systems with local nearest-neighbor  interactions allow a more general correlation mechanism between periodic orbit actions then one which is known for one particle systems. By using the  one-particle interpretation of classical dynamics it is easy to demonstrate that   standard partner orbits (e.g.,  Sieber-Richter pairs)   exist for an arbitrary $N$. As expected,  these pairs of periodic orbits  closely approach each other in the large $Nd$ dimensional configuration space.   It turns out, however, that completely  different classes of periodic orbit pairs with close actions exist, as well. In particular, these non-standard  pairs     traverse completely different  points  of the  large configuration space. 
As we show in the body of the paper,  they can be interpreted as  two-dimensional extensions of
standard Sieber-Richter pairs, where periodic orbits sweep  surfaces rather then  lines in the small configuration space of the corresponding single-particle system.  
 From the semiclassical point of view  the contributions from new families of correlating orbits  should   become dominant over the standard one  whenever the number of particles $N$ exceeds certain threshold value which depends logarithmically on $\hbar^{-1}_{eff}$. We believe that in this   regime it is essential to take them into account in order to construct a proper semiclassical theory.  

The paper is structured in the following order. In the next section we present our main ideas on the heuristic level. The emerging mechanism of correlations between periodic orbits is then discussed in the framework of the particular model of \textit{coupled cat maps}. This model is introduced in Section 3 where we also describe its main dynamical properties. In Section 4  a special duality relation between periodic orbits is introduced. Implications for periodic orbit correlations and few remarkable  mathematical identities stemming from this duality are discussed here.  In Section 5 we describe general mechanism of correlations
between periodic orbits and provide exhausting numerical evidence of its existence. We explicitly construct a number  of periodic orbit pairs with small action differences. A general formula for action differences within such pairs of orbits is derived here.
The implications of symmetries on the orbit correlations are studied in Section 6. In Section 7 we extend our results to the perturbed  maps. Finally, in Section 8 we discuss the relevance of our results for generic many-particle systems with local homogeneous interactions and its implications for a semiclassical quantum theory.

\section{Main Ideas \label{secgoals}}
In the present work our attention is  focused on  chaotic systems of $N$-particles with local  homogeneous interactions. To be more specific,  let us  write a prototype  Hamiltonian which governs the dynamics of suchlike systems: 
\begin{equation}
\H=\sum_{n=1}^N \frac{p_n^2}{2m} +\V_{\mathrm{in}}(x_{n\!\!\!\!\mod{N}},x_{n+1\!\!\!\!\mod{N}})+\V(x_{n}). \label{hamiltonian} 
\end{equation}
Here $x_n$ is the particle  coordinate  belonging  to a target space $\X$ e.g., $\X=R^d, S^d$ and $p_n$ is the corresponding momentum. 
\begin{figure}[htb]
\begin{center}{
\includegraphics[height=3.5cm]{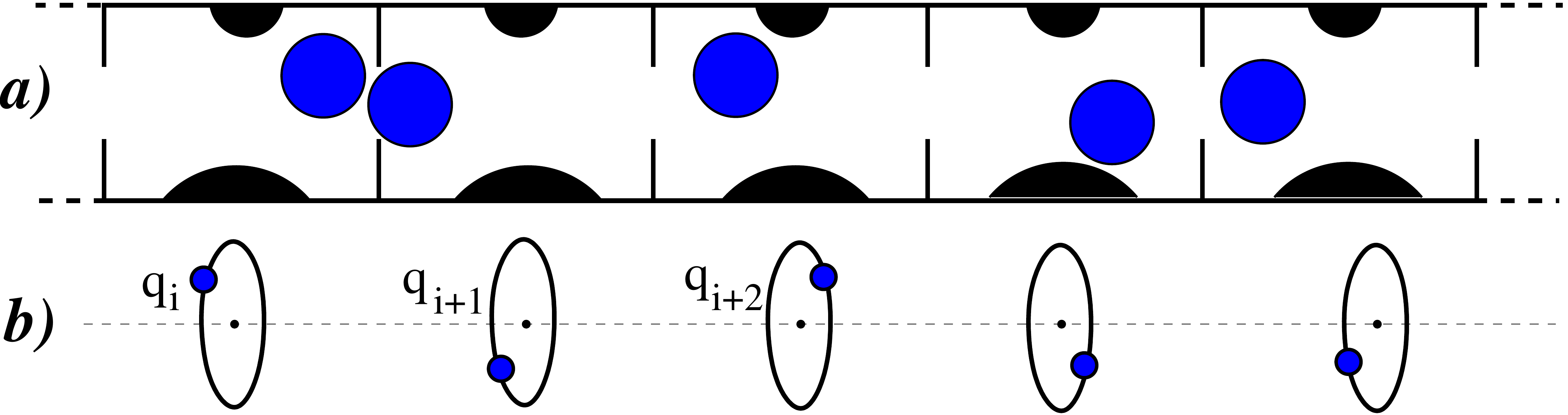}
}\end{center}
\caption{ \small{In the upper figure (a)  is shown the model of Boltzmann gas with the Hamiltonian of the form \ref{hamiltonian}. Each ball of the radius $a$ is constraint to a particular cell of the chain  and  interacts with the two neighbors through the openings of the size $L< a$. On the bottom figure (b) is shown the model of coupled cat map. Each particle lives on a circle of unit length and linearly  interacts with the two neighbors at the discrete moments of time. The corresponding time evolution of its coordinate $q_i$ is described by eq.~(\ref{newtoneq}). } }\label{bolzmann}
\end{figure}
Since the energy of the system is preserved, the dynamics of  (\ref{hamiltonian}) is restricted to the $2dN-1$ energy shell $\Omega_E$, determined by the conditions $\H(x_1\dots x_N,p_1\dots p_N)=E$.  Note that $\H$  is manifestly invariant under the cyclic shift \begin{equation}
     \sigma_1: n\to n+1\!\!\!\mod N\label{sigma1}                                                                                                                                                                                                                                                                                                                                                                                                                                                                                                                                           \end{equation}
 of the particle numbers. In addition to the shift invariance we will assume that   the dynamics induced by  $\H$ on $\Omega_E$ are fully hyperbolic. The last requirement implies that the derivative of the Hamiltonian flow has exactly $dN-1$ positive Lyapunov exponents in almost every point of $\Omega_E$. Informally speaking, the above Hamiltonian  describes  a closed string of  particles  uniformly interacting with their  neighbors and an external potential $\V$ such that the resulting dynamics are chaotic. One possible realization of such Hamiltonian  is provided by  a  Boltzmann gas type model   shown in fig.~\ref{bolzmann}a. 
{\rem
The  Hamiltonians of the type  (\ref{hamiltonian}) also appear naturally
  in connection with the spin chain models and    interacting bosons on a circular lattice. For instance,  the Bose-Hubbard Hamiltonian written in the second quantisation  language:
\[\widehat{\H}= -t\sum_{i=1}^N \hat{b}^{\dagger}_i \hat{b}_{i+1}+\frac{U}{2}\sum_{i=1}^N \hat{n}_i( \hat{n}_i-1) -\mu\sum_{i=1}^N \hat{n}_i, \quad \hat{n}_i=\hat{b}^{\dagger}_i \hat{b}_{i}\]
can be cast into the form similar to  (\ref{hamiltonian}) after substitution $\hat{b}_{i}=\frac{1}{\sqrt{2}}(\hat{x}_i+i\hat{p}_i)$, where $\hat{x}_i$ and  $\hat{p}_i$ should be interpreted as the   coordinate and  the momentum operators  of the  $i$-th harmonic oscillator.
}
 
\subsection{Time evolution}
Given some initial conditions the time evolution of a  $N$-particle system can be represented  as  collection
\[\Gamma(\tau)=\{\gamma_1(\tau), \gamma_2(\tau), \dots , \gamma_N(\tau)\}, \qquad \tau\in \mathbb{R}\]
  of one  particle trajectories $\gamma_n(\tau)=(x_n(\tau),p_n(\tau))\in V$, $n=1,\dots N$  in the one-particle $2d$-dimensional phase space $V:=T^*\X$.
In what follows we will refer to $\Gamma(\tau)$ as \textit{many-particle orbit}  and as  \textit{many-particle periodic orbit} (MPO) if $\Gamma(\tau)$ is closed i.e., $\gamma_n(\tau)=\gamma_n(\tau+\tau_0)$, $n=1,\dots N$ after a period $\tau_0$. 
Rather then treating evolution as a continues flow, it is convenient  to discretize dynamics  by introducing a discrete Poincare map acting on  a section  of the phase space. To this end we fix a  Poincare section $\P$  which is a  $2dN-1$ dimensional hypersurface invariant under the shift $\sigma_{1}$, see eq.~(\ref{sigma1}), and observe the system at the discrete moments of time $\tau_t$, $t\in \mathbb{Z}$  when many particle orbit $\Gamma(\tau)$ pierces $\P$. For the system shown in fig.~\ref{bolzmann}a one possible  choice for the Poincare section would be to set it in accordance with the pairwise collisions of the balls. In this case $\tau_t$ would correspond to the moments of time when   balls collide with each other.  The above procedure  defines the Poincare map $\Phi_{\Nn}$ acting on the $2dN-2$ dimensional  phase space
  $V_{\Nn}=\P\cap\Omega_E$, obtained by the intersection  of $\P$ with the  the energy shell. 

Under the action of   $\Phi_{\Nn}$ any  MPO  $\Gamma$ of the system (\ref{hamiltonian}) can be  naturally seen in two different ways.
 First,   $\Gamma$ can be thought  as collection of the  points in the $N$-particle phase  space $V_{\Nn}$ registered at the discretized moments of  time:
\[ \S^1(\Gamma)=\{\Gamma(\tau_1), \Gamma(\tau_2)\dots \Gamma(\tau_T)\}.\]
In other words  the evolution of the whole system can be seen as propagation of one particle in the ``large'' phase space $V_{\Nn}$. Accordingly  $\S^1(\Gamma)$ is naturally interpreted as a discretized  one dimensional line. 
Such a reduction  to one particle system  is a widely used tool in  dynamical systems. For example it allows to treat the Boltzmann gas as a  many-dimensional billiard \cite{arnold}. 

The second   conceptually different way  is to look at  $\Gamma$  as a set of $N\cdot T$ points in the ``small'' single-particle phase space $V$:
\[ \S^2(\Gamma)=\{ \gamma_{n,t}| \quad n=1,\dots,N ,\quad  t=1,\dots,T\},\] 
where $\gamma_{n,t}:= \gamma_n(\tau_t)$ marks the position  of the n'th particle  at  discrete moment of times $ t=1,\dots ,T$. For the sake of convenience of presentation it is  instructive  to introduce  the map $\Psi_\Gamma: \Zz\to V$ which attaches to each point $(n,t)$ of the discrete ``particle-time'' space
$\Zz:=\{1,\dots ,N\}\times\{1,\dots ,T\}$ the corresponding point  $\gamma_{n,t}$ of the phase  space $V$. With such notation we have
\[\S^2(\Gamma)=\Psi_\Gamma(\Zz). \]

In a (properly scaled) continuous limit $N, T\to \infty$ the  time evolution of the above many-particle system can be seen on an intuitive level   as
 the propagation  of a one dimensional closed string  $\gamma(\ell, \tau)=\gamma(\ell+\ell_0, \tau)$, where $\ell\in[0, \ell_0]$ measures the length along it  and $\tau$ is continues time parameter.   Accordingly, 
any  periodic orbit of a period $\tau_0$ can be represented by a  two dimensional torus swept by  $\gamma(\ell, \tau)$ in $V$ during the time  $\tau\in[0, \tau_0]$. This contrasts with the single-particle interpretation where periodic orbits of the system are naturally represented by one dimensional lines.
It should be emphasized that we actually neither  require the interaction  between particles  be attractive nor we consider the continues limit.  So the resulting time evolution  of the system  might be very different  from  a string-like motion where all particles   keep a linear order in the target space.  Nevertheless,  regarding  the correlation mechanism  between periodic orbits  it turns out to be    quite useful to think about a MPO $\Gamma$  as  of discretized two dimensional  surface $\S^2(\Gamma)$ in the one-particle phase space $V$ (rather then one dimensional line $\S^1(\Gamma)$ in the many-particle  space $V_{\Nn}$).

\subsection{One particle partner orbits.} 
\begin{figure}[htb]
\begin{center}{
a)\includegraphics[height=2.7cm]{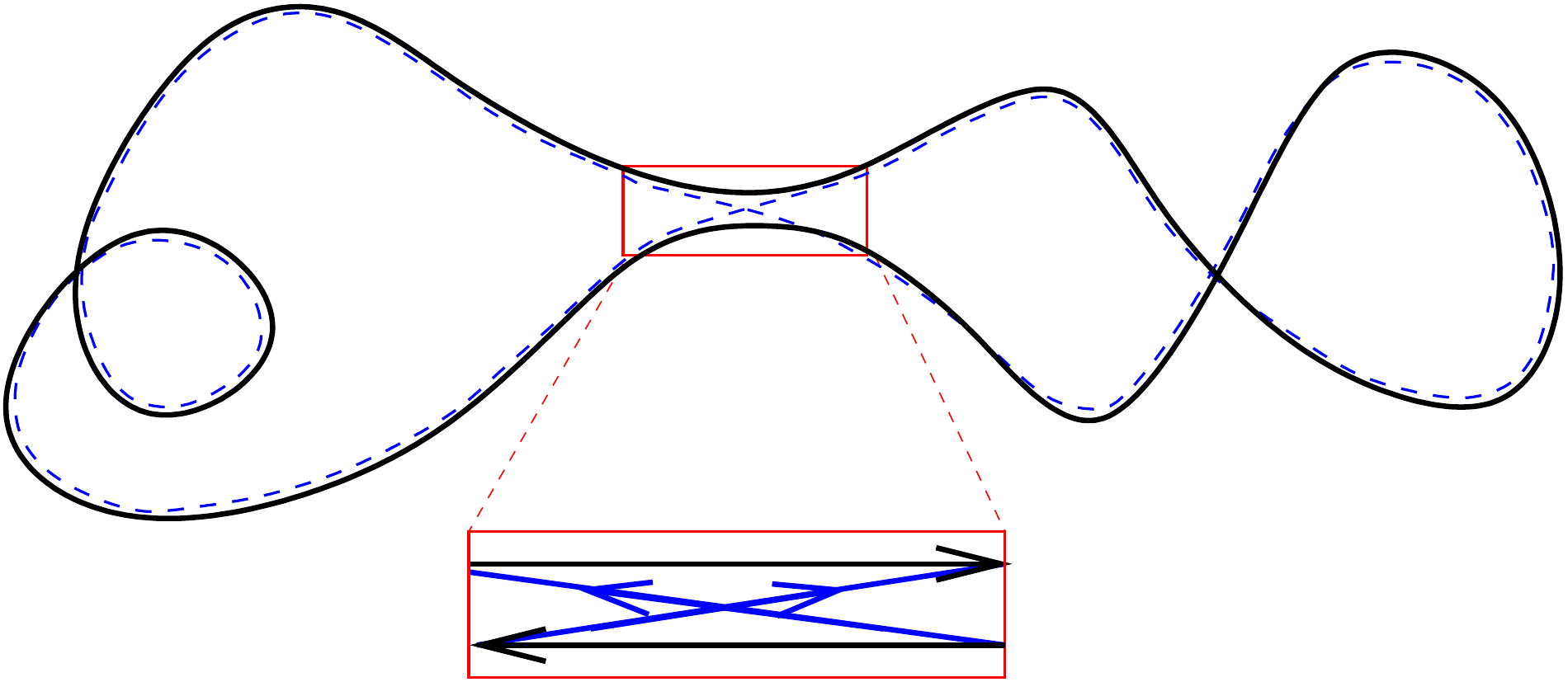}\hskip 1.5cm b)\includegraphics[height=2.4cm]{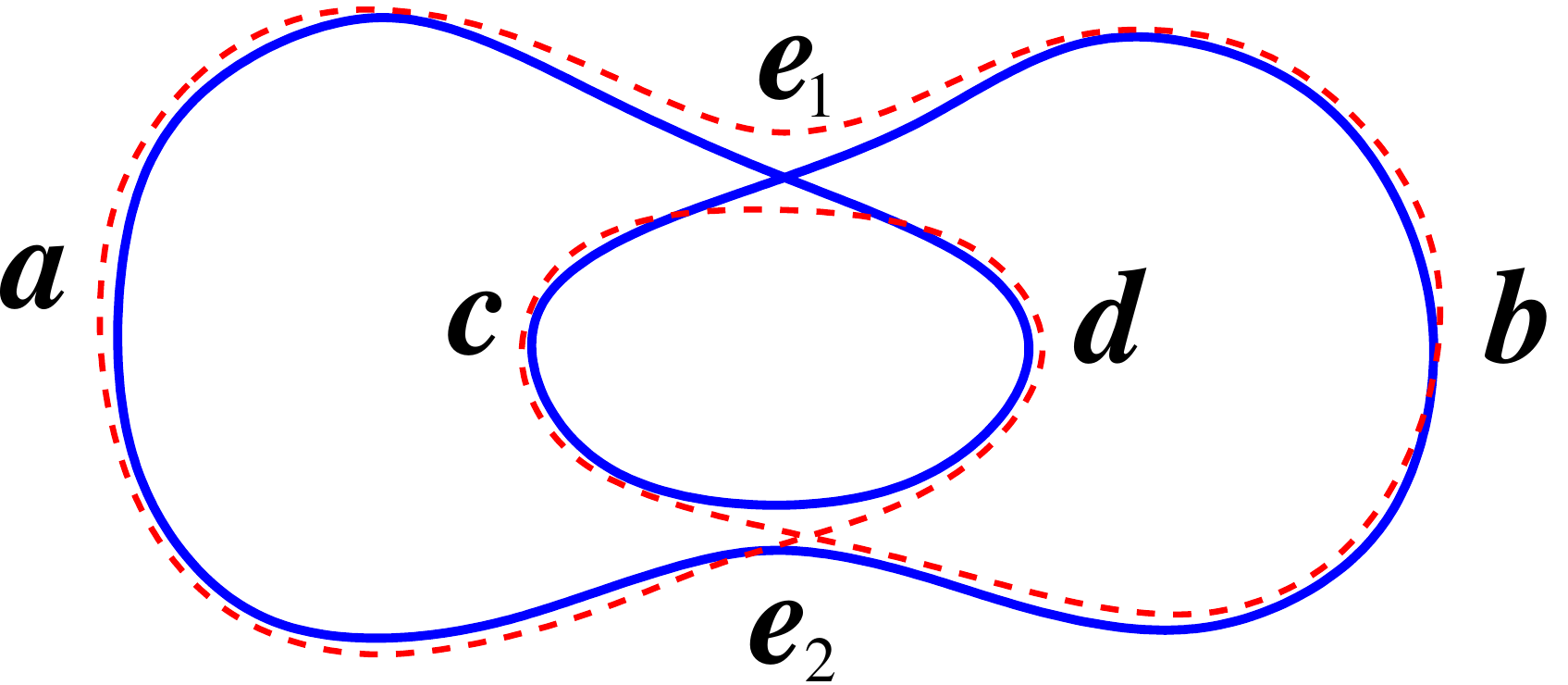}
}\end{center}
\caption{ \small{On the left is  shown the caricature of a Sieber-Richter pair  with one encounter  shown as a (red) box. Such partner orbits exist only in  systems with time reversal symmetry. On the right it is shown the caricature of a periodic orbit  (solid line) and its partner (dashed line)  with two encounters $e_1$, $e_2$. Their symbolic representations are given by (\ref{symbrepres}).  }  }\label{encounter}
\end{figure}

Before focusing on the action correlations between periodic orbits in many-particle systems,  let us briefly   recall the correlation mechanism in  one particle Hamiltonians \cite{haake, sieber}. A sufficiently long periodic orbit typically has a number of self-encounters. These are the stretches of the trajectory where it closely approaches itself in the configuration  space, see fig.~\ref{encounter}. For a trajectory possessing sufficiently long encounters  the hyperbolic  nature of the dynamics guarantees  existence of several  (the exact number depends on the encounter structure) of partner periodic orbits which traverse approximately the same points of the phase space, but  make  different  switches   at the encounters.  The action differences between the partners are accumulated primarily at  the encounter stretches and decrease exponentially with their lengths.
Accordingly,  the periodic orbits of a hyperbolic system can be 
organized into   families,  where all members have  approximately the same actions.
\begin{figure}[htb]
\begin{center}{
a)\includegraphics[height=2.7cm]{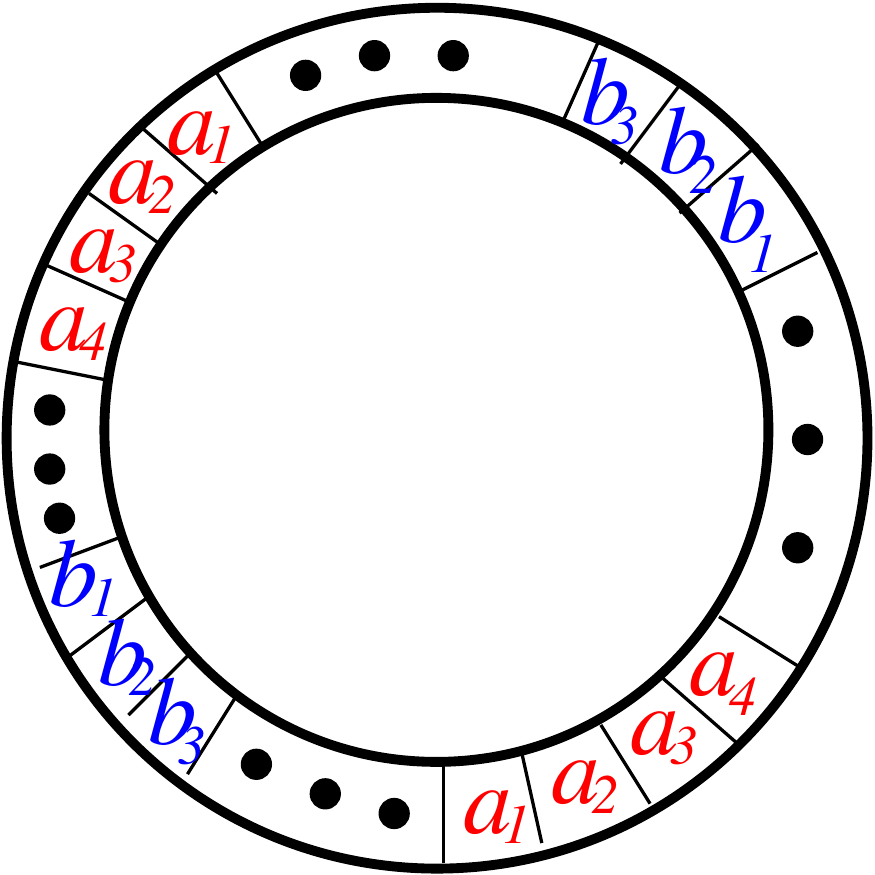}\hskip 1.5cm b)\includegraphics[height=2.7cm]{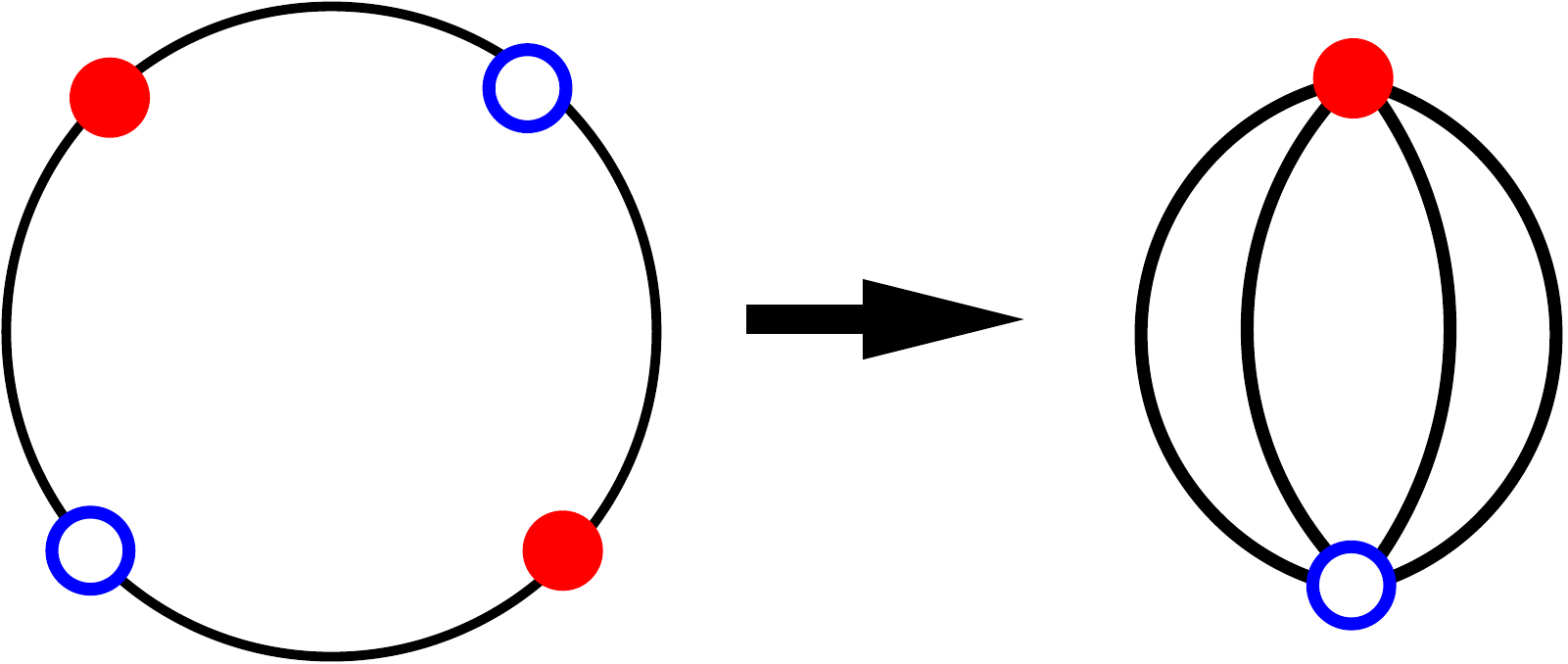}
}\end{center}
\caption{ \small{a) Schematic depiction of 1D symbolic representation of  the periodic orbit  on fig.~\ref{encounter}b. Two encounter regions correspond to the identical sequences of symbols labeled by $a$ and $b$ letters respectively  b) The structural diagram of the same periodic orbit. The encounter regions  correspond to two pairs of points on the circle $T^1$.    After ``gluing'' the circle at these points we obtain the graph shown in the second figure. This graph can be also interpreted as   Feynman diagram for a zero-dimensional sigma model which describes universal spectral correlations in quantum systems with classically chaotic dynamics, see \cite{haake}.       }  }\label{encounter_str1}
\end{figure}

\subsubsection{Symbolic dynamics} The intuitive description of the above correlation mechanism between periodic orbits can be  formalized    by making use of symbolic dynamics \cite{GO2013,GO2013b}. Within this approach  the first step is to fix a Poincare section and consider the discretized  time evolution.  Assuming that the reduced phase space of the  system allows a finite Markov partition $\bigsqcup_{\alpha \in\A} V_\alpha $, any  trajectory $\gamma$ can be encoded by a doubly infinite  sequence of symbols from some alphabet $\A$ of a finite size:
\[ \dots a_{-2}a_{-1}a_{0}.a_{1}a_{2}\dots\]
  Here each symbol $a_t$ registers the  position of the particle with respect to the partition at the discrete time $t$, i.e.,  $a_t$ equals to $\alpha\in \A$ if   $\gamma(t)\in V_\alpha $.
Correspondingly, a $n$-periodic orbit  $\gamma$ can be associated with a cyclic sequence of the length $n$:
 \[ a_\gamma=[a_{1}a_{2}\dots a_{n}], \qquad a_i\in\A.\]
On the level of symbolic dynamics each \textit{$l$-encounter} of  the length $p$ in $\gamma$ is
  a string of $p$ symbols  $e=\epsilon_1\epsilon_2\dots \epsilon_p$ which appears in $a_\gamma$  a number $l>1$ of times, see fig.~\ref{encounter_str1}a. Given a periodic sequence $a_\gamma$ its pair   sequence $a_{\bar{\gamma}}$  is constructed by rearranging  the  encounters and stretches of symbols connecting them in a way that any subsequence of consecutive $p$ symbols existing in $a_\gamma$ reappears in 
$a_{\bar{\gamma}}$ exactly the same number of times and vice versa. 
Since the approximate position of a point in the phase space is determined by a finite string of symbols, the above property guarantees that the corresponding orbits traverse approximately the same points of the phase space. 
For instance, the following two sequences: 
\begin{equation}
a_\gamma=[ae_1de_2ce_1be_2], \qquad a_{\bar{\gamma}}=[ae_1be_2ce_1de_2],\label{symbrepres}
\end{equation}
with the  encounters $e_1$, $e_2$  of some length $p$ and stretches $a$, $b$, $c$, $d$ of an arbitrary length connecting them   provide the symbolic representation of  partner orbits shown on fig.~\ref{encounter}b.

Structurally different families  of partner orbits $\{\gamma\}$ can be distinguished by their   diagrams $\F_{\gamma}$ accounting for different order of encounters.    Each of the diagrams is obtained by substituting encounters of $\gamma$ with the points and 
placing them (in the same linear  order) on a circle $T^1$.  
The points belonging to the same encounters are then identified (i.e.,  $T^1$ is ``glued'' to itself at these points), see fig.~\ref{encounter_str1}b. The resulting  graph  $\F_{\gamma}$ carries all structural information on the correlations between the partner orbits $\{\gamma\}$. In particular, it defines the number of partner  orbits  within the family. It worth mentioning  that in the semiclassical theory   based on the correlations between periodic orbits   such graphs  can be associated with Feynman diagrams of the  corresponding sigma model \cite{haake}.

\subsection{Many particle partner orbits.} 
\begin{figure}[htb]
\begin{center}{
\includegraphics[height=5.7cm]{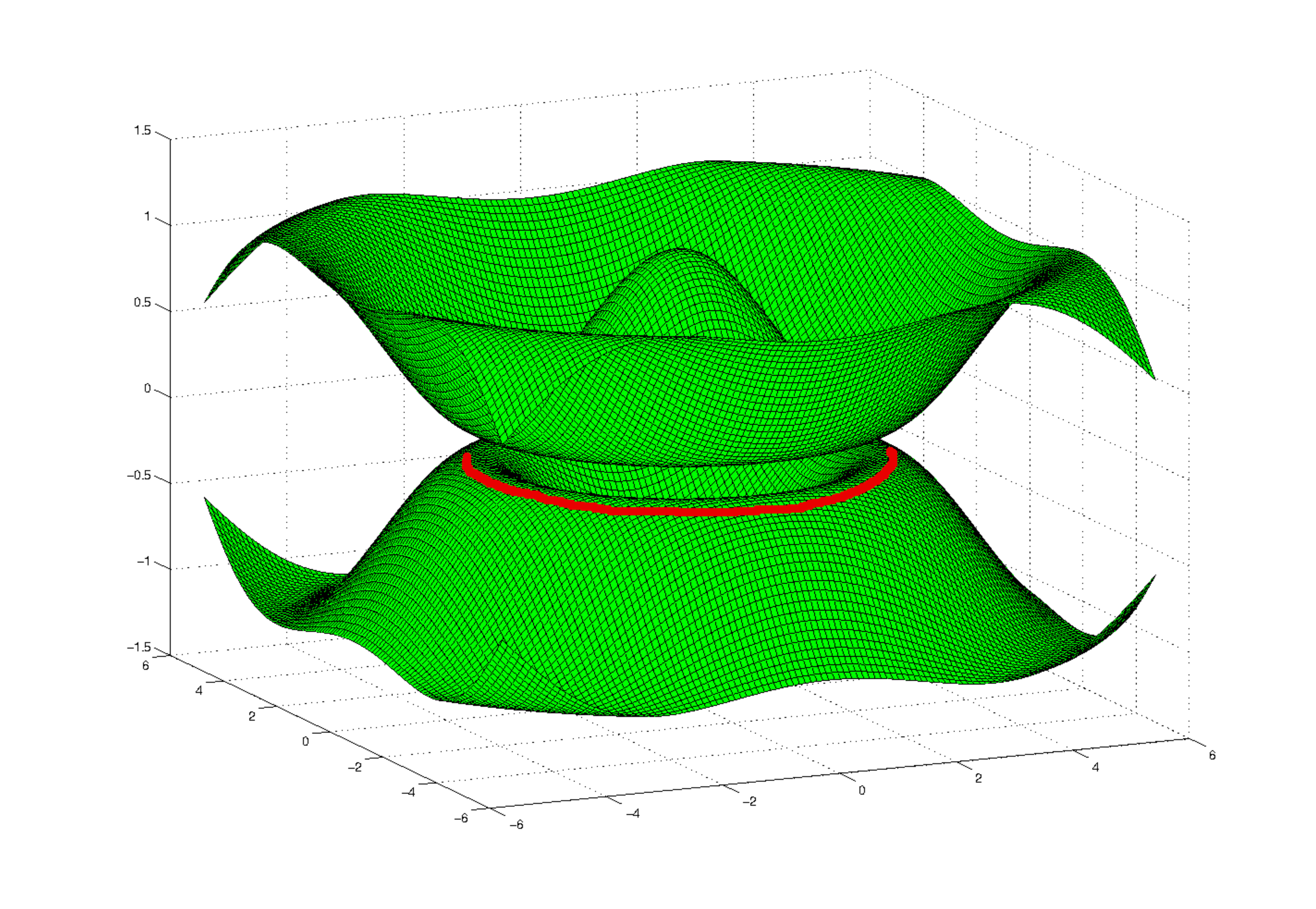}\hskip -0.5cm \includegraphics[height=4.7cm]{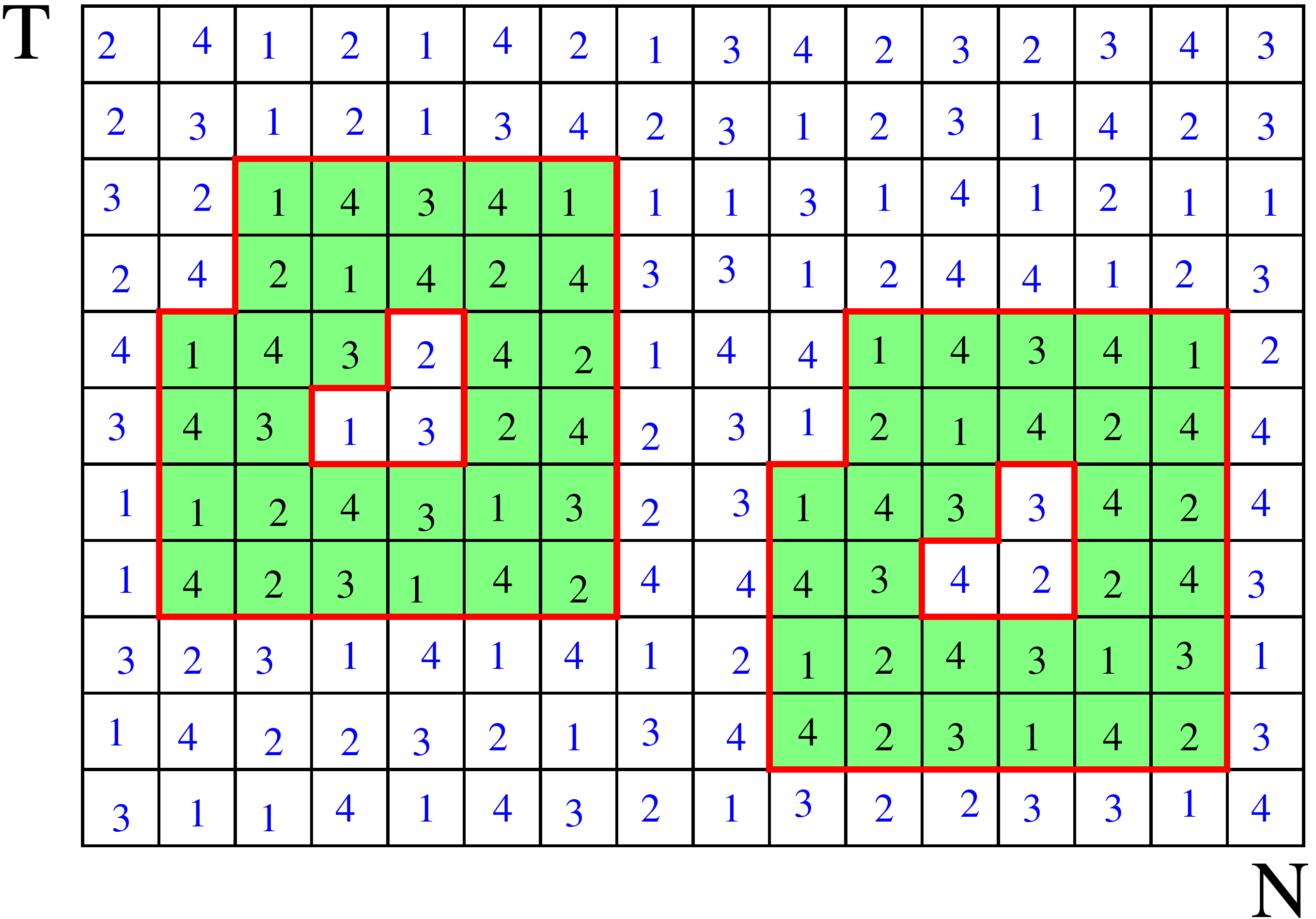}
}\end{center}
\caption{ \small{On the left is  a geometric representation of  MPO with an encounter (shown in red). On the right is shown symbolic representation of suchlike  MPO. Here the alphabet is composed of four symbols $\{1,2,3,4\}$. The   encounter region is encoded by 2D sequence of symbols (shown in green) which reappears in two different places of $\Zz$. (color on line)   }  }\label{surface}
\end{figure}

We turn now our attention to the $N$-particle systems with the local type of interactions (\ref{hamiltonian}).  At the beginning of this section we argued that in the limit of continuous $N$ the  periodic orbits of such a system can be regarded  as two-dimensional surfaces in the phase space.  Taking this as a guiding  principle let us first analyze  on an intuitive  level  its  implications    on  the correlation mechanism between periodic orbit actions. By the analogy with the single-particle case, we demand  that the two surfaces corresponding to some pair of partner periodic orbits $\Gamma, \bar{\Gamma}$ pass through approximately the same points of the phase space i.e.,
\[S^2(\Gamma)= \Psi_\Gamma(\Zz)\approx\Psi_{\bar{\Gamma}}(\Zz)=S^2(\bar{\Gamma}).\]
 For this to happen each periodic orbit must have at least one $l$-encounter, which is an element of the configuration  space where the surface approaches itself sufficiently closely for a number $l\geq 2$ times, see fig.~\ref{surface}. The existence of the encounter implies   that there are $l$ disjoint domains  $E^{(i)} \subset\Zz, i=1,\dots ,l$  related to each other by the translation shift $\sigma_2^{(n', t')}:(n,t)\to (n+ n'\mod N, t+ t'\mod T)$:
\begin{equation}
 E^{(i)}=\sigma_2^{(n_i, t_i)} \cdot E^{(1)}, \mbox{ for }  i=1,\dots ,l, \qquad   E^{(i)}\cap E^{(j)}= \emptyset, \mbox{ for } i\neq j\label{encounterdomains}
\end{equation}
such that  all  domains are mapped into approximately the same points of the target space $V_{\Nn}$:
\[\Psi_\Gamma(E^{(1)})\approx\Psi_\Gamma(E^{(2)})\approx\dots \approx\Psi_\Gamma(E^{(l)})\approx \Psi_{\bar{\Gamma}}(E^{(1)})\approx\Psi_{\bar{\Gamma}}(E^{(2)}) \approx\dots \approx\Psi_{\bar{\Gamma}}(E^{(l)}).\]
Conversely, whenever a closed surface possesses encounters it might be possible to find a pair surface  traversing  almost the same points of the phase space.    Therefore, the problem of classifying  structurally different families of correlating MPOs  can be on a geometrical level understood as one of finding   all topologically non-equivalent   ways to ``glue'' a two-dimensional surface to itself.  This is very much similar to  the single-particle case where we need to classify the non-trivial ways to ``glue'' one-dimensional lines.
\begin{figure}[htb]
\begin{center}{
a)\includegraphics[height=3.1cm]{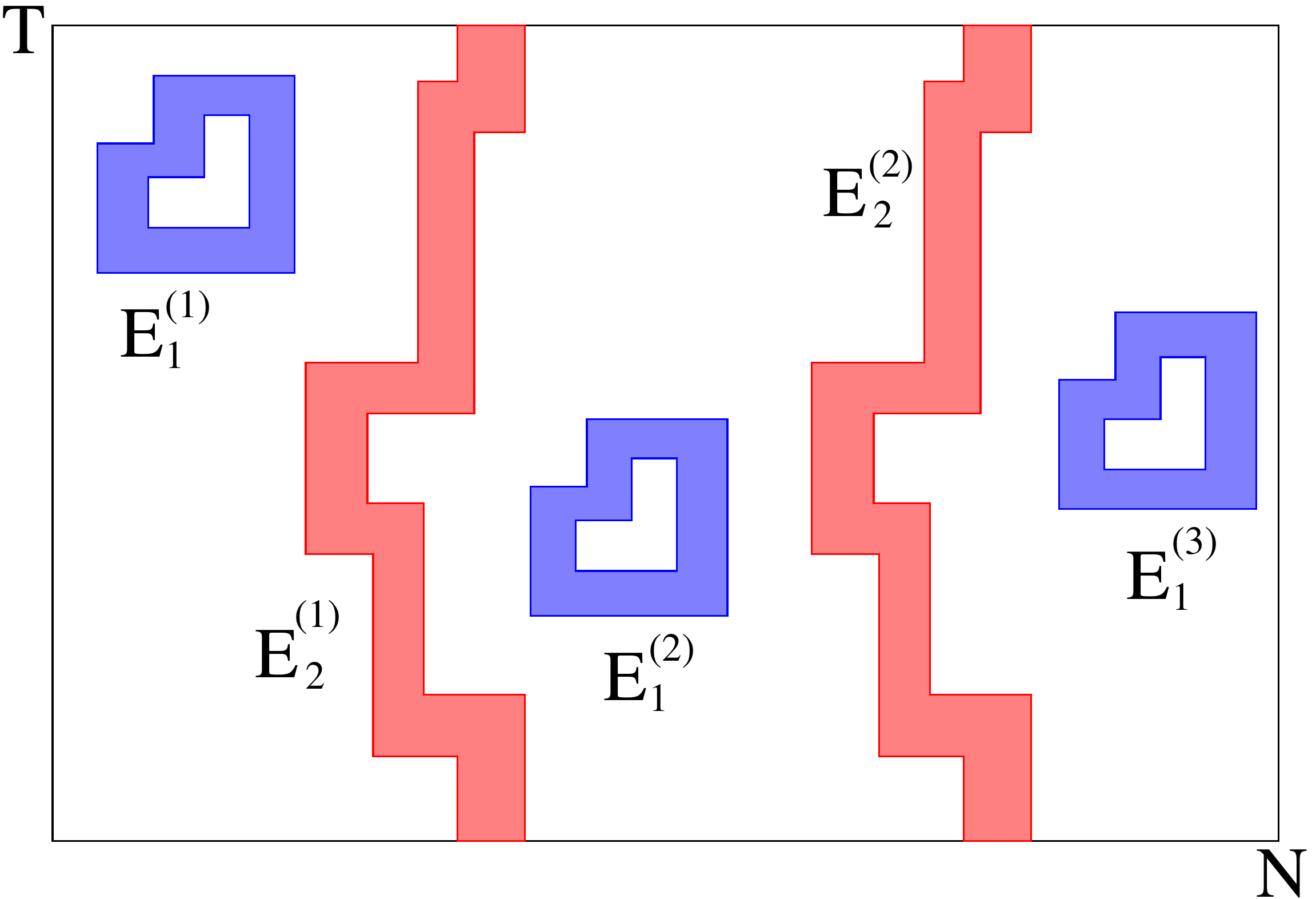}\hskip 1.5cm b)\includegraphics[height=3.1cm]{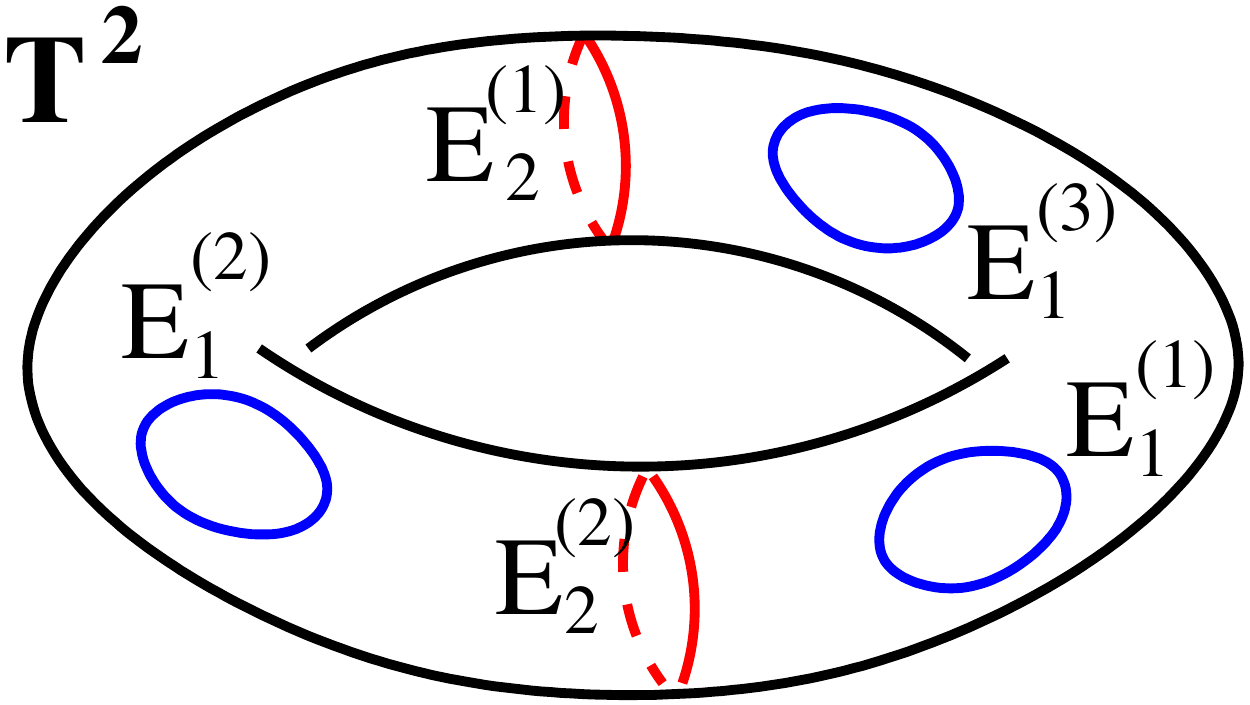}
}\end{center}
\caption{ \small{a) Schematic depiction of 2D symbolic representation for a MPO $\Gamma$ with two encounters. The first 3-encounter  has winding numbers $(0,0)$.  It is represented by three identical sets of symbols at the frame-like domains  $E^{(i)}_1, i=1,2,3$ shifted with respect to each other on the torus $\Zz$. The second 2-encounter  is represented by  two identical sets of symbols  at the strip-like domains  $E^{(i)}_2, i=1,2$ and has winding numbers $(0,1)$.   b) The structural diagram $\F_\Gamma$  of the MPO on the left figure. The two encounter regions correspond here to 3 identical (blue) oval lines and 2 (red) lines winding around  the torus.  The two-dimensional surface $\F_\Gamma$ is obtained by  ``gluing''  the torus along the identical lines. Compare with the one particle structural diagram shown in fig.~\ref{encounter_str1}.       }  }\label{encounter_str}
\end{figure}

\subsubsection{2D symbolic representation\label{SecSymb}} The above  intuitive geometrical picture  of partner  MPOs  can    be  made  precise through introduction of  two dimensional (2D) symbolic dynamics. Within this approach, it is assumed   that  each many particle (periodic)  orbit  $\Gamma$ can be   encoded by a two dimensional (periodic) lattice of symbols $a_{n,t}$,  $(n,t)\in\mathbb{Z}^2$, where symbols  $a_{n,t}$ belong to some alphabet $\A$ of a small size. Naturally, each MPO $\Gamma$ is represented then  by $N\times T$ toroidal  array of symbols: 
\[\Aa_{\Gamma}=\{a_{n,t}| \, (n,t)\in \Zz \}.\]

{\rem One formal way to  construct such a symbolic representation,  would be   to use  some partition $\bigsqcup_{\alpha\in \A} V_\alpha$ of the one-particle phase space $V$. 
At a  time  $t$  a symbol $a_{n,t}\in \A$ is attached to  the $n$'th particle, if the point  $\gamma_{n,t}$ belongs to the corresponding part of the partition: $V_{a_{n,t}}$. A symbolic representation of this type has been constructed for a model of coupled lattice maps in \cite{symbol}}\\

 To facilitate further discussion  we introduce the map $\psi_\Gamma:\Zz\to \A$ which assigns  to  each point  $(n,t)\in\Zz$ the corresponding  symbolic representation $a_{n,t}$ of    $\Gamma$ i.e., 
\[\psi_\Gamma(n, t)=a_{n,t}.\]
In what follows the array $\Aa_{\Gamma}\equiv \psi_\Gamma(\Zz)$ is  referred to as 2D symbolic representation  of $\Gamma$. By the analogy with  one-dimensional symbolic dynamics of hyperbolic systems we  require few  basic  properties  for such 2D encoding.

\begin{itemize}
\item 
 \textit{Small alphabet}: The number of symbols in $\A$ is independent of $N$.
 \item 
 \textit{Locality}:  The   symbols in a neighborhood of each point $(n,t)$ define an approximate position in the phase space of the $n$'th particle    at the time  $t$.  More precisely, let $p$ be an odd integer and let 
\begin{align*}\R^{(n,t)}_p= \{ (i,j)| i=n-(p-1)/2\!\!\!\!\mod \!N,\dots, n+(p-1)/2\!\!\!\!\mod \!N, \\
 j=t-(p-1)/2\!\!\!\!\mod \!T,\dots, t+(p-1)/2\!\!\!\!\mod \!T \}
\end{align*}
 be a square-like set of $p\times p$ points around   $(n,t)$. We require that for any two MPOs $\Gamma=\{\gamma_{n,t}| (n,t)\in\Zz\}$, $\Gamma'=\{\gamma'_{n,t}| (n,t)\in\Zz\}$  having the same symbolic representation   at the set  
$\R^{(n,t)}_p$ i.e., $\psi_\Gamma(\R^{(n,t)}_p)= \psi_{\Gamma'}(\R^{(n,t)}_p)$,  the distance between the corresponding points in the phase space  is bounded in an exponential way:
\begin{equation}
 |\gamma_{n,t}- \gamma'_{n,t}| <\Lambda^{-p},
\end{equation}
 where  $\Lambda$ is some constant.
\item \textit{Uniqueness}: For each   MPO there is a unique 2D symbolic representation. In other words if  $\psi_\Gamma(\Zz)= \psi_{\Gamma'}(\Zz)$, then necessarily $ \Gamma=\Gamma'$. Note that we do not require the opposite i.e., there might be 2D sequences of symbols  which do not correspond to any real MPO. 
\end{itemize}

Whether or not there exists  symbolic representation of periodic orbits  with the above properties is a highly non-trivial question,  which  should be, in principle,  explored for each system individually. It is known that such 2D symbolic representation  can be constructed for some   coupled lattice maps \cite{symbol}. In the body of the paper we also demonstrate  the existence of suchlike symbolic dynamics for the model of coupled cat maps.  Leaving for a while the problem of existence  aside let us describe  how the 2D symbolic  representation  can be utilized in order to find partner  MPOs. 

\subsubsection{Partner orbits}
First recall that by definition  partner  MPOs  $\Gamma, \bar{\Gamma} $ pass through approximately the same points of the phase space.  
The locality property of the symbolic dynamics   implies   that the  symbolic representation $\Aa_{\Gamma}$ in a neighborhood of a point $(n,t)$ must coincide with  one of its partner  $\Aa_{\bar{\Gamma}}$ in the  neighborhood of some  another point $(\bar{n},\bar{t})$. 
This motivates the following definition.

{\definition 
 Let $p$ be a fixed positive integer. We call two symbolic representations  $\Aa_{\Gamma},\Aa_{\bar{\Gamma}}$  as \textit{p-close} iff  any $p\times p$  square of symbols:  

\begin{equation}
 [\alpha]_p= \begin{pmatrix}
      \alpha_{1,1}     & \alpha_{1,2}    &    \dots &\alpha_{1,p} \\
      \alpha_{2,1}     & \alpha_{2,2}    &    \dots &\alpha_{2,p} \\
      \vdots           & \vdots          &    \ddots &\vdots   \\
      \alpha_{p,1}     & \alpha_{p,2}    &    \dots &\alpha_{p,p} \\
     \end{pmatrix}, \qquad \alpha_{i,j}\in \A
\end{equation}
  appears the same number of times (which might be also zero) in both $\Aa_{\Gamma}$ and $\Aa_{\bar{\Gamma}}$.}\\

\noindent Clearly if  two MPOs  $\Gamma, \bar{\Gamma} $ have  p-close 2D symbolic representations $\Aa_{\Gamma},\Aa_{\bar{\Gamma}}$   then $\Gamma$ and $\bar{\Gamma} $ are partner orbits, with
  the parameter $p$ controlling   how close the  sets  $S^2(\Gamma), S^2(\bar{\Gamma})$ approach  each other  in the phase space. 

Given a 2D symbolic  representation  $\Aa_{\Gamma}$   a p-close 2D sequence  might only exist if $\Aa_{\Gamma}$   contains at least one  $l$-encounter of a ``width'' $p$. In other words, if $\Gamma$ has a partner orbit $\bar{\Gamma}$ there exist $l$ disjoint domains  $E^{(i)} \subset\Zz, i=1,\dots ,l$    satisfying (\ref{encounterdomains})   such that 2D subsequences of symbols at all  $E^{(i)}$ coincide (see fig.~\ref{surface}):
\begin{equation}
 \psi_{\Gamma}(E^{(1)})=\psi_{\Gamma}(E^{(2)})=\dots=\psi_{\Gamma}(E^{(l)}).
\end{equation}
A general encounter region  can be defined  by fixing   closed (translationally related) discrete line contours  $\varUpsilon_{\scriptscriptstyle E^{(i)}}, i=1,\dots ,l$  and expanding them  into  ``strips'' of  finite  width. The resulting sets   $E_{i}, i=1,\dots ,l$ 
must  fulfill the  eq.~(\ref{encounterdomains}) and satisfy the following  two requirements: 1) The interior of each $\varUpsilon_{\scriptscriptstyle E^{(i)}}$ 
should contain at least one point which does not belong to $E^{(i)}$ i.e., $E^{(i)}$ has to be 
  multi-connected. 2) For any $(n,t)\in \varUpsilon_{\scriptscriptstyle E^{(i)}}$ one has  $\R^{(n,t)}_p\subseteq E^{(i)}$ . Structurally different encounters  can  be distinguished  by their winding numbers $w=(w_1,w_2)$ specifying how many times the path  $\varUpsilon_{\scriptscriptstyle E^{(i)}}$  goes around the torus $\Zz$ in the particle and the time  directions, respectively.

Let $\Gamma$ be a MPO and let  $\Aa_\Gamma$ be its symbolic representation.
In general $\Gamma$ might contain a number $\N$ of encounters  with different multiplicities $l_n$ and winding numbers $w_n$,  $ n=1,\dots \N$. For each  encounter   there exist $l_n$ corresponding sets of points $E^{(i)}_n\subset\Zz, i=1,\dots l_n$ related by shift translation such that symbolic representation of $\Gamma$ on all these sets coincide, see fig.~\ref{encounter_str}a. 
To distinguish between  different classes of MPOs let us introduce for each $\Gamma$ the corresponding diagram $\F_\Gamma$ keeping the  topological information about its encounters structure.  To this end substitute  $\Zz$ with a 2-dimensional torus $T^2$ and each set $E^{(i)}_n$ with a line $\varUpsilon^{(i)}_{n}\subset T^2$ (whose exact shape of no importance) having  the same winding numbers and  the same geometrical order as the corresponding encounters. We then identify the points of $T^2$ which belong to the  lines $\varUpsilon^{(i)}_{n}, i=1,\dots l_n$, see fig.~\ref{encounter_str}b. (Informally speaking, we glue the torus to itself  along the lines  from the same encounter). The resulting surface    $\F_\Gamma$ will be referred to as \textit{structural diagram} of $\Gamma$. It has the the same role as the structural graph  for one-particle periodic orbits, see fig.~\ref{encounter_str1}. In particular, $\F_\Gamma$ defines the number of $\Gamma$'s partners.

The symbolic representation $\Aa_{\bar{\Gamma}}$ of a partner orbit $\bar{\Gamma}$ of $\Gamma$ can be recovered from $\Aa_\Gamma$ by exchanging the patches of symbols  outside the encounter regions. The  exact rules for such exchange and  the  number of partner orbits obtained in this way  depend only on the structural diagram $\F_\Gamma$ of the MPO.   Below  we  illustrate this procedure for  periodic orbits with the minimal possible number  (one and two) of 2-encounters  having winding numbers: $w=(0,1)$, $w=(1,0)$ and $w=(0,0)$. As will be  argued in Section~\ref{conclusions}, MPOs with such  encounters   are the  most relevant from the semiclassical point of view.  \\
\begin{figure}[htb]
\begin{center}{
a)\includegraphics[height=3.7cm]{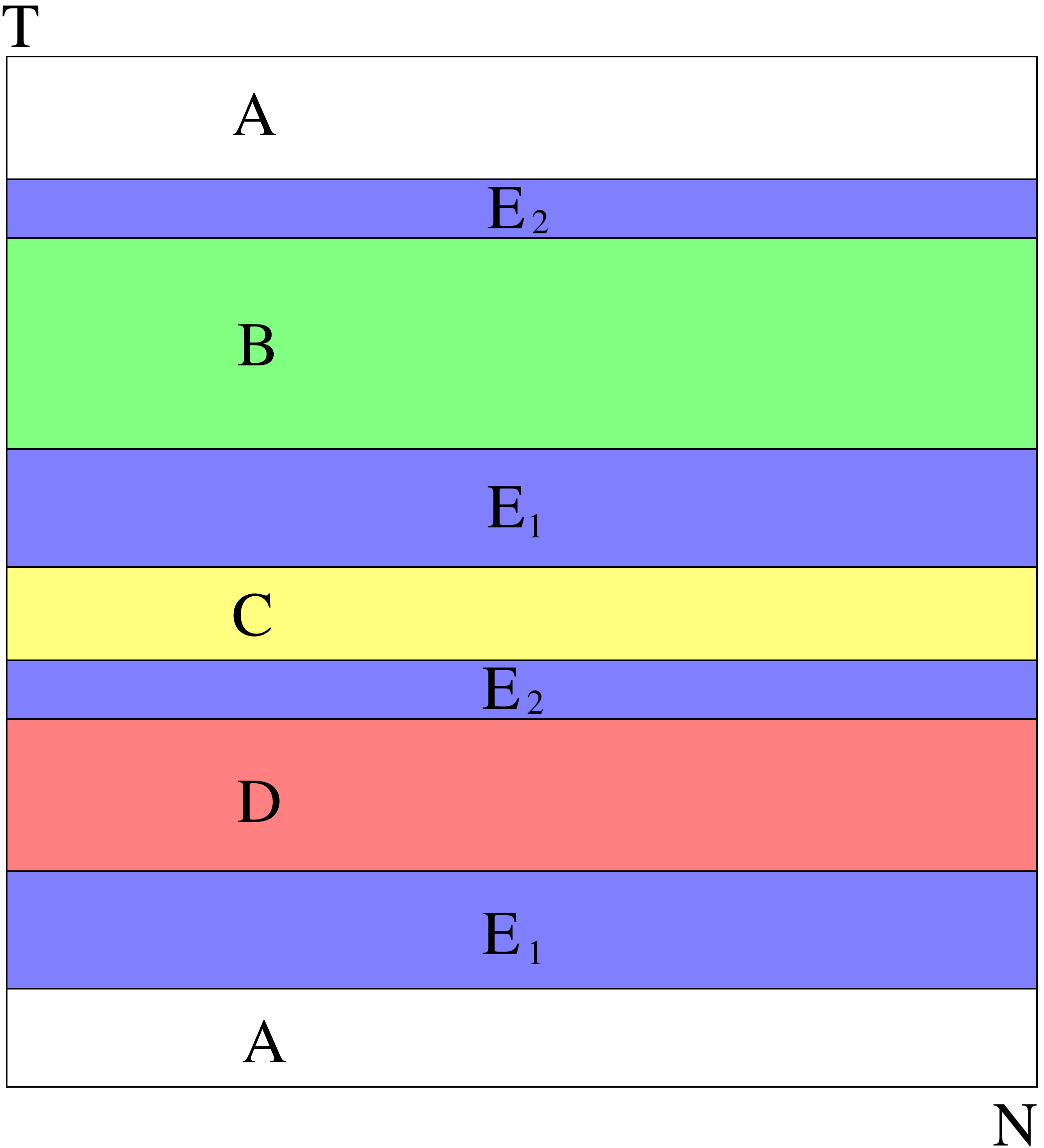}\hskip 0.1cm \includegraphics[height=3.7cm]{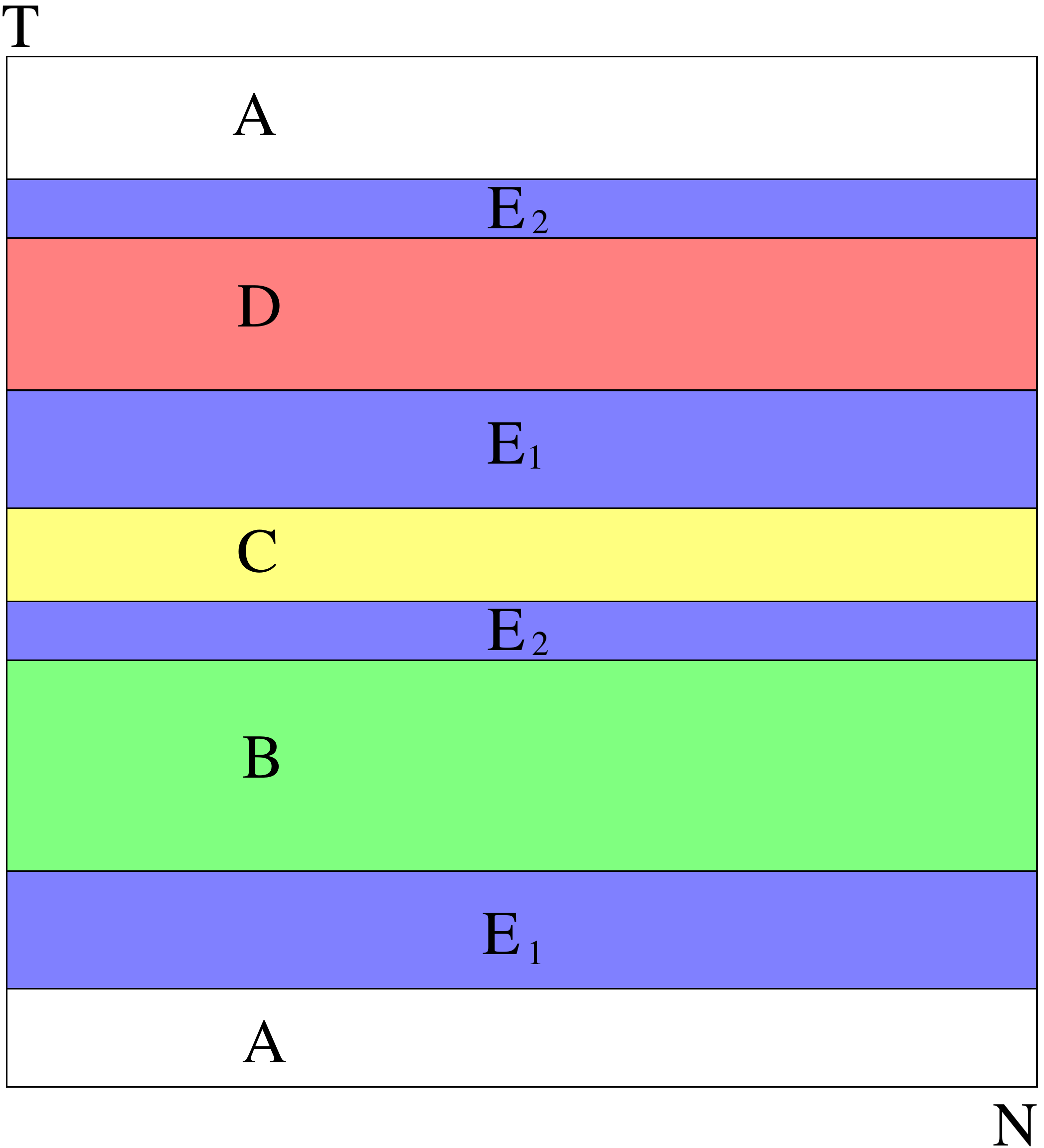}\hskip 0.2cm b)\includegraphics[height=3.7cm]{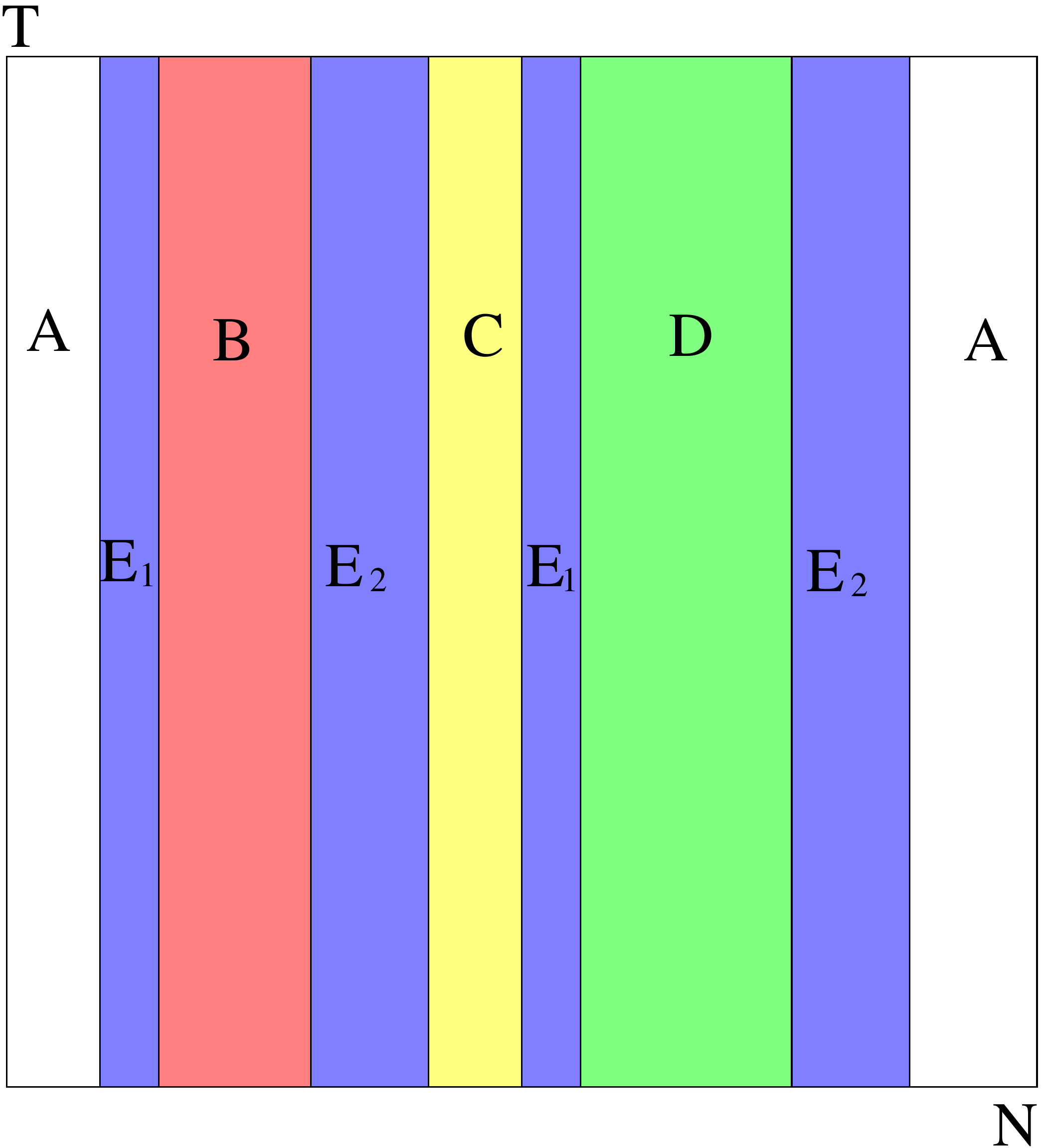}\hskip 0.1cm \includegraphics[height=3.7cm]{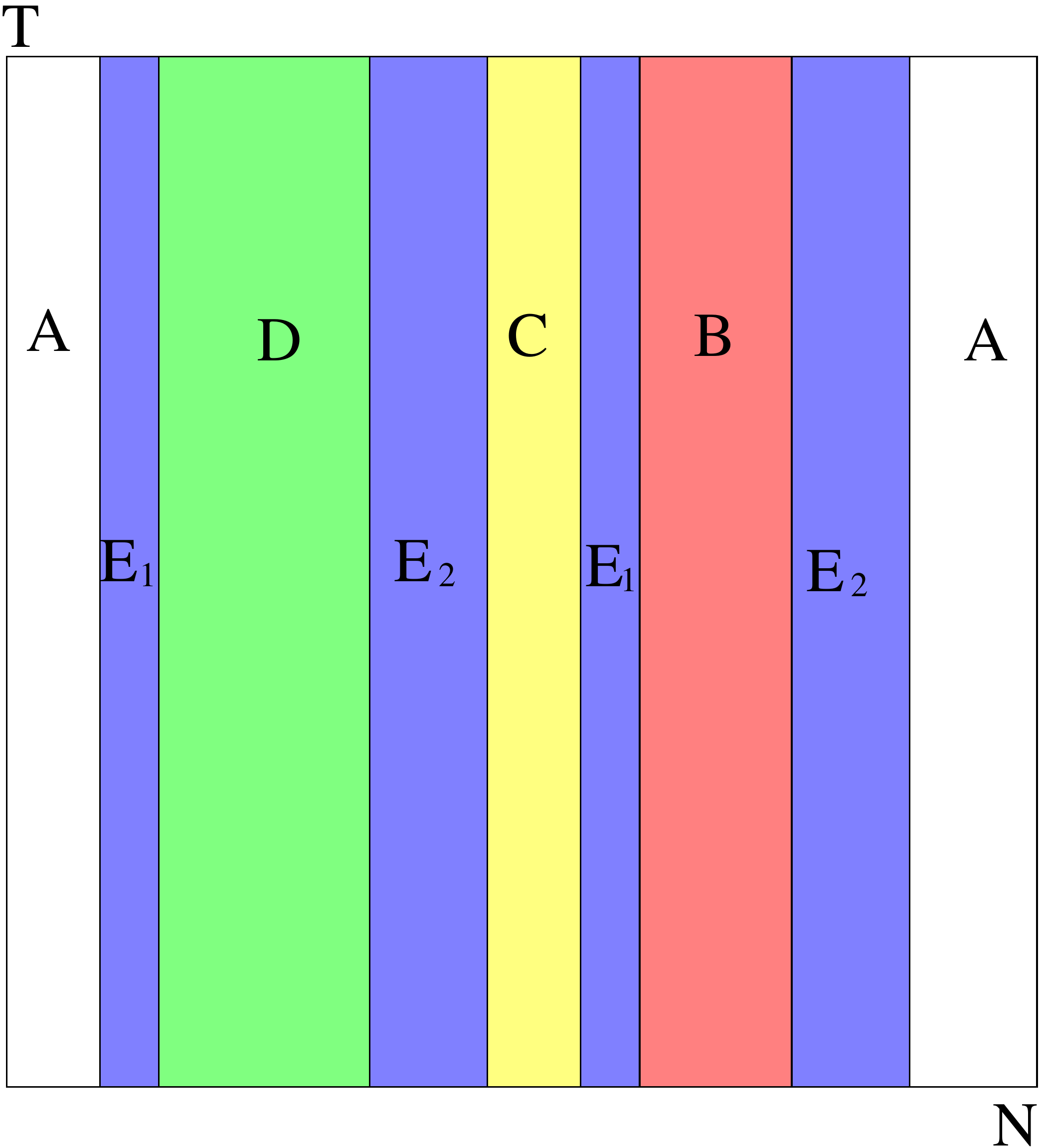}
}\end{center}
\caption{ \small{On the left  figure  (a) are  shown  two partner   MPOs with two  encounters winding around the torus $\Zz$ in the particle direction. The 2D symbolic representation of the second MPO is obtained from the first one by exchanging symbols in the domains $B$ and $D$. In one particle interpretation this is precisely the pair of trajectories shown in fig.~\ref{encounter}b. On the right figure (b) are shown two partner orbits with encounters winding around the torus in the time direction. Their construction is precisely the same as on the left figure with the  switched particle and time directions.    }  }\label{encounter2}
\end{figure}

\begin{figure}[htb]
\begin{center}{
\includegraphics[height=3.7cm]{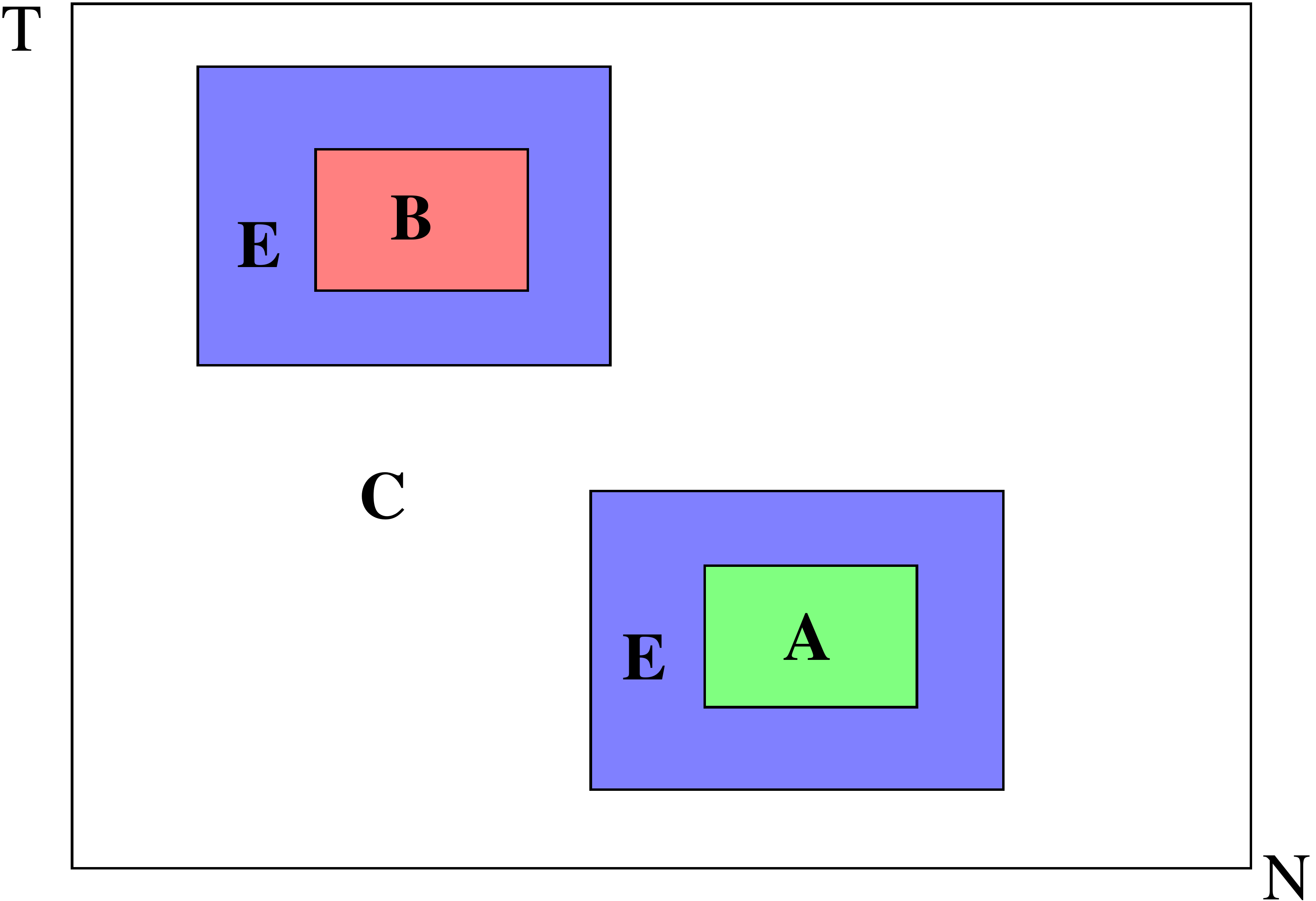}\hskip 0.1cm \includegraphics[height=3.7cm]{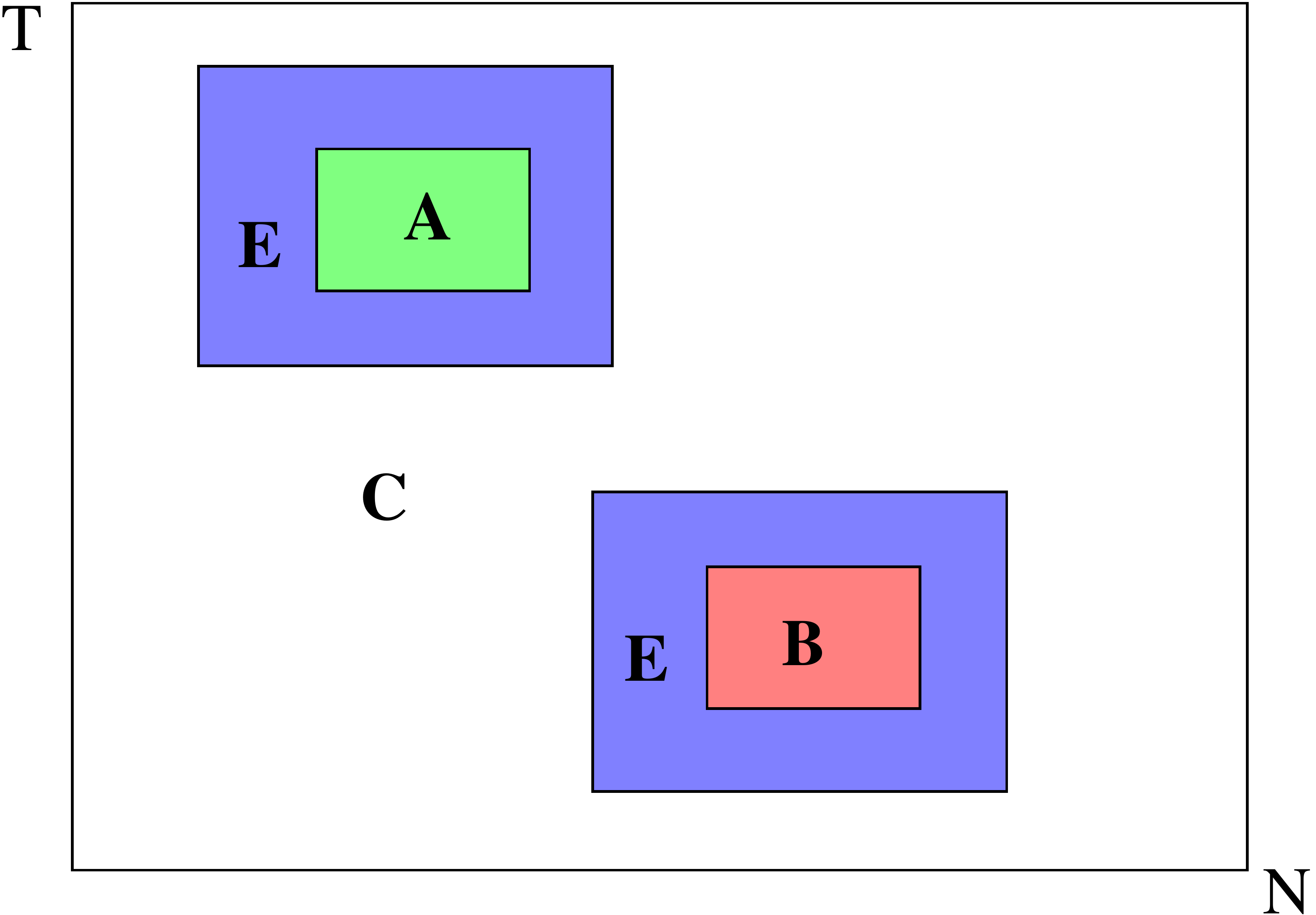} 
}\end{center}
\caption{ \small{ The pair  of  MPOs with one  encounter having zero  winding numbers. The 2D symbolic representation of the second MPO (right) is obtained from the first  one (left) by exchanging symbols in the domains $A$ and $B$.      }  }\label{encounter3}
\end{figure}
\noindent\textbf{a) ``Horizontal''  encounters:  $w=(1, 0)$.} Encounters of this type wind up torus in the particle direction. A simple example is provided by the two horizontal   strips:  
\[ E=\{t_1,\dots t_1+p\}\times \{1,\dots, N\}, \quad E'= \{t_2,\dots t_2+p\}\times \{1,\dots, N\},\]
 $|t_2-t_1|>p\in \mathbb{N} $, with the    symbolic sequences coinciding on these sets:
\[ \mathrm{E}=\psi_{\Gamma}(E)=\psi_{\Gamma}(E').\]  If  $\Gamma$ has  two  such encounters with  the symbolic representations $\mathrm{E}_1$ and  $\mathrm{E}_2$, respectively    the symbolic representation of its partner $\bar{\Gamma}$ can be constructed simply  by using the same procedure as in the single-particle case, see eq.~(\ref{symbrepres}) and fig.~\ref{encounter2}a. 

 \noindent\textbf{b)  ``Vertical'' encounters: $w=(0,1)$.} These encounters wind up around torus in the time direction. For instance the following pair of  vertical strips
\[ E=\{1,\dots, T\}\times\{n_1,\dots n_1+p\}, \quad E'= \{1,\dots, T\}\times\{n_2,\dots n_2+\Delta n\},\]
$|n_2-n_1|>p\in \mathbb{N}$ caring the same symbolic representation defines an encounter with the required winding numbers.
The construction of partner orbits here   is completely analogous   to the previous case up to the  switch between   particle and time directions, see  fig.~\ref{encounter2}b. 

\noindent\textbf{c) ``Annular''  encounters:  $w=(0, 0)$.}  For the construction  of partner orbits in this case, it is sufficient  to have a periodic orbit with just one encounter. Let $A$  and $B$ be two sets of the points from $\Zz$ corresponding to the interiors of the encounter sets $E,E'$, as shown in fig.~\ref{encounter3}.  If $\Aa_\Gamma$ is the symbolic representation of  MPO $\Gamma$, the symbolic representation $\Aa_{\bar{\Gamma}}$ of its partner orbit $\bar{\Gamma}$ is obtained by  exchanging  the symbols   $\mathrm{A}=\psi_{\Gamma}(A)$ with $\mathrm{B}=\psi_{\Gamma}(B)$,  see fig.~\ref{encounter3}.

 \subsection{Comparison between one-particle and $N$-particle partner orbits}

Recall that a MPO might have a partner orbit only if it poses an encounter. 
  What does the existence of  an encounter in $N$-particle systems implies on the level of individual particles?  
Regarding a MPO as single-particle closed orbit in many-dimensional phase space the standard definition of encounter would imply that  all particles  simultaneously reaper at  (approximately) the same positions at least twice during the period of motion. This is indeed the correct interpretation for horizontal type of encounters having winding numbers $(1,0)$.    
However,  encounters with other winding numbers  have a different interpretation  in the many particle systems (\ref{hamiltonian}). For instance,  if the two encounter domains have the shapes of  horizontal strips, see fig.~\ref{encounter2}b,
 only a number of the   particles $n_1,n_1+1, \dots, n_1+p$ 
perform approximately the same motion in $V$ as  the   particles $n_2,n_2+1,\dots, n_2+p$, but they do it for all times between $0$ and $T$. Furthermore, having an encounter of  an  annular type  shown in fig.~\ref{encounter3}  implies that  a group  of  particles retrace positions of another group  of  particles in a different point of time.
 We would like to emphasize that only the horizontal strips which wrap the torus in the particle direction would qualify as encounters in the   one particle interpretation. So the 2D representation of MPOs is absolutely  necessary in order to account for partner orbits with  other  types of encounters. \\

So far the whole discussion  of the correlation mechanism between MPOs has been restricted to a pure conceptual  level. In the next section we introduce a particular  model of the type  (\ref{hamiltonian}).  Within this model we are able to  demonstrate existence of the  partner MPOs  and estimate the differences between their actions.   

\section{Model of coupled  cat maps}
From the technical point of view  it is often convenient to consider  dynamical systems whose time evolution    is given from the start by a discrete  map rather then by a continues flow. One of the best studied and understood examples of systems  with  chaotic dynamics  are provided by    \textit{cat maps}  which are  the hyperbolic automorphisms of the unit 2-torus $V=T^2$ \cite{Berry1980,keating1}. 
To introduce a many particle set up we  couple $N$   cat maps in a linear way, such that the resulting $2N$ dimensional map:
\[\Phi_{\Nn}:V_{\Nn}\to V_{\Nn}, \qquad V_{\Nn}=\underbrace{T^2 \times T^2 \dots \times T^2}_N \]
 is again a hyperbolic automorphism of the unit 2N-torus $V_{\Nn}\cong T^{2N}$, see fig.~\ref{bolzmann}. 
Specifically, the generating function of the map $\Phi_{\Nn}$  is defined as
\begin{multline}
 S(\q_t, \q_{t+1})=d\sum_{n=1}^N q_{n,t} q_{1+(n\!\!\!\!\!\mod{N}),t} +c\sum_{n=1}^N q_{n,t} (q_{n,t+1}+\mq_{n,t+1})+\\
+\frac{a}{2}\sum_{n=1}^N q^2_{n,t}+\frac{b}{2}\sum_{n=1}^N (q_{n,t+1}+\mq_{n,t+1})^2-\mp_{n,t+1}q_{n,t+1} +\V(q_{n,t}), \label{model}
\end{multline}
 where $\q_t=\{q_{n,t}\}^N_{n=1}$, with  $q_{n,t}$ being  the coordinate of $n$-th particle $n=1\dots N$ at the moment of time $t=\mathbb{Z}$, and $\mq_{n,t+1}, \mp_{n,t+1}$ are integer numbers which stand for  winding numbers along the $q$ and $p$ directions of the 2N-torus. The coefficients 
 $ a,b, c, d $ are constants which will be specified below and $\V(q)$ is a smooth periodic function $\V(q+1)=\V(q)$. In the following we refer to the map generated by the action (\ref{model}) as non-perturbed (resp. perturbed) \textit{coupled cat map} in the  case when $\V(q)=0$ (resp. $\V(q)\neq 0$).

The map $\Phi_{\Nn}$ shares the most important properties with the model (\ref{hamiltonian}). In particular, it is defined in such a way that all  particles are coupled in a uniform way   with its nearest  neighbors. Models  of this type are  known
as coupled lattice maps. Their dynamical properties have  been extensively studied in the last three decades \cite{coupledmap}. Of special interest for us,  is the result from \cite{symbol}, where existence of  2D symbolic dynamics  was demonstrated for a particular model of coupled lattice map. In the following we show that 2D symbolic dynamics with the properties described in Section~\ref{SecSymb} can be constructed also for the map $\Phi_{\Nn}$.

\subsection{Dynamics} The equation of motion is generated using 
$p_{n,t}=-\partial S/\partial q_{n,t}$,  $p_{n,t+1}=\partial S/\partial q_{n,t+1}$:
 \begin{align} q_{n,t+1}&=\frac{1}{c}\left(-p_{n,t}-aq_{n,t}\right)- \frac{d}{c}\left(q_{n+1,t}+q_{n-1,t}\right)-\mq_{n,t+1}-\frac{1}{c}\V'(q_{n,t}) \label{eqmotion1}\\ \nonumber
 p_{n,t+1}&=\frac{1}{c}\left(-bp_{n,t}+(c^2-ab)q_{n,t}\right)- \frac{db}{c}\left(q_{n+1,t}+q_{n-1,t}\right)-\mp_{n,t+1}-\frac{b}{c}\V'(q_{n,t}).
\end{align}
Under the condition that $\V(q)=0$ these equations can  also be written in the matrix form:
\[ Z_{t+1}=\B_{\scriptscriptstyle N} Z_{t}\mod 1, \qquad Z_{t}= (q_{1,t},p_{1,t}, \dots  q_{{\scriptscriptstyle{N}},t},p_{{\scriptscriptstyle{N}},t})^T,\]
with $2N\times 2N$  matrix  $\B_{\scriptscriptstyle N}$ given by:
\begin{equation}
 \B_{\scriptscriptstyle N}= \begin{pmatrix}
      A     & B     &   \bm 0  & \dots  &\bm 0 & B\\
      B     & A     &     B      &\dots  & \bm 0 &\bm 0\\
      \bm 0 & B     &    A      &\dots  & \bm 0 &\bm 0\\
      \vdots&\vdots &   \vdots & \ddots &\vdots  &\vdots \\
      \bm 0 & \bm 0 &   \bm 0  & \dots  &A  & B\\
      B     & \bm 0 &   \bm 0  & \dots  &B  & A\\
     \end{pmatrix},
\end{equation}
\[ 
A= -\frac{1}{c}\begin{pmatrix} a& 1\\
ab-c^2 & b
\end{pmatrix}, \qquad 
B= -\frac{1}{c}\begin{pmatrix} d& 0\\
db & 0
\end{pmatrix}.\]
For the map to be continues the elements of the matrix $\B_{\Nn}$ must be integers. To satisfy this requirement  we set $c=-1$ and $a, b, d \in \Z$.    
  
By excluding  the momentum from the  system of  eqs.~(\ref{eqmotion1}) the dynamical equations  for the  time evolution can be  cast into the  Newtonian form:
\begin{align}
 &c(q_{n,t+1}+q_{n,t-1}) +d(q_{n+1,t}+q_{n-1,t})=-(a+b)q_{n,t}+m_{n,t}-\V'(q_{n,t}),  \label{newtoneq}\\
 &m_{n,t}=-b\mq_{n,t}-c\mq_{n,t+1}+\mp_{n,t}.\nonumber 
\end{align}

\subsection{Spectrum of $\B_{\scriptscriptstyle N}$} In order to establish the type of dynamics generated by $\Phi_{\Nn}$ we need to determine the spectrum of the matrix $\B_{\scriptscriptstyle N}$. Since $\B_{\scriptscriptstyle N}$ is a circular matrix which  commutes with the shift operator its spectrum can be found explicitly. After substitution   $\bar{Z}=(\bar{q}_{1},\bar{p}_{1}, \dots, \bar{q}_{{\scriptscriptstyle{N}}},\bar{p}_{{\scriptscriptstyle{N}}})$ into the spectral equation 
\[\lambda\bar{Z}=  \B_{\scriptscriptstyle N}\bar{Z}\]
we obtain the following condition  
\[\gamma \bar{q}_{n}=\bar{q}_{n+1}+\bar{q}_{n-1}, \qquad  d\gamma=-s+\lambda+\frac{1}{\lambda}, \quad s=a+b,\]
which is equivalent to the iterative equation:
 \begin{equation}
\begin{pmatrix} \bar{q}_{n+1}\\ \bar{q}_{n}
\end{pmatrix}
=
  \begin{pmatrix} \gamma& -1\\
1 & 0
\end{pmatrix}
\begin{pmatrix} \bar{q}_{n}\\ \bar{q}_{n-1}
\end{pmatrix}.
 \end{equation}
In order to close this equation the eigenvalue of the matrix above must be of the form $e^{2\pi i k/N}$, $k=1,\dots N$.
This yields to the following quadratic  equation for  eigenvalues of $\B_{\scriptscriptstyle N}$:
\begin{equation}
 \lambda+\lambda^{-1}=s+2d\cos(2\pi k/N), \qquad k=1,\dots N.\label{qudreq}
\end{equation}
Accordingly, $\Phi_{\Nn} $ is fully hyperbolic iff $ |s|>2|d|+2$. In this case all solutions $\lambda_\pm(k)$ of this equation  are paired such that $\lambda_+(k)= \lambda_-^{-1}(k)$ and  $|\lambda_+(k)|>1$ for all $k$. In what follows we will assume that this condition  is always satisfied.

\subsection{Periodic orbits} In this paper we primarily  focus  on the periodic orbits of $\Phi_{\Nn}$. In the case of fully hyperbolic dynamics  the total  number $\mathcal{N}(T)$ of MPO with period $T$  can be  easily obtained by the following formula:
\begin{equation}
  \mathcal{N}(T)=\left|\det(I-\B_{\scriptscriptstyle N}^T)\right|\approx\prod_{k=1}^N \lambda_+^T(k),\label{numberpo}
 \end{equation}
where $\lambda_+(k)$ are solutions of eq.~(\ref{qudreq}) with the largest absolute value.
It is straightforward to see that  $\mathcal{N}(T)$ grows exponentially both with $N$ and $T$.

 The action of   periodic orbits  can be, in principle, obtained by summing up all the terms  in eq.~(\ref{model}) along the trajectory. Note, however that the generating function  $S(\q_t,\q_{t+1})$   also contains half-integer factors $b(\mq_{n,t})^2/2$ which are in fact irrelevant, as far as, the  dynamical equations  of concern. For convenience we  omit these factors from the definition of    actions. Such omission also does not affect  the semiclassical theory, where only a fractional part of the actions plays a role, see e.g., \cite{keating1, saraceno}. Accordingly if $\Gamma$ is a MPO with a period $T$,  its action is defined by:
\begin{multline}
 S_{\Gamma}=\sum_{t=1}^T\sum_{n=1}^N \left(dq_{n,t} q_{1+(n\!\!\!\!\!\mod{N}),t} +cq_{n,t} q_{n,t+1}
+\frac{a}{2} q^2_{n,t}+\frac{b}{2}q_{n,t+1}^2+\V(q_{n,t})\right)+\\
+\sum_{t=1}^T\sum_{n=1}^N \left(cq_{n,t} \mq_{n,t+1}
+bq_{n,t+1}\mq_{n,t+1}-\mp_{n,t+1}q_{n,t+1} \right), \label{actionaction}
\end{multline}
where $q_{n,t}, m^{(q,p)}_{n,t}$, are coordinates and winding numbers along $\Gamma$. 
Using eq.~(\ref{newtoneq}) this expression can be further simplified leading to:
 \begin{equation}
S_{\Gamma}=-\frac{1}{2}\sum_{t=1}^T\sum_{n=1}^N  m_{n,t}q_{n,t}+ q_{n,t}\V'(q_{n,t})-2\V(q_{n,t}) \label{actionaction1}
\end{equation}
{\rem It worth mentioning that in most of the works on cat maps  the actions of  periodic orbits  are defined in a somewhat different way, see e.g., \cite{keating1}. Namely, if a periodic
orbit $\Gamma$ has a period $T$  one looks for a fixed point $(\q,\p)$ of the map $\Phi_{\Nn}^T$,  and then defines the  action of $\Gamma$ as $\tilde{S}_\Gamma(\q)=S(\q,\q)$, where the generating  function $S$ is given by eq.~(\ref{model}) with the parameters $a,b,c,d$ defined by the map $\Phi_{\Nn}^T$, rather then by $\Phi_{\Nn}$. It can be shown that the difference between two actions $\tilde{S}_\Gamma$ and $S_{\Gamma}$ is always an integer. Since in quantum mechanics only fractional part of actions are of a relevance  one can use both definition equivalently for semiclassical calculations. We, however believe that (\ref{actionaction1}) is a more natural one from a classical point of view as its properties  resemble ones of periodic orbit actions in generic Hamiltonian systems.   Note, that in contrast to $S_{\Gamma}$,   $\tilde{S}_\Gamma(\q)$ depends   
on the choice of the initial point $\q$ at $\Gamma$ i.e., $\tilde{S}_\Gamma(\q)\neq\tilde{S}_\Gamma(\Phi_{\Nn}\cdot\q) $ and  it grows exponentially with $T$.}

\section{Time-particle duality}

 Remarkably, for $c=d$ the equation  (\ref{newtoneq}) becomes symmetric under the exchange of times and particle numbers: $n \leftrightarrow t$.  This immediately leads to the following duality relationship between periodic orbits of the system.

{\lemma{ Let $\Phi_k$  be a sequence of maps  (\ref{model}) with some fixed parameters $a,b,c=d$ and varied number of particles $k=1,2,\dots$.  For each MPO  $\Gamma$ of $\Phi_{\Nn}$ with a  period  $T$ there exists corresponding MPO $\Gamma'$ of the map $\Phi_{\Tt}$ with the period $N$, such that both trajectories run through the same set of points  in the configuration space. Furthermore, the actions of two orbits coincide $S_\Gamma=S_{\Gamma'}$.}\label{lemma1} }\\

\textit{Proof:} 
The proof is straightforward. Given a MPO $\Gamma$  the set of points  $\{q_{n,t}, m_{n,t}| n=1,\dots N; t=1,\dots T\}$ traversed by $\Gamma$   in the configuration space must satisfy eq.~(\ref{newtoneq}).  By defining $q_{n,t}=q'_{t,n}$, $m_{n,t}=m'_{t,n}$   one can easily see that the new set  $\{q'_{n,t}, m'_{n,t}| n=1,\dots T; t=1,\dots N\}$ satisfies eq.~(\ref{newtoneq}), as well. It remains to notice that the set $\{q'_{n,t}, m'_{n,t}\}$ uniquely defines the trajectory $\Gamma'$ of the period $N$ for the map $\Phi_{\Tt}$. The equality between actions then follows immediately from eq.~(\ref{actionaction}).\hfill $\square$

{\rems 1) The above lemma holds also for the case of  perturbed cat maps when $\V(q)\neq0$.  2) Note that although the two trajectories $\Gamma$, $\Gamma'$ traverse the same points of the configuration space, they do not have the same set of momenta. 2) For $N=T$ the set of  periodic trajectories becomes self-dual. This means that each MPO $\Gamma$ either has   a pair MPO $\Gamma'$ satisfying   $q_{n,t}= q'_{t,n}$   or  satisfies the self-dual constraint $q_{n,t}= q_{t,n}$ for each $t$ and $n$. The first option would actually imply that  $\Gamma$ and $\Gamma'$ traverse through exactly  the same points of the configuration space. We have not observed such pairs in our numerical simulations. } \\

\noindent An immediate consequence of this lemma is the following connection between numbers of periodic orbits.
{\lemma The number of MPOs with a period  $T$ of the map $\Phi_{\Nn}$ is the same as the number of  MPOs with the period  $N$ of the map $\Phi_{\Tt}$. Equivalently:
 \begin{equation}
  (-1)^N\det(I-\B_{\scriptscriptstyle N}^T)= (-1)^T\det(I-\B_{\scriptscriptstyle T} ^N).\label{numberpo2}
 \end{equation}
}
\textit{Proof:} Straightforwardly follows from Lemma~\ref{lemma1} and eq.~(\ref{numberpo}).\hfill $\square$\\

\noindent Since the spectrum of matrix $\B_{\Nn}$  is known  explicitly,  eq.~(\ref{numberpo2}) allows to extract   a couple of interesting mathematical identities.  

{\col Let $\cosh\beta_k=s/2-\cos(2\pi k/N), \cosh\tilde{\beta}_k=s/2-\cos(2\pi k/T)$, with $ T, N$ being integers and $s\geq 4$ then:
\begin{equation}
 \prod_{k=1}^N 4\sinh^2\left(\frac{T\beta_k}{2}\right)=\prod_{m=1}^T 4\sinh^2\left(\frac{N\tilde{\beta}_m}{2}\right).
\end{equation}
}
After  taking the limit $s\to 0$ this equation leads to:
\begin{equation}
 T^2 \prod_{k=1}^{N-1}4\sin^2\left(\frac{\pi k T}{N}\right)=N^2 \prod_{m=1}^{T-1}4\sin^2\left(\frac{\pi m N}{T}\right).
\end{equation} 
In particular, for $T=1$ we obtain the well known identity:
\[ \prod_{k=1}^{N-1}\sin\left(\frac{\pi k }{N}\right)=\frac{N}{(-2)^{N-1}}.\]

\subsection{Duality between partner orbits} The  duality relationship has very important implication on the  correlation mechanism between periodic orbits of the system.  In particular, it implies that       pairs of partner orbits of a period $T$ with the  encounter structure given by winding numbers  $(w_1,w_2)$ for  the $N$-particle system are  in one-to-one correspondence with the      pairs of partner orbits of the period $N$ with the  encounter structure $(w_2,w_1)$ for  the $T$-particle system. For a small number of particles $N$ and large times $T$ it is natural to expect that most of the partner orbits poses  $(0,1)$ type of encounters. The duality relationship shows that the opposite regime when $T$ is large and  $N$ is small is dominated by the partner orbits with inverse winding numbers $(0,1)$. In a sense these  two regimes  are dual to each other. For instance, the number of partner orbits and  their action differences  must coincide. Although the exact duality relationship i.e.,
 Lemma~\ref{lemma1} 
holds only under specific condition $c=d$, we believe that in its weaker sense, as connection between correlation mechanism in the above regimes   it makes sense for general parameters $a,b,c,d$.

\section{ Many particle partner orbits}
The main goal of the present section is an  explicit construction of partner MPO's with close actions. Here we restrict our attention to the case of the non-perturbed maps $\Phi_N$ with   $\V=0$.  A remarkable property  of the non-perturbed cat maps  is that their  periodic orbits traverse the  points of the phase space with  rational coordinates and  can be found  exactly rather then as an approximation.  Furthermore, their actions are given by rational numbers and can be explicitly  evaluated,  as well. 

\subsection{ Symbolic dynamics}
As it was explained in the section~(\ref{secgoals}),  a convenient representation  of partner MPO's can be achieved by  means of  2D symbolic dynamics with a finite alphabet.
 To introduce the  symbolic representation  we use as the alphabet the  set of winding numbers $m^{p}_{n,t}, m^{q}_{n,t}$  entering the dynamical equations of motions (\ref{eqmotion1}). For each moment of time we  define the  vector
\[\m_{n,t}= \begin{pmatrix}
      m^{q}_{n,t}\\  m^{p}_{n,t}  \end{pmatrix}, \qquad \Mmm_t=\begin{pmatrix}
      \m_{1,t}& \m_{2,t}  & \dots & \m_{N,t}\end{pmatrix}^T.
\]
Note that the  whole set of these  vectors from $t=1$ up to $t=T$  uniquely  defines the corresponding periodic orbits.
Indeed, the global winding number vector $\M_\Gamma$ of a periodic trajectory $\Gamma$ can be restored from $\M_t$ by:
  \begin{equation}
   \M_\Gamma= \sum_{t=1}^{T}\B_N^{T-t}\Mmm_t. \label{winding1}
  \end{equation}
This vector can be  used then in order to restore the initial conditions for the MPO:
\begin{equation}
 Z_0=(\B_N^T-I)^{-1} \M_\Gamma.\label{winding3}
\end{equation}
As a result  the following matrix of winding numbers, 

\begin{equation}
 \Mm_{\Gamma}= \begin{pmatrix}
      \m_{1,1}   &  \m_{2,1}     & \dots   & \m_{N,1}\\
      \m_{1,2}   &   \m_{2,2}    &\dots    & \m_{N,2}\\
      \vdots    &   \vdots     & \ddots  &\vdots \\
      \m_{1,T}   &  \m_{2,T}     & \dots   & \m_{N,T}\\
     \end{pmatrix},
\end{equation}
can be seen as the unique symbolic representation of  $\Gamma$.

Importantly,   because of the local nature of the interactions the integer vectors  $\m_{n,t}$ can take only a finite number of  possible values. Accordingly we obtain 2D-symbolic representation of periodic orbits with a small  alphabet, whose size is independent of $N$.  

\subsection{Construction of partner MPOs \label{protocol}}

\begin{figure}[htb]
\begin{center}{
\includegraphics[height=2.8cm]{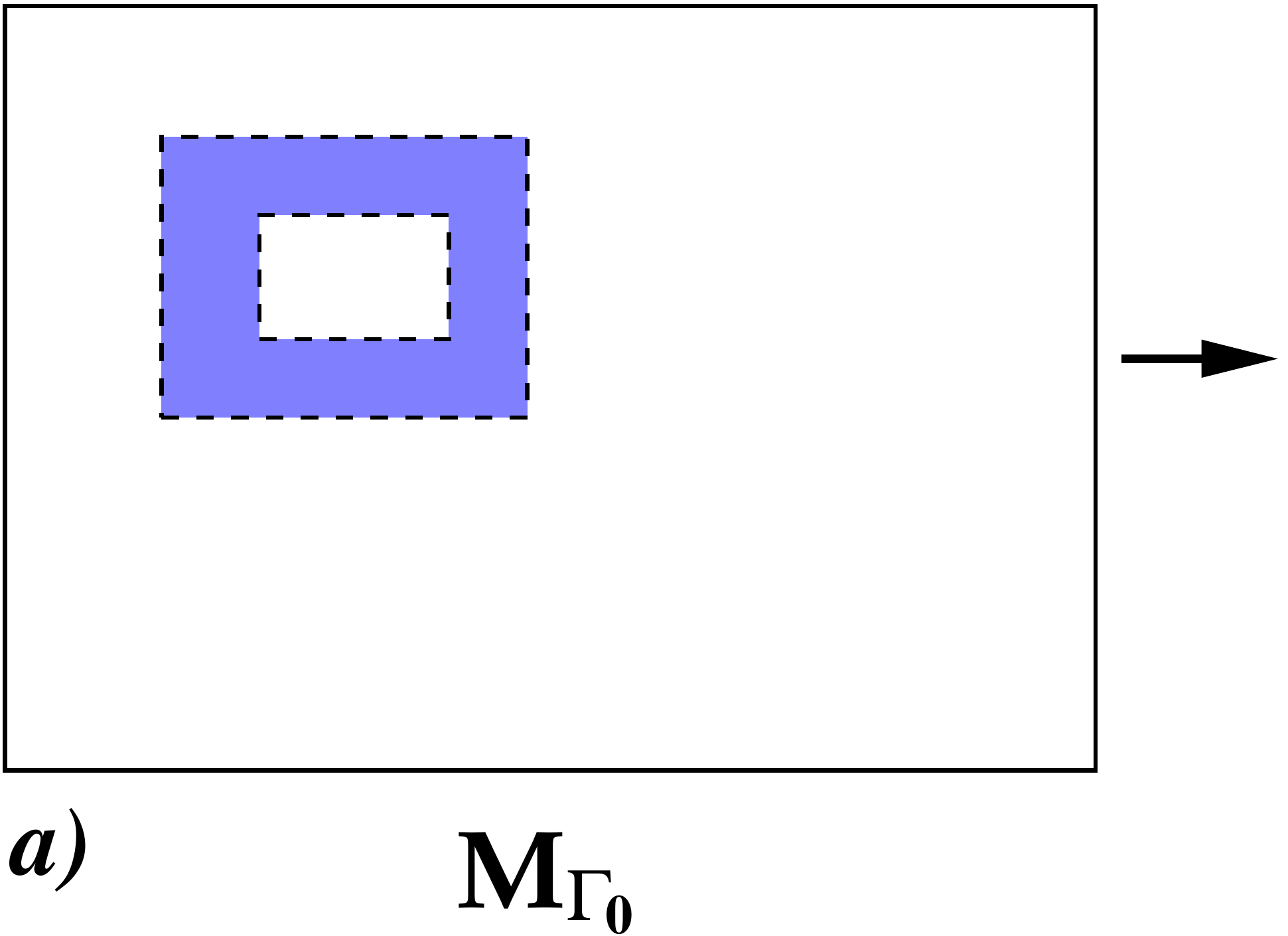}\hskip 0.0cm \includegraphics[height=2.8cm]{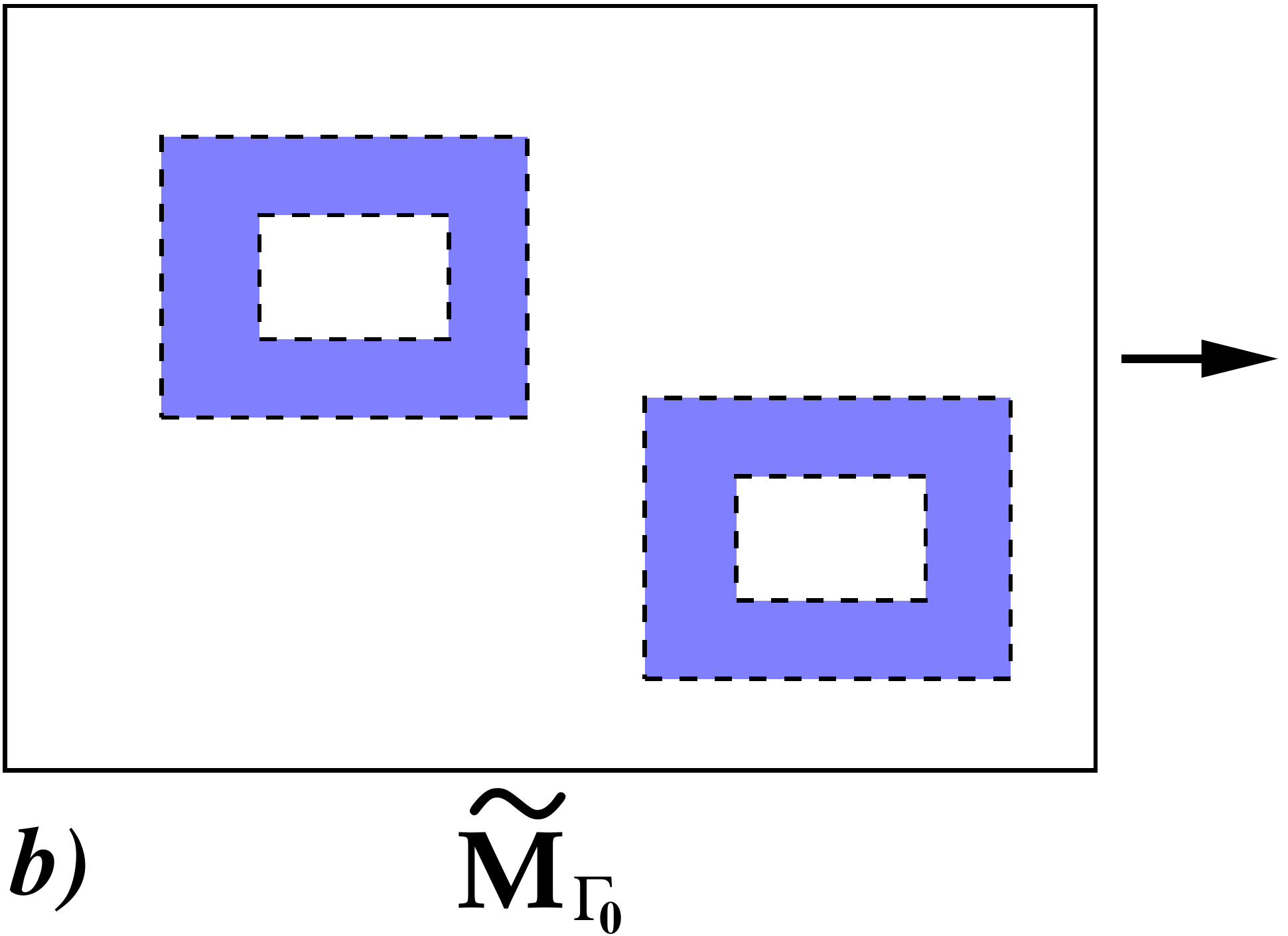}\hskip 0.0cm \includegraphics[height=2.8cm]{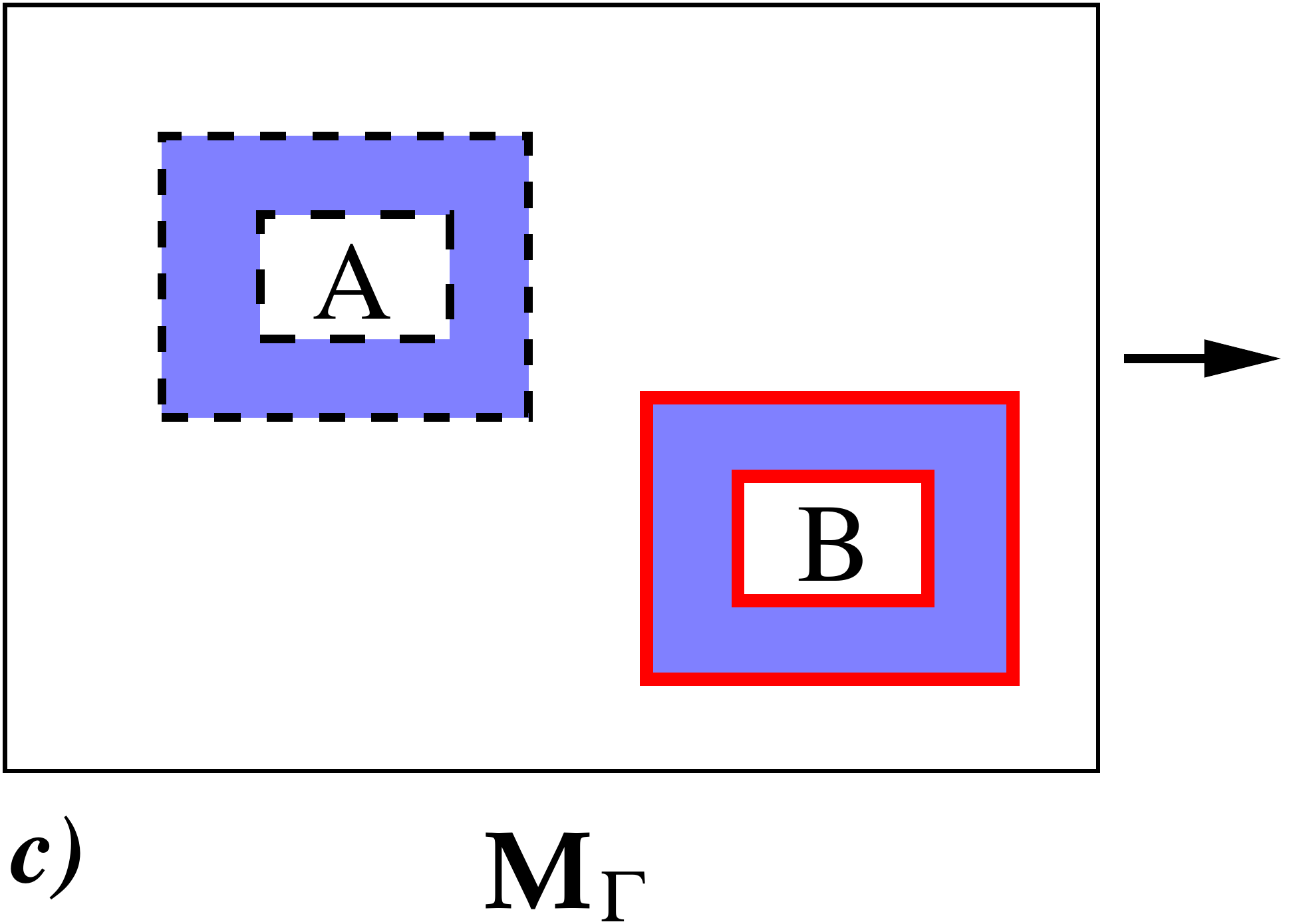}\hskip 0.0cm \includegraphics[height=2.8cm]{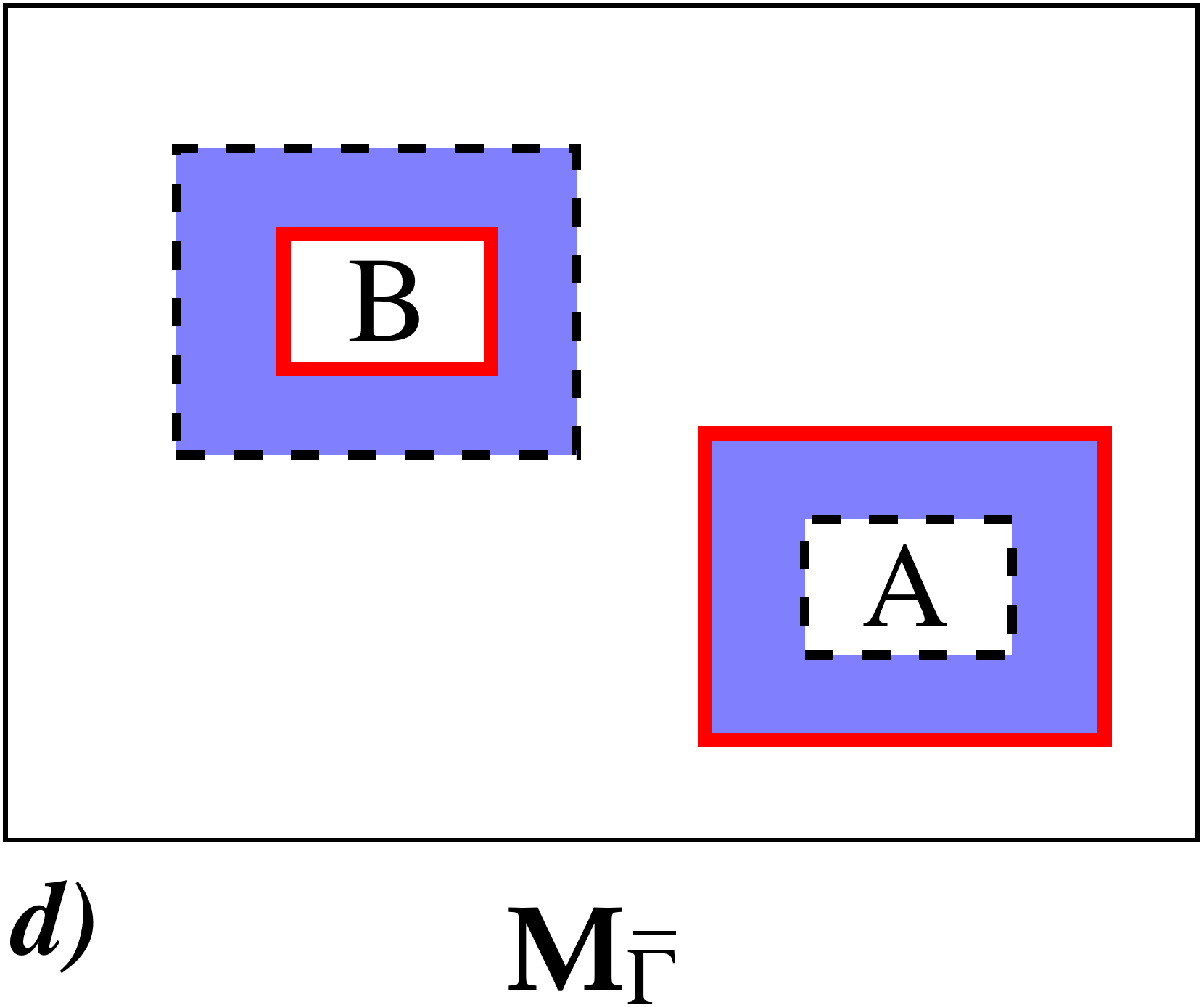}
}\end{center}
\caption{ \small{4-Step  construction   of partner  MPOs. }  }\label{superconstruction}
\end{figure}

Our construction of partner MPO's with the prescribed encounter structure can be divided into few steps which are described below and schematically summarized in the fig.~\ref{superconstruction}.\

\textit{Step 1 (fig.~\ref{superconstruction}a).} As the first step,   we generate a random periodic trajectory $\Gamma_0$ and its symbolic representation $\Mm_{\Gamma_0}$ by means of the following procedure. Let $\M_0$  be an arbitrary vector of integers. Then  the initial conditions   for the corresponding  MPO can be  obtained by
\begin{equation}
 Z_0=(\B_N^T-I)^{-1} \M_0\mod \bm 1,\label{winding2}
\end{equation}
where $Z_0$ is the $2N$ dimensional vector of coordinates and momenta   at the initial moment of time. The rest part of the orbit is
obtained by iterative action of the matrix $\B_N$ on the vector $Z_0$.
Note, that since  the entries of the matrix $\B_N$ are integers,
 the coordinates of the MPO are rational numbers. This property of
the map allows to calculate the orbits explicitly avoiding any  numerical approximation. 
\

\textit{Step 2 (fig.~\ref{superconstruction}b).} 
Recall that a
partner orbit for any MPO $\Gamma$ might exist only if it has (at least)
one encounter. In other words certain pattern of symbols in symbolic
representation of $\Gamma$ must reaper for a number of times.
Naturally, for small $N$ and $T$ a generic MPO   does not satisfy this property and the
purpose of this step is to prepare such an orbit. For that we  choose
some region of a given shape in the original $\Mm_{\Gamma_0}$, which
is then  copied and pasted into another location of $\Mm_{\Gamma_0}$.
The resulting sequence of symbols  $\widetilde{\Mm}_{\Gamma_0}$ has
the correct encounter structure, but, in general,  does not correspond
to  any real MPO.\

\textit{Step 3 (fig.~\ref{superconstruction}c).}  In order to generate 2D sequence  corresponding to a valid periodic orbit  we now use  $\widetilde{\Mm}_{\Gamma_0}$ as an initial data for eq.~(\ref{winding1}). More specifically, let   $\widetilde{\Mmm}_t:=\widetilde{\Mm}_{\Gamma_0}(t)$ be the $t$-th column of the  matrix $\widetilde{\Mm}_{\Gamma_0}$. We define the integer  vector:
\begin{equation}
\widetilde{\M}_0= \sum_{t=1}^{T}\B_N^{T-t}\widetilde{\Mmm}_t, \label{validmonodromy}
\end{equation}
  which is then used as the input for the right hand side  of eq.~(\ref{winding2}) in order to generate the corresponding  MPO.  Denote this periodic orbit  as $\Gamma$ and  its  2D symbolic representation as  $\Mm_{\Gamma}$. The crucial observation is that  $\Mm_{\Gamma}$ and 
$\widetilde{\Mm}_{\Gamma_0}$ differ only locally at the places of encounter boundaries. As a result, we obtain the MPO ${\Gamma}$ with   the desired encounter structure.\

\textit{Step 4 (fig.~\ref{superconstruction}d).} At this stage we can construct the partner orbit $\bar{\Gamma}$ of ${\Gamma}$ by rearranging the symbols $\Mm_{{\Gamma}}$ outside of the encounter regions, into a new 2D sequence  $\Mm_{\bar{\Gamma}}$  as has been explained in   Section~\ref{secgoals}. Note that this new sequence of symbols is in general symbolic representation of a real MPO $\bar{\Gamma}$ which can be straightforwardly recovered from $\Mm_{\bar{\Gamma}}$ by eqs.~(\ref{winding1},\ref{winding3}). By construction $\bar{\Gamma}$ and ${\Gamma}$ are partner MPO's  traversing approximately the same points of the phase space.\\

{\rem In principle it might happen that  the 2D symbolic sequence obtained from $\Mm_{{\Gamma}}$ by exchange of the sequences $A$ and $B$ (see fig.~\ref{superconstruction})  differs from  2D symbolic sequence  $\Mm_{\bar{\Gamma}}$ of the partner orbit $\bar{\Gamma}$ by a small number of symbols. This should happen whenever one of the two points ${\Gamma}_{t,n}$,  $\bar{\Gamma}_{t',n'}$  closely approaching  each other in the phase space turn out to be encoded by two locally different symbolic subsequences. We however have not observed such a phenomena in our numerical constructions of partner  MPO. \textbf{think again} } 
 
{\rem One can expect that  the 2D symbolic sequence obtained from
$\Mm_{{\Gamma}}$ by exchange of the sequences $A$ and $B$ (see
fig.~\ref{superconstruction})  sometimes  differs from the 2D symbolic sequence
$\Mm_{\bar{\Gamma}}$ of the partner orbit $\bar{\Gamma}$ by a small
number of symbols. It might  happen, for instance, if some two points
${\Gamma}_{t,n}$ and $\bar{\Gamma}_{t',n'}$ are closely approaching
each other in the phase space, but encoded by two different symbols.
In our numerical calculations performed for the non-perturbed maps,
see Appendix~A, we have not found evidence of such phenomena.}\\

\noindent\textbf{Example.} In the  Appendix~A  we illustrate the above
four step procedure by particular calculation of the partner orbits for the case
of non-perturbed coupled cat maps.

\subsection{Action differences} Keeping in mind applications to quantum mechanics it is of great  importance to evaluate the  action differences
$\Delta S= S_{\Gamma} -S_{\bar{\Gamma}}$ between partner orbits $\Gamma, \bar{\Gamma}$.  Indeed, in a semiclassical theory  $\Delta S$ provide   the weights with which periodic orbits of a given  encounter structure  contribute  to spectral correlations \cite{haake}. 
\begin{figure}[htb]
\begin{center}{
a)\includegraphics[height=3.5cm]{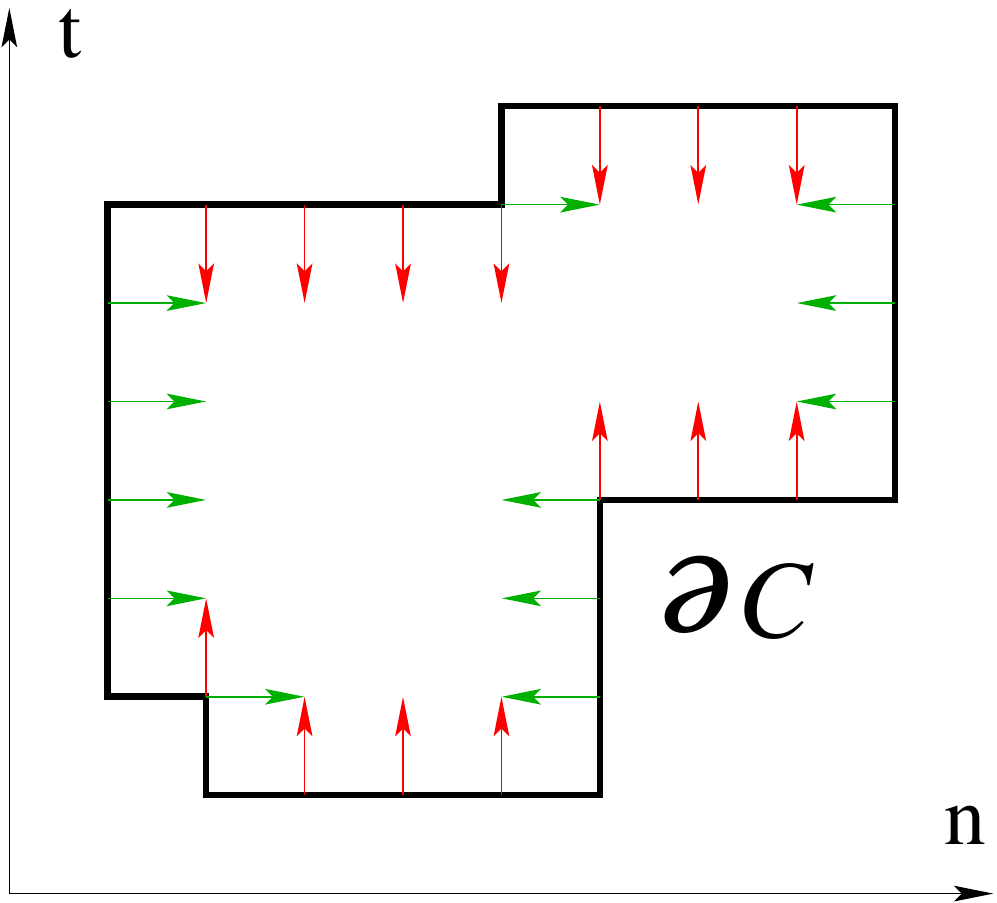}\hskip 0.5cm b)\includegraphics[height=3.5cm]{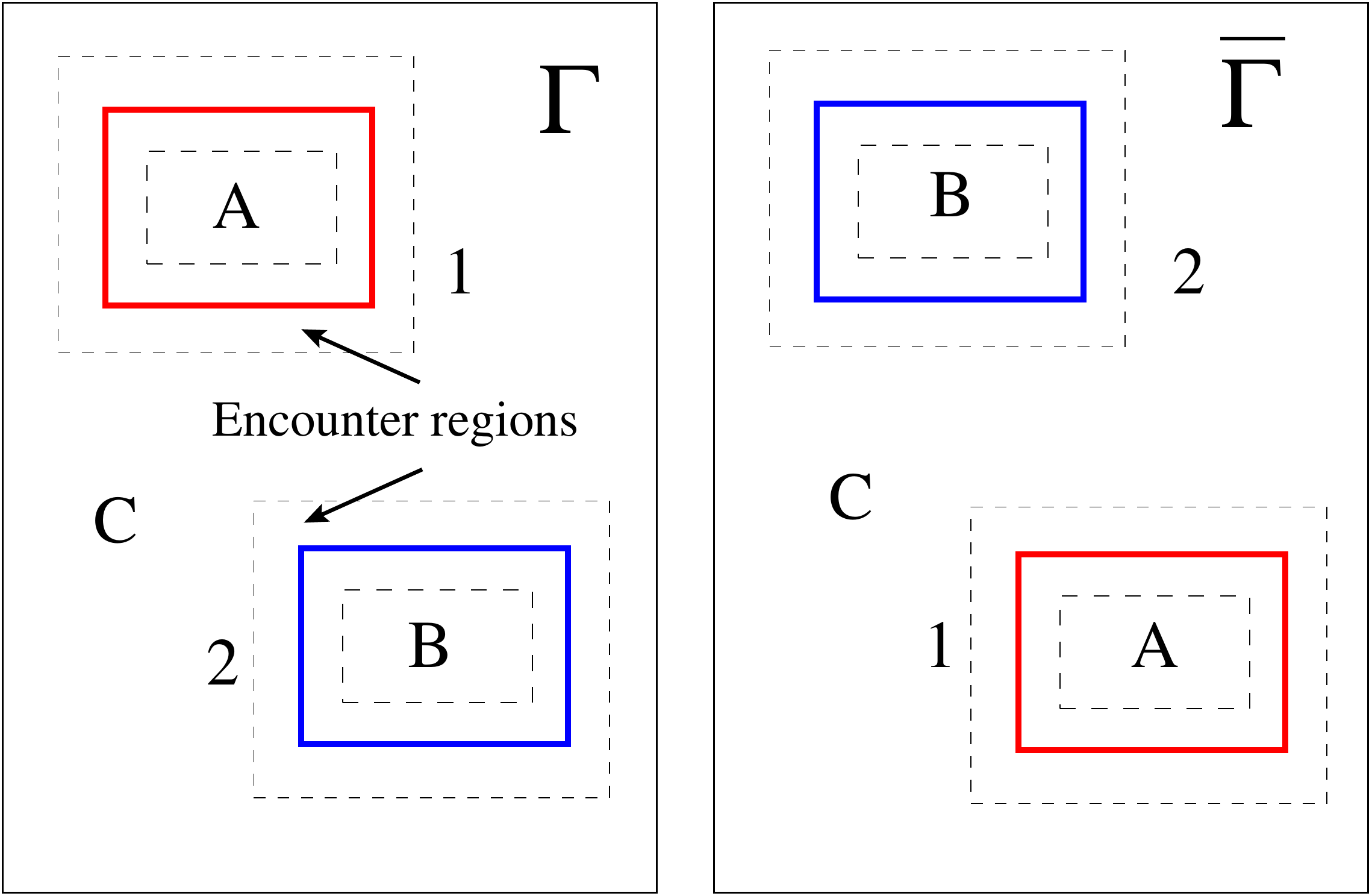}\hskip 0.5cm c)\includegraphics[height=3.5cm]{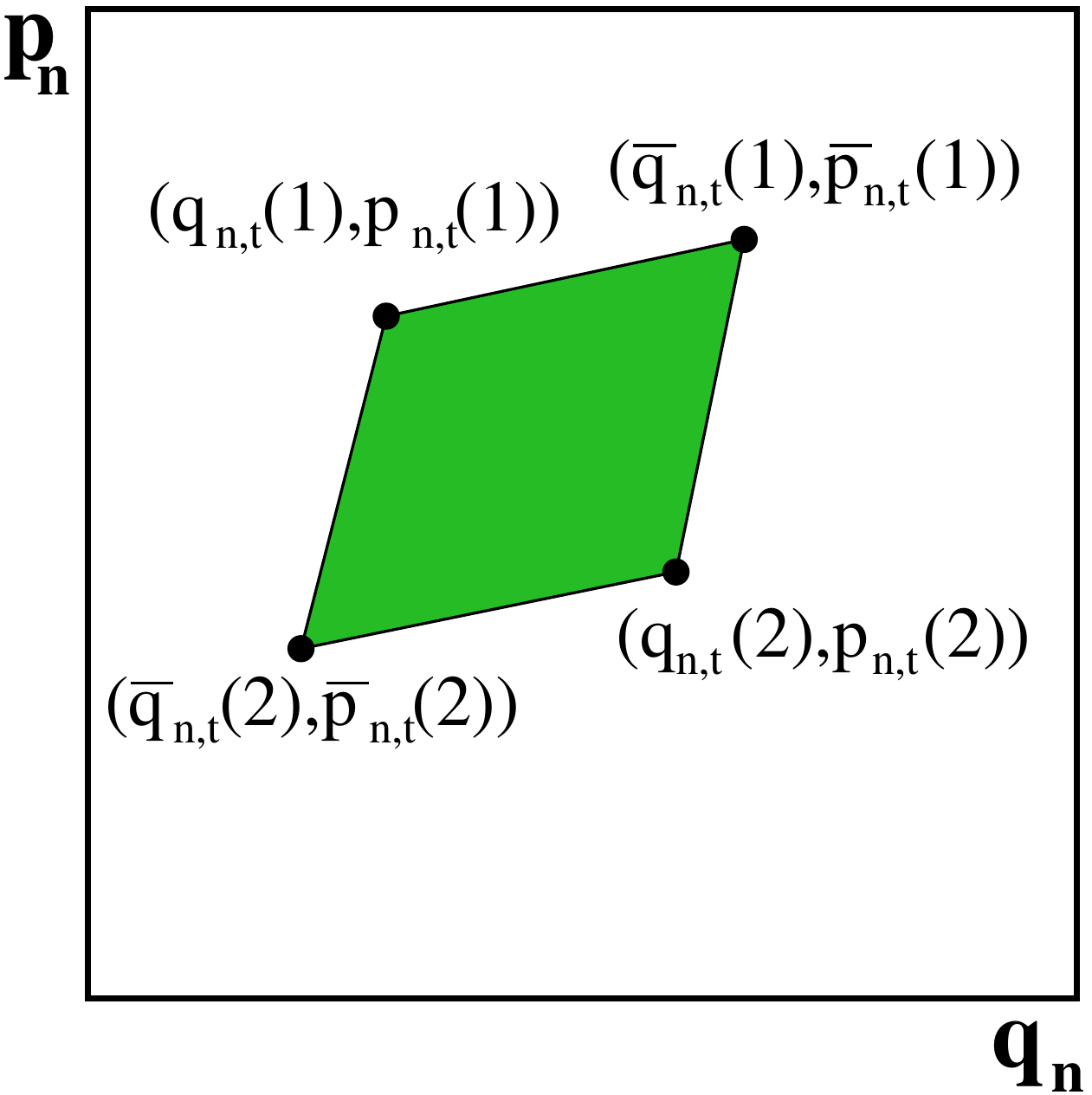}
}\end{center}
\caption{ \small{The  figure (b) in the middle schematically shows    two partner MPO together with the encounter domain and a contour inside. On the left figure (a)  is shown  the magnified caricature  of the contour $\partial C$ where the  difference between actions  is calculated. The right figure (c)  shows the region in the single-particle phase space whose area  contributes to eq.~(\ref{ActionDif})}  }\label{construction}
\end{figure}
It is known that in the case of single-particle chaotic systems the differences between actions of partner orbits accumulate at the encounters, where the distance between  points is maximal. Below we show that a similar result holds  for many particle systems, as well. 

 By setting in eq.~(\ref{actionaction1}) $\V=0$ we obtain for the total action  of  $\Gamma$
\begin{equation}
 S_\Gamma=\frac{1}{2}\sum_{(n,t)\in \Zz}q_{n,t}m_{n,t}.
\end{equation}
Let  $\Gamma$,  $\bar{\Gamma}$  be two different MPO such that  $\{q_{n,t}, m_{n,t}|(n,t)\in\Zz\}$, $\{\bar{q}_{n,t}, \bar{m}_{n,t}|(n,t)\in\Zz\}$  are  the corresponding solutions of eq.~(\ref{newtoneq}). Let us calculate  the difference between their  actions   accumulated in a certain region $C$ of $\Zz$:
  \begin{equation}
   S^{(C)}_\Gamma=\frac{1}{2}\sum_{(n,t)\in C}q_{n,t}m_{n,t}, \qquad S^{(C)}_{\bar{\Gamma}}=\frac{1}{2}\sum_{(n,t)\in C}\bar{q}_{n,t} \bar{m}_{n,t}.
  \end{equation}
We will assume that inside of the domain $C$ the two MPO are close to each other and the corresponding winding numbers coincide i.e.,  $ m_{n,t}=\bar{m}_{n,t}$ for $(n,t)\in C$. 
Denote by  $\partial C$  the boundary of $C$. Note that if  $C$ is a multiconnected domain  then $\partial C$ is  composed of   a  number of disjoint  components.  
Multiplying both sides of eq.~(\ref{newtoneq}) for $\Gamma$   by the  coordinates $\bar{q}_{n,t}$, and similarly eq.~(\ref{newtoneq}) for $\bar{\Gamma}$   by the coordinates $q_{t,n}$ we obtain two equations containing  bilinear forms in $\bar{q}_{n,t}, q_{n,t}$. After subtraction of these  equations we obtain:
  \begin{multline}
  2( S^{(C)}_\Gamma-S^{(C)}_{\bar{\Gamma}})=\sum_{(n,t)\in \partial C_\uparrow} q_{n,t}\bar{q}_{n,t+1}-\bar{q}_{n,t}q_{n,t+1}+\sum_{(n,t)\in \partial C_\downarrow} q_{n,t}\bar{q}_{n,t-1}-\bar{q}_{n,t}q_{n,t-1}\\
+\sum_{(n,t)\in \partial C_\rightarrow} q_{n,t}\bar{q}_{n+1,t}-\bar{q}_{n,t}q_{n+1,t}+\sum_{(n,t)\in \partial C_\leftarrow} q_{n,t}\bar{q}_{n-1,t}-\bar{q}_{n,t}q_{n-1,t}, \label{diference1}
  \end{multline}
where $\partial C_\uparrow,\partial C_\downarrow$ and $\partial C_\rightarrow, \partial C_\leftarrow$ are pieces of the boundary going in the vertical  and  horizontal directions respectively, see fig.~(\ref{construction}a). Note that the points $(n,t)$ at the corners of the domain $C$ belong to the both sets simultaneously. Remarkably the right hand side of the above equation contains  only the sum over  the boundaries of  $C$.

 Let us now evaluate the difference between total actions of $\Gamma$ and $\bar{\Gamma}$. For simplicity of exposition we will assume that there is only one encounter of the order two. In such a case the whole set $\Zz$ can be divided into three regions $A, B, C$  separated by the boundary  $\partial C = \partial A \cup \partial B$  which passes  inside of encounter regions, see fig.~(\ref{construction}b). Accordingly the action difference can be written as a sum of three therms
\begin{equation}
 \Delta S= S^{(A)}_\Gamma -S^{(A)}_{\bar{\Gamma}} + S^{(B)}_\Gamma -S^{(B)}_{\bar{\Gamma}} + S^{(C)}_\Gamma -S^{(C)}_{\bar{\Gamma}} .\label{ActionDif0}
\end{equation}
 By eq.~(\ref{diference1}) the differences between actions  depend only on  the  boundary values of $q_{n,t}, \bar{q}_{n,t}$ at $\partial C$. As a result    the above expression  for $\Delta S$ can be rewritten in a compact form as the sum over the boundary $\partial C$:
\begin{equation}
 \Delta S=\sum_{(n,t)\in \partial C_\parallel}\Delta S^{\parallel}_{n,t}+\sum_{(n,t)\in \partial C_\perp}\Delta S^{\perp}_{n,t},\label{ActionDif}
\end{equation}
where $\partial C_\parallel:=\partial C_\uparrow\cup \partial C_\downarrow $,  $\partial C_\perp :=\partial C_\leftarrow\cup \partial C_\rightarrow$ are pieces of the boundary directed in the vertical and horizontal directions respectively. The local action differences are defined here as  
\begin{multline}
 \Delta S^{a}_{n,t}= \left[q_{n,t}(2)-q_{n,t}(1) \right]\left[\bar{p}^{a}_{n,t}(2)-\bar{p}^{a}_{n,t}(1) \right]\\
+\left[\bar{q}_{n,t}(2)-\bar{q}_{n,t}(1) \right]\left[p^{a}_{n,t}(2)-p^{a}_{n,t}(1) \right], \quad a\in\{ \parallel, \perp\}, \label{LocalActionDif}
\end{multline}
with $q_{n,t}(k)$ (resp.  $\bar{q}_{n,t}(k)$)  being the coordinate at the encounter $k=1,2$ of $\Gamma$  (resp. $\bar{\Gamma}$) and 
 $p^{a}_{n,t}(k)$ (resp.  $\bar{p}^{a}_{n,t}(k)$) are the corresponding generalized momenta, see fig.~(\ref{construction}b). The relationship  of these momenta to the coordinates depends on the direction $a$ of the boundary at the point $(n,t)$:
\begin{equation}
 p^{\parallel}_{n,t}(k):=q_{n,t+1}(k)-q_{n,t}(k), \qquad p^{\perp}_{n,t}(k):=q_{n+1,t}(k)-q_{n,t}(k).
\end{equation}
It worth noticing that the expression  (\ref{LocalActionDif}) for  $\Delta S^{a}_{n,t}$ can be interpreted as a local symplectic area of the region  formed by the four points $(q_{n,t}(k), p^{a}_{n,t}(k))$, $(\bar{q}_{n,t}(k), \bar{p}^{a}_{n,t}(k))$, $k=1,2$ in the phase space, see fig.~(\ref{construction}c). The representation (\ref{LocalActionDif}) is essentially an extension of the similar  result for the single-particle case. 
There the difference between actions of partner orbits is  determined by  an analogous   symplectic area evaluated at an arbitrary  point of their encounter \cite{haake}.

\section{Symmetries \label{symmetries}}
So far we have ignored the possibility that the map $\Phi_N$ might posses a discrete symmetry. Such symmetries are  of crucial importance for the associated quantum problem, where they determine which universality class system belongs to.  On the classical side of the problem the presence of symmetries gives rise  to  partner orbits with close actions which do not traverse  approximately the same points of phase space, but rather two  sets of points related by the symmetry operation. For instance, in the  systems with the time reversal invariance  there exist  also partner orbits which traverse (approximately) the same points of the configuration  space but have opposite momenta at some pieces of the trajectories.  For this to happen it is sufficient to have just one encounter, which is the case of the original Sieber-Richter pairs \cite{sieber}.

All symmetries of Hamiltonian systems can be divided into two classes.  The ones which preserve the form of Hamiltonian's equations are called canonical. For $2N$ dimensional cat maps they are represented by $2N\times 2N$ integer matrices $\C$, $Z \to \C Z$, $Z=(q_1,p_1,q_2,p_2,\dots ,q_N,p_N)^\intercal$ satisfying 
\begin{equation} 
\C\B_{N} \C^{-1} = \B_{N}, \qquad \C^2 = \bm I.
 \end{equation}
For instance, all maps (\ref{eqmotion1}) commute with the matrix $\C=-I$ which effects the transformation $\q\to -\q$, $\p\to -\p$.
On the other hand a Hamiltonian system might also have anticanonical symmetries, 
    which reverse the signs of Hamiltonian's equations and the Poisson brackets. The most prominent example of the last one is  time reversal symmetry. 
The
anticanonical symmetries of $\Phi$ are represented by $2N\times 2N$ integer matrices $\C$,  that satisfy
\begin{equation} 
\C\B_{N} \C^{-1} = \B_{N}^{-1}, \qquad \C^2 = \bm I, \qquad \det \C=-1.
 \end{equation}
Its easy to check that for $b=0$ the map $\Phi$ has the anticanonical symmetry:
\[\C (q_1,p_1,q_2,p_2,\dots ,q_N,p_N)^\intercal =(p_1,q_1,p_2,q_2,\dots ,p_N,q_N)^\intercal,\] 
which amounts to  exchanging of the corresponding  moments and coordinates.

Assuming that the map $\Phi_{N}$ posses  a symmetry $\Sym_\C$ (either canonical or anticanonical) defined by a $2N\times 2N$ matrix $\C$ with  integer  entries let us describe the structure of partner orbits with one encounter. First observe that  each MPO $\Gamma$ has a pair orbit  $\Gamma^*=\Sym_\C\cdot\Gamma$ which is obtained by action of the linear map $\C$ on the set of points   $\Gamma$. Both orbits have exactly the same actions. The corresponding  partner orbits are obtained then by a recombination of different patches from $\Gamma$ and $\Gamma^*$.

To be more specific consider an  orbit $\Gamma$ having one non-trivial encounter with zero winding numbers $(0,0)$.
Such  an encounter   corresponds   to two frame-like regions $E_1$, $E_2$ of  $\Zz$   where 2D symbolic representations of $\Gamma$ and its conjugate  $\Gamma^*$ are related  by the symmetry operation $\Sym_\C$ in the following way:
\begin{equation}
 \Ee:=\psi_\Gamma(E_1)=\psi_{\Gamma^*}(E_2), \qquad  \Ee^*:=\psi_\Gamma(E_2)=\psi_{\Gamma^*}(E_1).\label{encountersym}
\end{equation}
  \begin{figure}[htb]
\begin{center}{
\includegraphics[height=3.0cm]{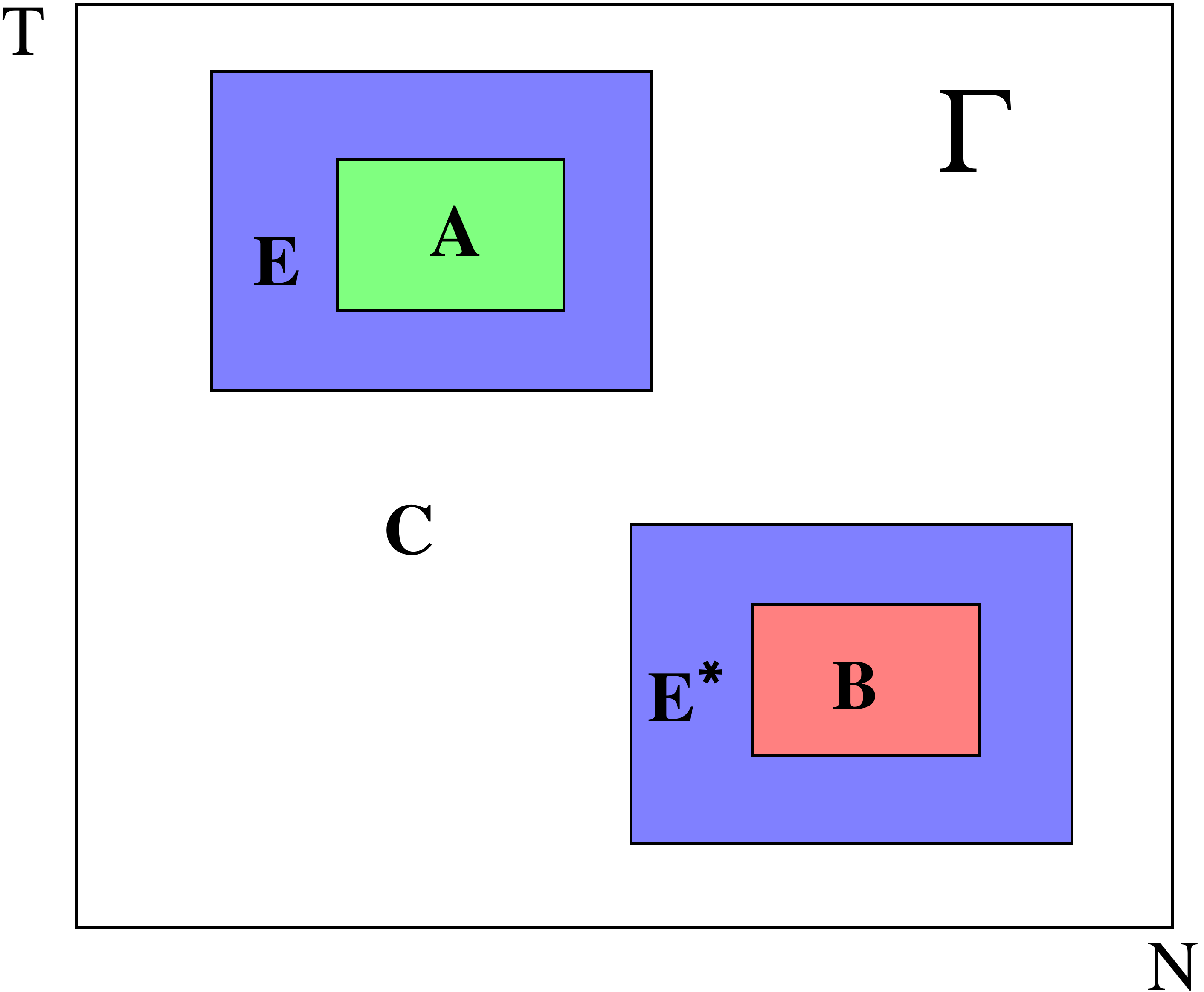}\hskip 0.0cm \includegraphics[height=3.0cm]{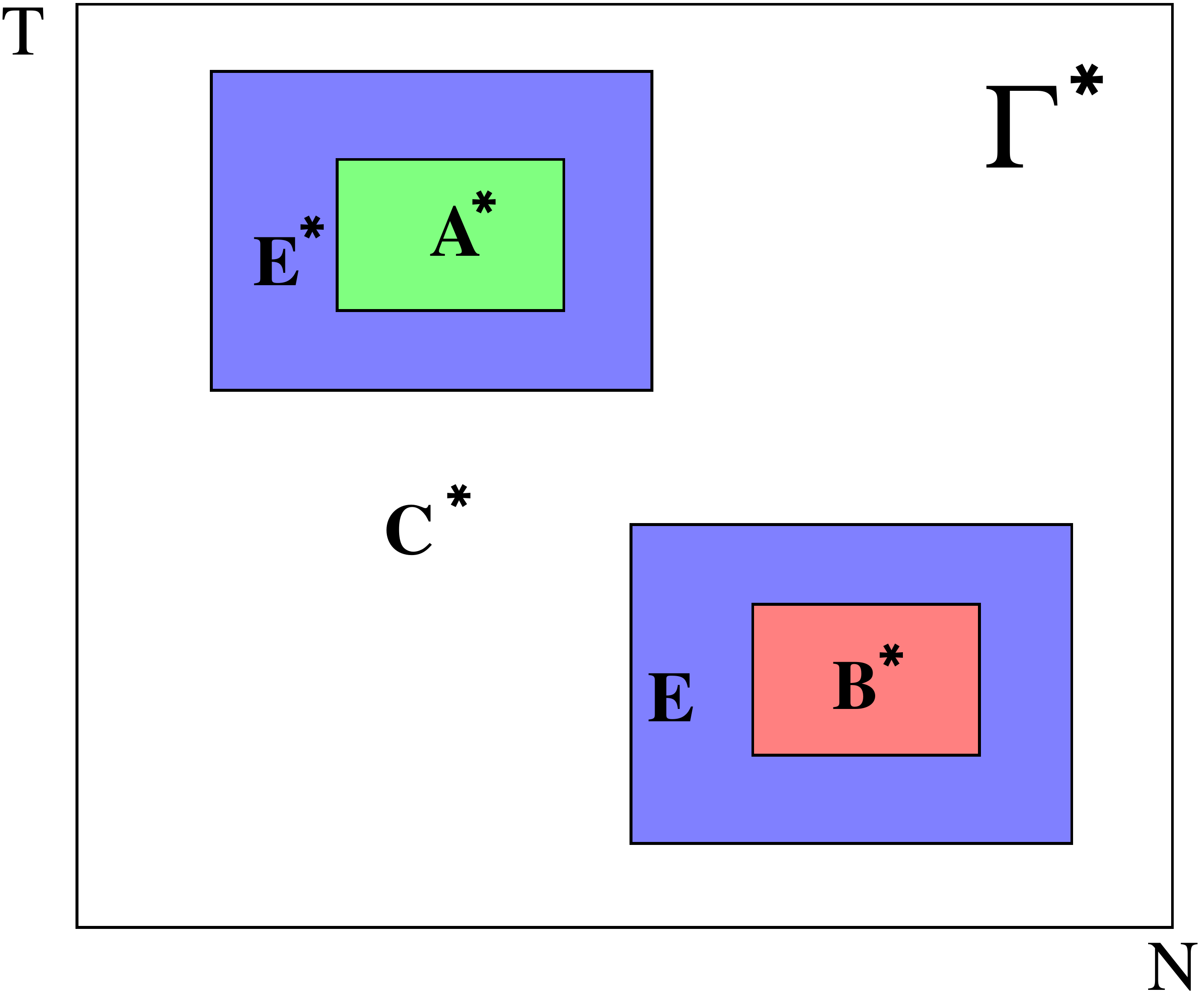}\hskip 0.3cm \includegraphics[height=3.0cm]{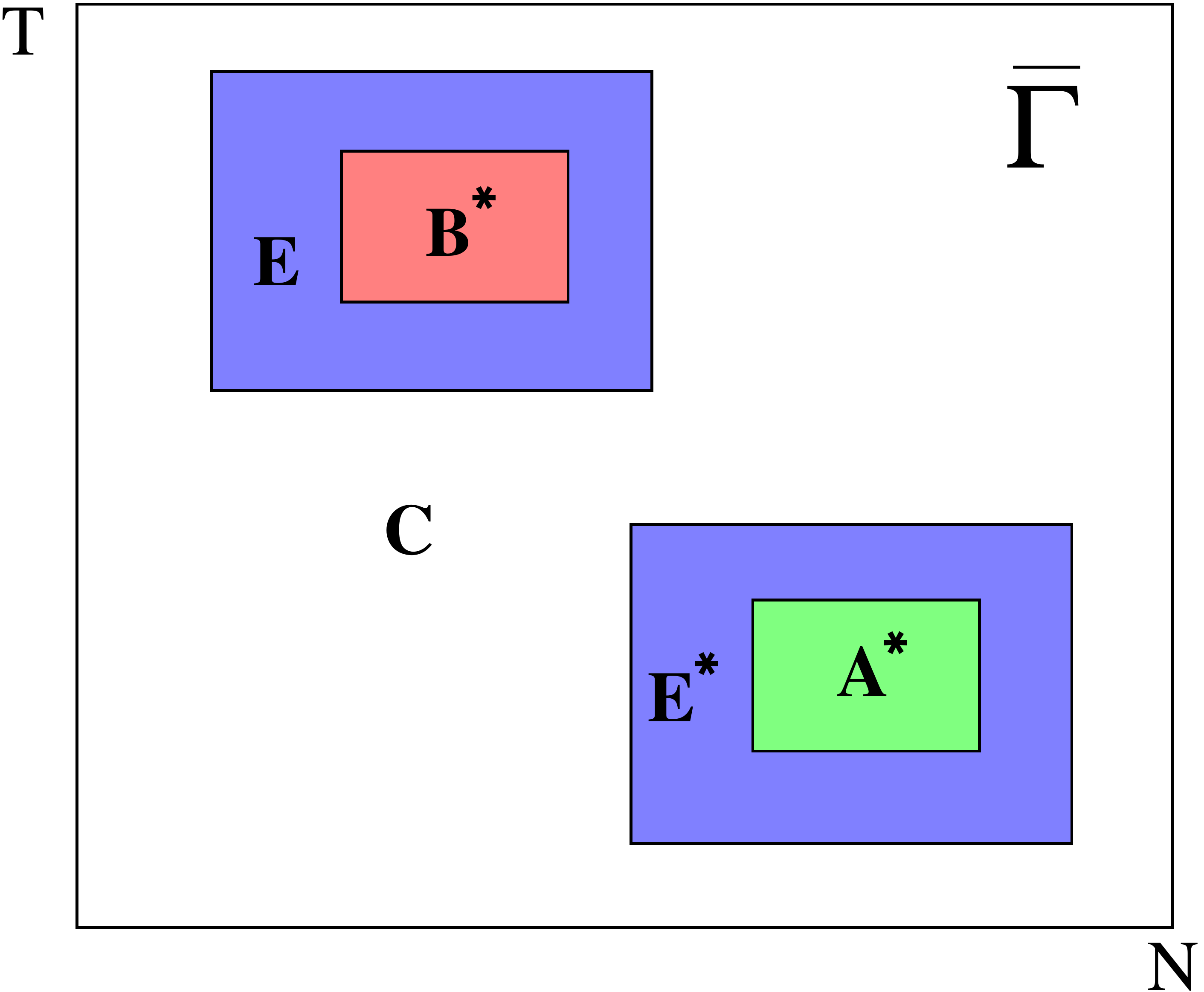}\hskip 0.0cm \includegraphics[height=3.0cm]{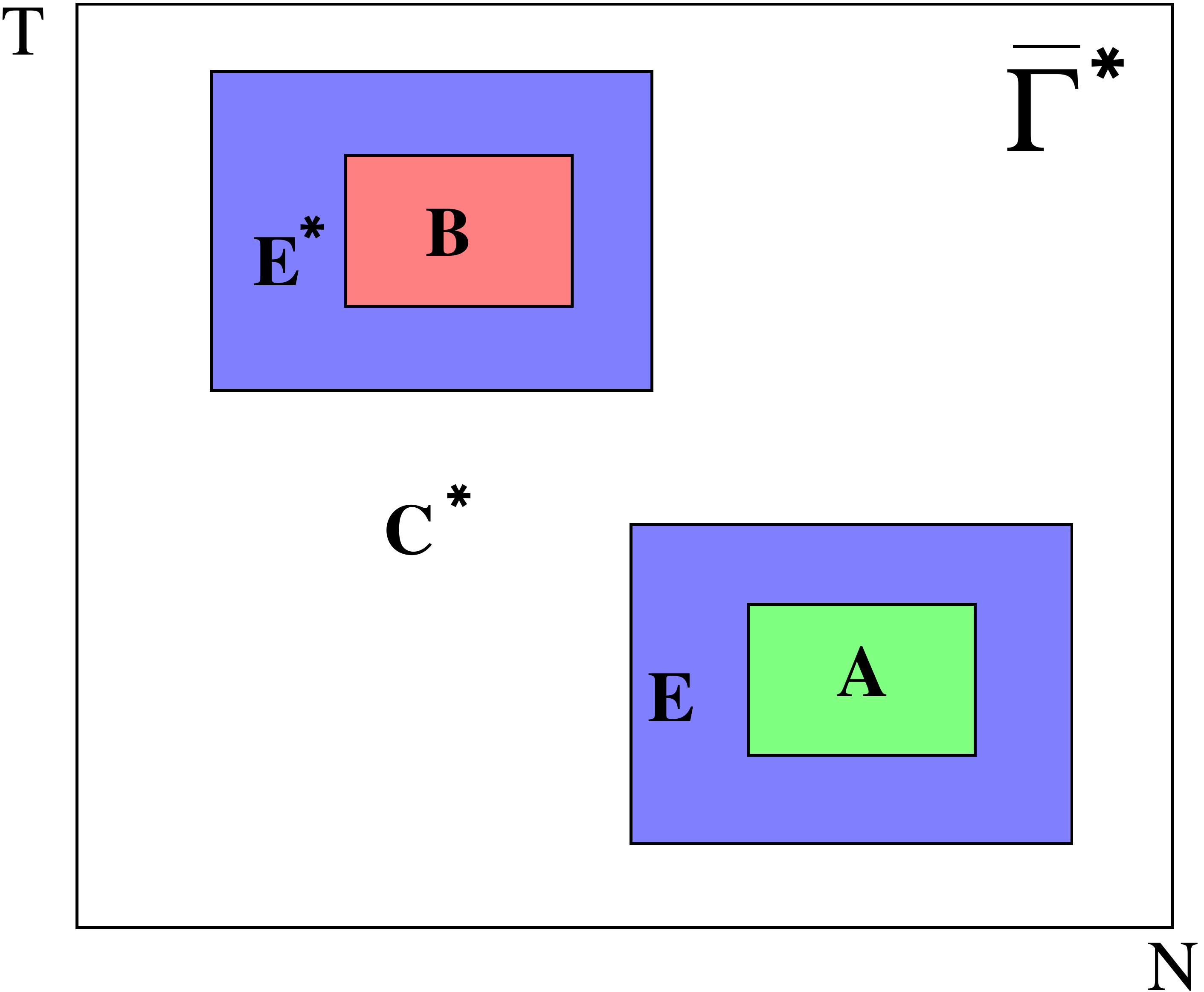}
}\end{center}
\caption{ \small{On the two left figures are schematically shown the 2D symbolic representations $\Mm_{\Gamma}$, $\Mm_{\Gamma^*}$ of some MPO $\Gamma$ and its symmetric counterpart $\Gamma^*$ having  one  encounter with $(0,0)$ winding numbers. Such conjugate pairs of MPO's with  exactly the same action exist whenever the system posses a  symmetry.   On the two right pictures are depicted the symbolic representations $\Mm_{\bar{\Gamma}}$ $\Mm_{\bar{\Gamma}^*}$ of the corresponding partner orbits $\bar{\Gamma}$, $\bar{\Gamma}^*$. They are obtained from  $\Gamma$ and  $\Gamma^*$ by copy pasting of the interior regions of encounters. In general $\bar{\Gamma}$, $\bar{\Gamma}^*$  (resp.  $\Gamma$,  $\Gamma^*$) have close but non-identical  actions.}  }\label{encounter_sym}
\end{figure}
\begin{figure}[htb]
\begin{center}{
\includegraphics[height=3.0cm]{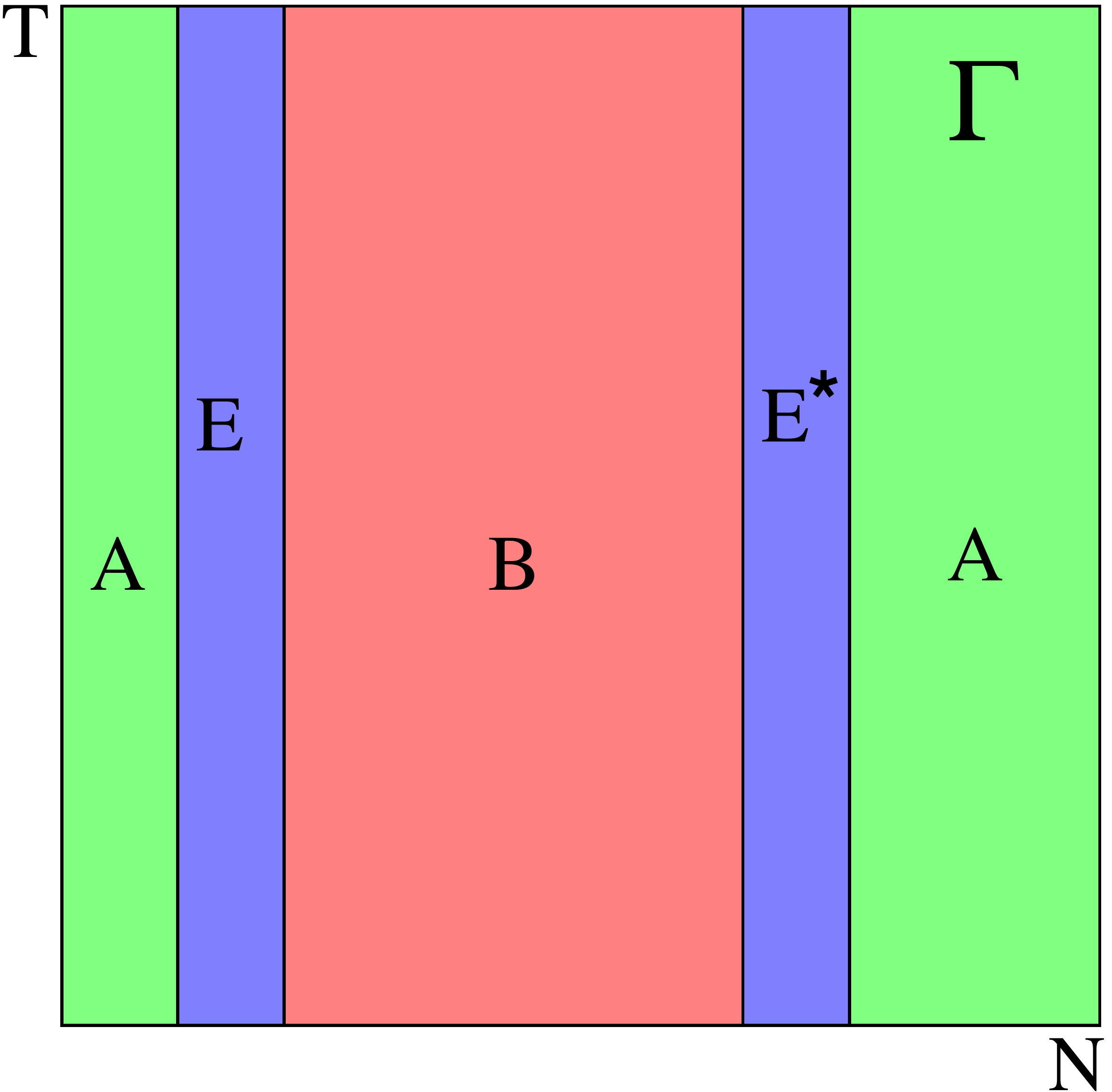}\hskip 0.3cm \includegraphics[height=3.0cm]{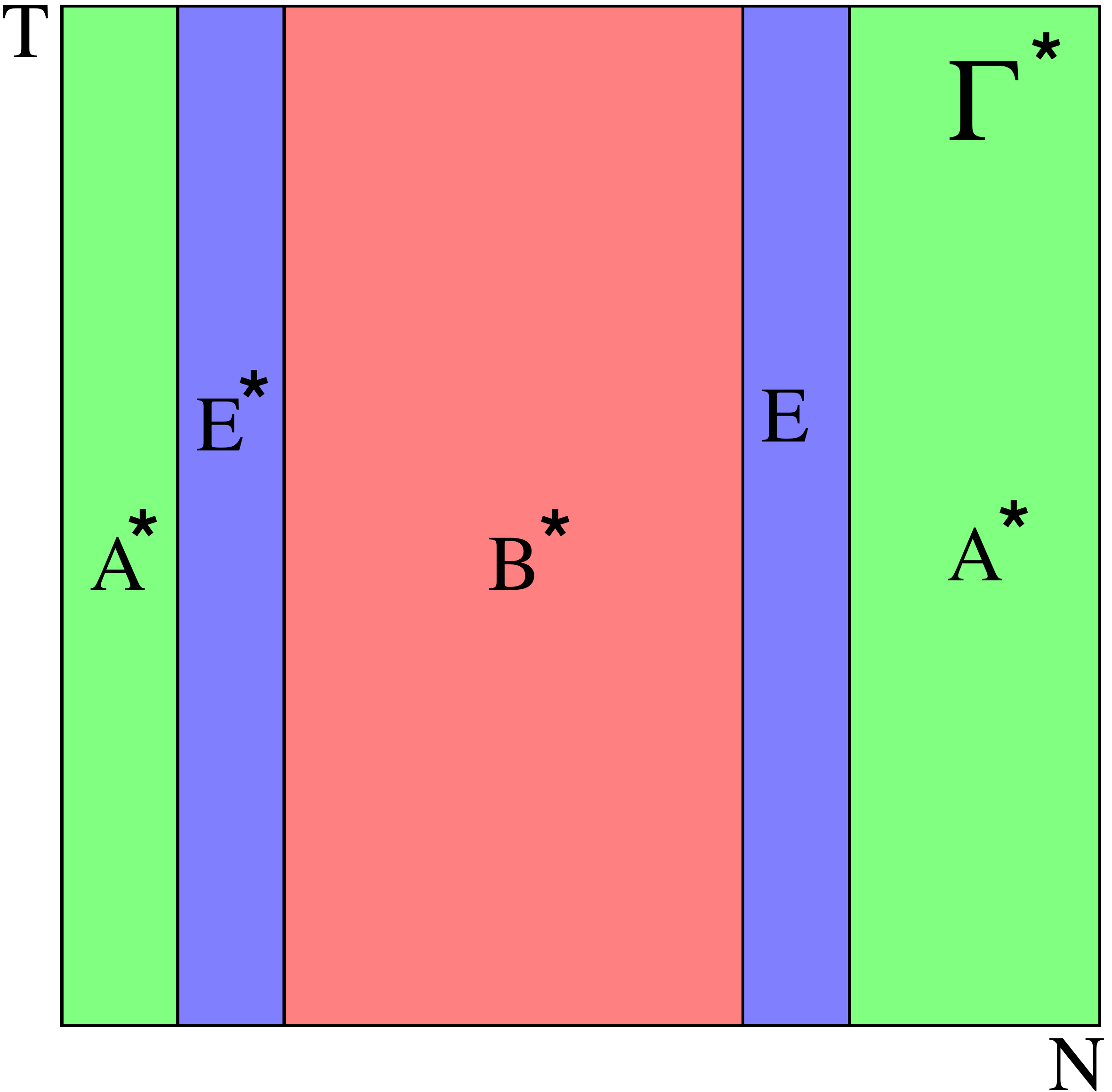}\hskip 1.3cm \includegraphics[height=3.0cm]{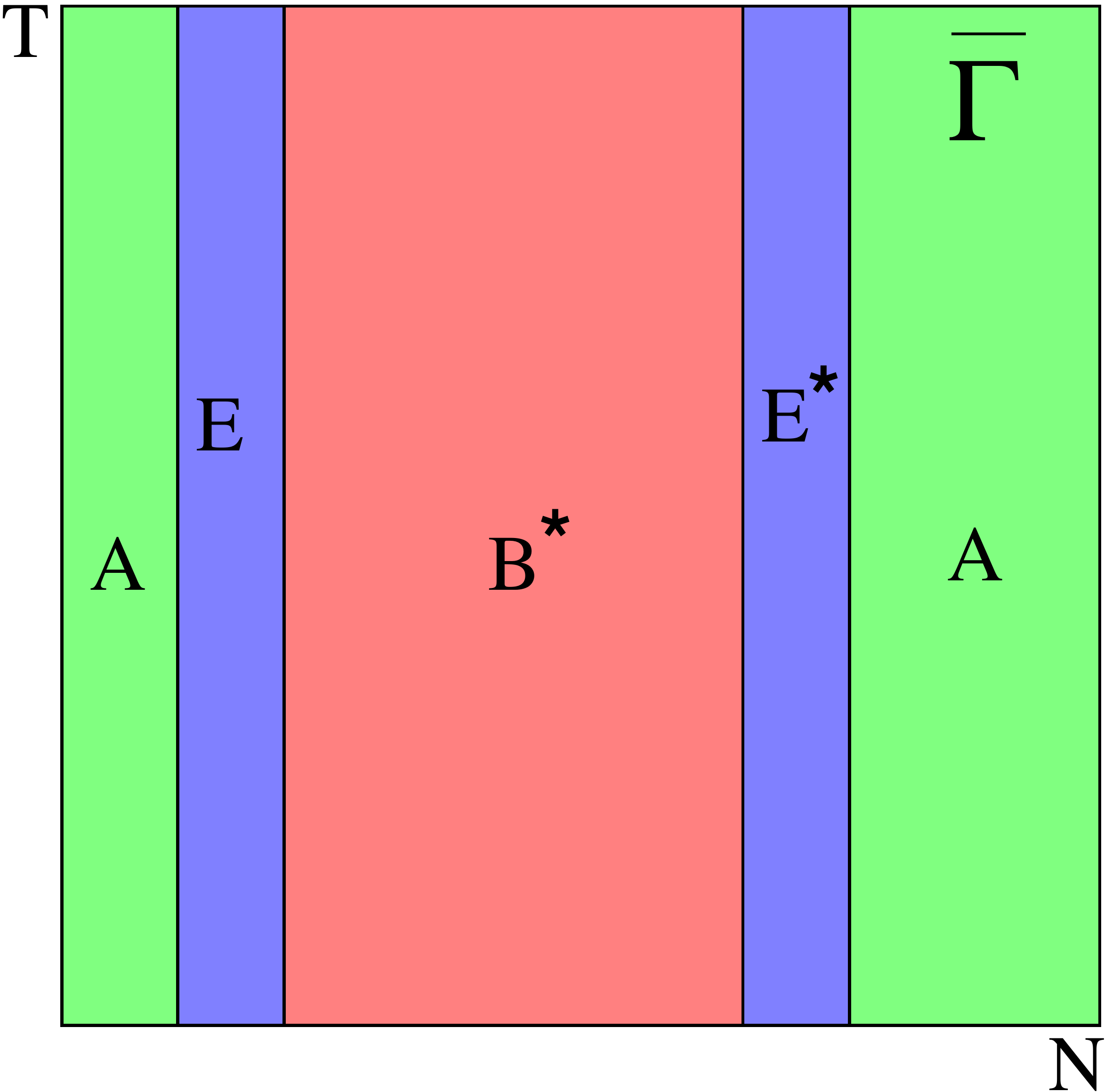}\hskip 0.3cm \includegraphics[height=3.0cm]{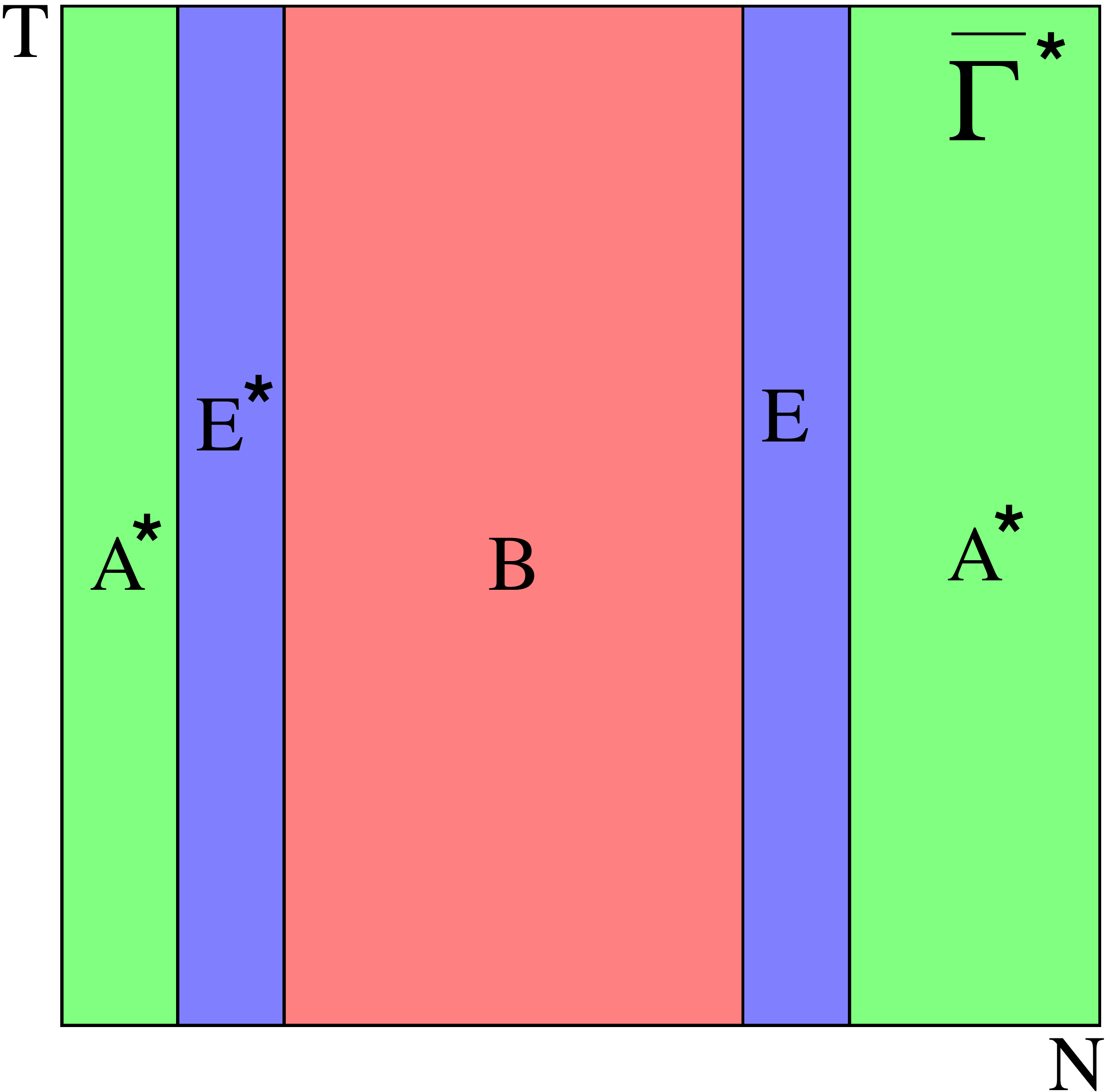}
}\end{center}
\caption{ \small{On the two left figures are schematically shown the 2D symbolic representations $\Mm_{\Gamma}$, $\Mm_{\Gamma^*}$ of   symmetrically  conjugate MPO's  with  one  encounter having $(1,0)$ winding numbers.   On the two right pictures are depicted the symbolic representations $\Mm_{\bar{\Gamma}}$ $\Mm_{\bar{\Gamma}^*}$ of the corresponding partner orbits $\bar{\Gamma}$, $\bar{\Gamma}^*$. They are obtained from  $\Gamma$ and  $\Gamma^*$ by copy pasting of the corresponding interior regions of $\Mm_{\Gamma}$, $\Mm_{\Gamma^*}$, as shown in the figure. The partner orbits for MPO's with encounters of the type  $(0,1)$ are constructed in the same way by exchanging particle and time directions.}  }\label{encounter_sym_sieber}
\end{figure}
The 2D symbolic representation $\Mm_{\bar{\Gamma}}$ of the partner orbit  $\bar{\Gamma}$ is then obtained by copying the  interiors $A$, $B$ of the encounter regions from $\Mm_{\Gamma^*}$ and pasting them into the corresponding places of $\Mm_{\Gamma}$, as shown in fig.~\ref{encounter_sym}. The partner orbit of $\Gamma^*$  can readily be obtained by applying symmetry operation  to $\bar{\Gamma}$ i.e.,  $\bar{\Gamma}^*=\Sym_\C\cdot\bar{\Gamma}$. In Appendix~B we illustrate the above procedure with an explicit construction of a quadrapole of periodic orbits $\Gamma,\Gamma^*,\bar{\Gamma},\bar{\Gamma}^*$.

For  MPO's with one  encounter  having   winding numbers $(1,0)$ and $(0,1)$  the construction is analogous. In this case the corresponding  encounter regions of $\Zz$  are two strips winding torus in the particle (resp. time) direction and separating it into two pieces. The symbolic representation of  partner orbits is obtained from  $\Mm_{\Gamma}$, $\Mm_{\Gamma^*}$ by copy pasting of the symbols from these regions, as shown in fig.~\ref{encounter_sym_sieber}. Note that for the encounters of the type $(0,1)$ the obtained partner MPO's are the original Sieber-Richter pairs in $2N$ dimensional phase space of the map $\Phi_N$.

\section{Partner MPOs for perturbed map \label{Sec7}} 
It is well known that the spectra of quantum cat maps essentially differ from  ones of generic chaotic systems~\cite{keating1}. From the semiclassical point of view this anomaly can be attributed to   the large degeneracies among the actions of the periodic orbits which are given by rational numbers for a cat map. These degeneracies    grow exponentially  with the  period of orbits~\cite{keating2}
 leading to  the distortion in the distribution of action differences.  The genericity, however, may be  recovered by adding a periodic nonlinear perturbation~\cite{keating3}. From this prospective it is natural to inquire
whether the correlation mechanism between  the partner MPOs described in the previous sections  carries over to the  perturbed coupled cat maps.  Below we demonstrate that  the answer to this  question is indeed affirmative.

Consider the  model of perturbed coupled  cat maps generated by the function~(\ref{model}), where some periodic potential $\V(q)\equiv\kappa\V_0(q)$ is turned on. We require that, $\max_{q}\abs{\V''_0(q)}=1$ and relate to $\kappa$ as the strength of the  perturbation. The particular choice of the parameter $\kappa$ is dictated by the requirement that  the perturbed map $\Phi_{\Nn}$ remains fully chaotic. By the structural stability  of classical chaos this is  guaranteed for sufficiently small $\kappa$. A maximum possible range for $\kappa$ can be estimated by considering the linearized form  of the map $\Phi_{\Nn}$. Let  $q_{n,t}, n=1,2,\dots, N$ be some solution of  eq.~(\ref{newtoneq}) and let $\delta q_{n,t}$ be a small deviation from it. After substituting $q_{n,t}+\delta q_{n,t}$ into  the equations of motion (\ref{newtoneq}) and leaving out only linear terms we obtain 
\begin{equation}
 \delta q_{n,t+1}+\delta q_{n,t-1} -d(\delta q_{n+1,t}+\delta q_{n-1,t})=(s+\V''(q_{n,t}))\delta q_{n,t}.  \label{linnewtoneq}
\end{equation} By comparison with the non-perturbed case analysis in  Section~3.2  we arrive to  the inequality $|\kappa| < \kappa_{\max}=|s|-2|d|-2$. It can be expected that   under this condition  the perturbed map $\Phi_{\Nn}$ is fully hyperbolic.

\subsection{Construction of MPO partners} In general, calculating  periodic orbits  of a  non-integrable system is a non-trivial task. To this end a number of methods have been developed, see e.g., \cite{BDM1988}. 
Our approach  is based on the minimization of the perturbed action, i.e., the generating function $S(q_{n,t})$ with respect to $q_{n,t}$.  The procedure goes as follows.   One starts from a MPO $\Gamma_0=\{q^{(0)}_{n,t}| (n,t)\in\Zz\}$ of a  non-perturbed map  and adds a small perturbation $\delta\kappa\V_0(q)$ to the action. By structural stability of chaotic dynamics  it can be expected that,  for a small $\delta\kappa$ the non-perturbed orbit lays in the vicinity of the perturbed one $\Gamma_1=\{q^{(1)}_{n,t}|(n,t)\in\Zz\}$ which can be easily found by  minimization of the action $S(q_{n,t})$ by using  the standard  gradient descent method.  At the next step one takes  $\{q^{(1)}_{n,t}\}$ as the initial data  for the gradient descent method and obtains  a new MPO $\Gamma_2$ for the doubled  strength of the perturbation  $2\delta\kappa$. By iterating this procedure  $m$  times one arrives to a periodic orbit $\Gamma_m$ which is a smooth deformation of $\Gamma_0$ for the 
perturbation  of the strength $\kappa=m\delta\kappa< \kappa_{\max}$. Importantly,   if we start from two partner orbits $\Gamma_0,\bar{\Gamma}_0$ with the  related    matrices $\Mm_{\Gamma_0}$,$\Mm_{\bar{\Gamma}_0}$ of winding numbers, this relationship  persists at each step of the iterations. This means that the resulting MPOs $\Gamma_m$, $\bar{\Gamma}_m$ are
 again partners with small action differences, whose symbolic representations are related in the same way as ones of the initial orbits. 

In Appendix~C we give an example of explicit calculation of a pair of MPOs $\Gamma,\bar{\Gamma}$  for the perturbed coupled cat map and demonstrate stability of the pairing  phenomena. Namely, we show that all the points of $\Gamma$ and $\bar{\Gamma}$ are paired and their  action differences  accumulate in the encounter regions.

\subsection{Action differences}
It is instructive to see how the formula~(\ref{ActionDif}) for action differences between the partner orbits is affected by the perturbation. In accordance with eq.~(\ref{actionaction1}) the actions get extra terms depending on $\V$. After repeating the same steps as in the derivation of the non-perturbed case we end up with the additional term in the right hand side of eq.~(\ref{diference1}):
\[\delta S^{(C)}= \sum_{(n,t)\in C} \frac{1}{2}\bigg(\V'(\bar{q}_{n,t})+\V'(q_{n,t})\bigg)\left(q_{n,t}-\bar{q}_{n,t}\right) -\V(q_{n,t})+\V(\bar{q}_{n,t}).\] 
Like  the main contribution  in eq.~(\ref{diference1}), the term $\delta S^{(C)}$ is accumulated mainly in the encounter region. It is straightforward to see, however, that $\delta S^{(C)}$ is proportional to  the third order of  $\delta q_{n,t}:=q_{n,t}-\bar{q}_{n,t}$: 
$$
\delta S^{(C)}=\sum_{(n,t)\in C}\frac{\delta q_{n,t}^3}{12}\;\V^{(3)}\bigg(\frac{q_{n,t}+\bar{q}_{n,t}}{2}\bigg)+\mathcal{O}(\delta q_{n,t}^5),
$$
while the principle term $S_{\Gamma}^{(C)} - S_{\bar{\Gamma}}^{(C)}$ is of the order $\delta q_{n,t}^2$. As a result, $\delta S^{(C)}$ can be dropped out for MPOs with  large encounters (where the distances between the partners are small). Since the same conclusion holds for all other terms in eq.~(\ref{ActionDif0}) it follows immediately that the formula~(\ref{ActionDif}) is still valid in an approximate sense. The larger the size of encounters, the better approximation gives the equation~(\ref{ActionDif}).

\section{Conclusions\label{conclusions}}

\subsection{ Quantum problem} So far we  have studied clustering of MPOs in the context of purely classical theory. While leaving  applications  to quantum mechanics for future investigations, let us briefly discuss the implications of our results for  a semiclassical theory of many-particle systems. Spectral correlations in  quantum systems with fully chaotic classical  dynamics   can be  related to correlations between periodic orbits.  Consider as an example  the  spectral  form factor $K_2$ which is defined as the Fourier transform of the  two-point spectral correlation function. Via the trace formula it can be  expressed as a double sum over pairs of partner periodic orbits. These pairs $\Gamma, \bar{\Gamma}$ can be clustered  in accordance to their   topological structures $\F_\Gamma$, such that:
\begin{equation}
 K_2(T,N)=\Re\sum_{\F}\,\,\,\sum_{\{\bar{\Gamma},\Gamma|\F_\Gamma=\F\}}\A_{\Gamma} \exp(i\Delta S_\Gamma/\hbar_{eff}),\label{formfac}
\end{equation}
where the first sum runs over all possible topological structures  $\F$ of MPOs \cite{haake,Daniel}. The factors $\A_{\Gamma}$, $\Delta S_\Gamma$  are determined by stabilities and by  differences between actions of partner orbits, respectively, with $\hbar_{eff}^{-1}$ being  the integer Hilbert space dimension of  the corresponding  quantum cat map, see \cite{keating1}.

 In order to clarify which topological structures of MPOs are of a relevance for given number of particles $N$ and time $T$, it is of crucial importance to estimate  how many periodic  orbits $\Gamma$ posses a given   topological structure $\F$. The amount of  such orbits is determined by the number of possibilities to choose  the encounter sets $E^{(k)}_n, k=1,\dots l_n$  within  the prescribed topological order. As we argue below  this,  in turn,  depends on the  winding numbers $w_n$ of  $E^{(k)}_n$'s  and  their widths $p_n$. Note  that systematic contribution into the sum (\ref{formfac}) comes from the terms for which  $\Delta S_\Gamma$ is of the same order as $\hbar_{eff}$. This fixes the  scale of encounter width  as $p_n\sim|\log\hbar_{eff}|$. Consider now an encounter set  having  the  winding numbers $w=(0,1)$, i.e.,   winding around  the torus in  
the particle  direction. Such a set  occupies at least $N \cdot p$ points of $\Zz$.
   Analogously, for  encounters with winding numbers  $w=(1,0)$, $w=(0,0)$ the minimal number of points  is given by  $T \cdot p$ and $p^2$, respectively.   It is easy to see that the number of periodic orbits with the fixed symbols at the encounter domains $E^{(k)}_n, k=1,\dots l_n$  grows exponentially with the total number of points of $\Zz$ outside of the encounter sets. From this we can immediately conclude  that for  $N\ll |\log \hbar_{eff}|\ll T$  most of the periodic orbits have  encounters of the type $w=(0,1)$, since they occupy the minimal ``area'' of $\Zz$. For such parameters $N,T$  the correlations between actions of  MPOs  can be understood just by interpreting  the system dynamics  as single-particle motion in many-dimensional phase space. In other words in this regime the semiclassical theory of many-particle systems should be  the same  as for single-particle systems with chaotic dynamics.
 In the regime $T\ll |\log \hbar_{eff}|\ll N$ the dominating encounter type of periodic orbits is $w=(1,0)$ and the clustering mechanism is dual to the previous one. This means that  correlations between actions of periodic orbits can be accounted by utilizing single-particle theory with the parameters $N$ and $T$ exchanged.  Finally in the case when both parameters are ``large'':  $T,N \gg |\log \hbar_{eff}|$ most of the periodic orbits posses encounters with zero winding numbers $w=(0,0)$. In this case the correlations  between actions of MPOs cannot be accounted on the basis   of  single-particle interpretation of the system dynamics. For this regime a new  genuine many-particle  semiclassical theory has to  be constructed. 

\subsection{Generalizations} The  theory proposed in this paper has been constructed and verified in the framework of a  specific model of coupled lattice maps. One might  wonder whether it carries over to the Hamiltonian systems with continues time evolution. We believe that the main results hold also for  generic    Hamiltonians of the form (\ref{hamiltonian})  provided that the  dynamics are chaotic. In order to clarify this point  consider, as an example, a many-particle periodic orbit  $\Gamma(t)=\{(x_n(t),p_n(t))\}$  where $t\in [0,T]$ now  continues time and the coordinates  $x_n(t)$ satisfy the system of $N$ Newtonian equations generated by the Hamiltonian  (\ref{hamiltonian}). Assume that $\Gamma(t)$ has an encounter  domain 
 with the frame-like structure shown in fig.~\ref{encounter3}. In the coordinate form this  means that for $(n,t)\in E$ we have an approximate equality  $x_n(t)\approx x_{n+n'}(t+t')$, where    $(n+n',t+t')\in E'$ are the points of the shifted encounter set. (For sufficiently large $T$ and $N$ such  encounters should exist  just by statistical reasons.) Let us now construct another trajectory following the rules for search of partner orbits. To this end  define:
\[\bar{x}_n(t):=\begin{cases} x_n(t) &\text{if $(n,t)\notin B\bigcup A$},\\
                 x_{n+n'}(t+t') &\text{if $(n,t)\in B$},\\
                 x_{n-n'}(t-t') &\text{if $(n,t)\in A$}.
                \end{cases}
\]   
  It is now a straightforward observation following from the locality of the Hamiltonian interactions  that the set of trajectories  $\bar{\Gamma}(t)=\{(\bar{x}_n(t),\bar{p}_n(t))\}$  approximately satisfies the  equations of motion. It follows  then from the shadowing lemma for flows of chaotic systems \cite{katok1995} that   a real periodic orbit  close to  $\bar{\Gamma}(t)$ should exist.  By construction this new orbit is  the  partner of  $\Gamma(t)$.  

Taking a step further one can also  question whether  discretization in the particle direction  might be left out as well. Making $n$ continues naturally leads to a field model, where the field $\phi(x,t)$ satisfies some homogeneous differential equation and the continues  parameter $x\in [0,L]$ plays the  role   of  $n$. Assuming that the resulting    dynamics are chaotic it  can be expected that periodic solutions $\phi_{per}(x,t)$ of the differential equation  posses similar properties as MPOs in the coupled lattice maps. In particular we expect that the  mechanism of correlations between MPOs carries over to  $\phi_{per}(x,t)$. We believe that this correlation mechanism can be of importance  for development  of the semiclassical approach to the corresponding Quantum Field Theory.  

Another natural extension of the present setup arises if the particles are ordered with respect  to the d-dimensional  lattice i.e., $n\in \mathbb{Z}^d$,  $d>1$ rather then to  the  one-dimensional $ \mathbb{Z}$. In this case  propagating MPOs sweep $d+1$-dimensional ``surfaces'' in the one-particle phase space. It would be of interest to study how the dimension $d$ affects the correlation mechanism between MPOs. 

\subsection{Summary}
 Periodic  orbits of chaotic systems can be organized into  clusters of partner orbits where all elements have approximately the same  actions. The clustering phenomena plays a crucial role for  the semiclassical theory. So far however, it has been studied only  on the level of  effectively ``single-particle''  systems.
In the present work we considered clustering mechanism among periodic orbits of fully chaotic many-particle  systems which dynamics are governed by translation-invariant Hamiltonians with local interactions. On the heuristic level   periodic orbits in such  systems can be  visualized by  discretized  two-dimensional ``surfaces'' in the one-particle configuration space. A more   rigorous description is  achieved by  using  symbolic dynamics.  Contrary to the single-particle formalism,  where periodic orbits are encoded by linear symbolic sequences, here the symbolic representation is provided by 2D toric arrays of  symbols.
A periodic orbit $\Gamma$ has a partner $\bar{\Gamma}$ when its symbolic representation $\Aa_{\Gamma}$ has at least one $l$-encounter - a 2D subsequence of symbols repeating itself for $l>1$ of times.  The symbolic representations $\Aa_{\bar{\Gamma}}$ of the partner orbits are then obtained by rearrangement of symbols $\Aa_{\Gamma}$ in accordance to certain prescription. The  differences  between actions of $\Gamma$ and its partners accumulate mainly in  the encounter regions and decrease exponentially with their width. 

Different families of partner orbits can be distinguished by  the corresponding topological structure diagrams. Each structure is represented by a toroidal surface glued to itself along a number of one-dimensional lines representing the  encounters. The diagrams with encounters  winding around torus torus only in the particle direction are the ones which  obtained in the single-particle interpretation of classical dynamics. The other  structural diagrams  are essentially different.  They represent partner periodic orbits whose correlation mechanism differs from one  found in  single-particle systems.  These new families of partner orbits seem to be essential for construction of a proper semiclassical theory in the case when the number of particles is sufficiently large.

\section*{ Acknowledgments:}
We are grateful to M.~Akila, P.~Braun, T.~Guhr and D.~Waltner  for useful discussions.
 Financial support by SFB/TR12 and  DFG research grant Gu 1208/1-1  is gratefully acknowledged.

\section*{Appendix A}
In this section we construct explicit example of two partner MPOs   for the map $\Phi_{\Nn}$ with the parameters:   $N=70$,  $a=3$, $b=2$, $c=d=-1$. The initial vector $\M_0$ in the equation~(\ref{winding2}) is generated as a set of random integers taken from the interval $[-10^{21},10^{21}]$. The two partner MPOs of the  period $T=50$ were generated following the protocol described in Section~5.2, see also the explanatory scheme on the figure~(\ref{superconstruction}). The resulting  symbolic representations $\Aa_\Gamma$, $\Aa_{\bar{\Gamma}}$ are presented on the picture~\ref{Fig12}, where each of 16 valid pairs of winding numbers   $\m=(m^{(q)}, m^{(p)})$ (see figure~\ref{Fig12})  is encoded by a symbol $a$ from the hexadecimal alphabet $\A$. The correspondence  between symbols of $\A$ and winding numbers  is provided by the table below:  
\begin{figure}
\centering
\includegraphics[height=7.8cm]{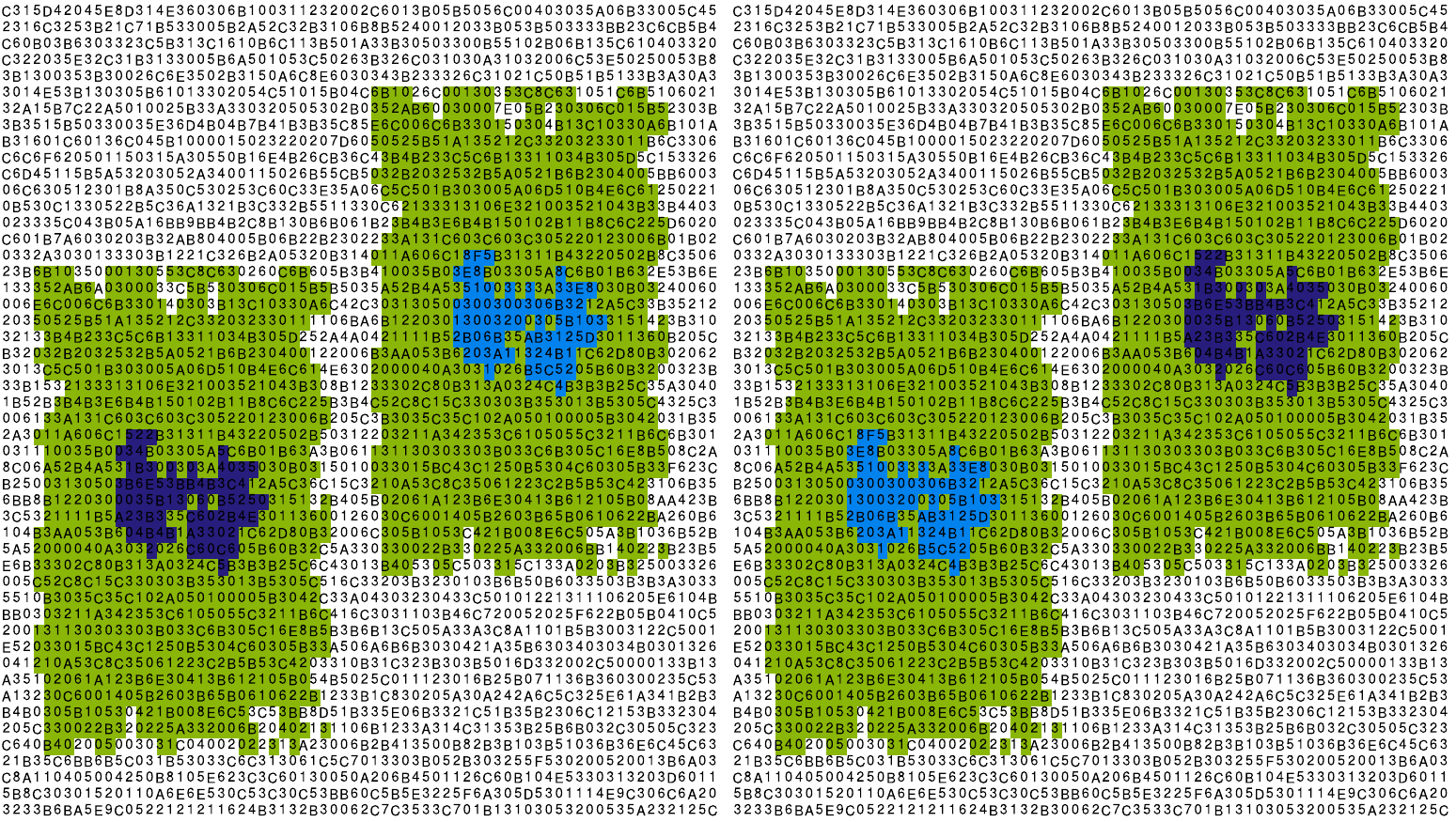}
	\caption{
Symbolic representations $\Aa_{\Gamma}$ (left) $\Aa_{\bar{\Gamma}}$ (right) of the two partner periodic orbits (${\Gamma}$, $\bar{\Gamma}$) for parameters $a=3$, $b=2$, $c=d=-1$, $T=50$, $N=70$. The corresponding elements of the  matrices $\Mm_{\Gamma}$ and $\Mm_{\bar{\Gamma}}$ are encoded by the hexadecimal numbers in accordance with the table~(\ref{table}). The identical encounter regions are highlighted by green (color on-line). The coinciding inner regions of the encounters (which are exchanged  in ${\Gamma}$ and  $\bar{\Gamma}$) are colored by the identical colors (blue and dark blue).
}\label{Fig12}
\end{figure}


\begin{equation}\label{table}
\begin{array}{l|cccccccccccccccc}
\m\!& \!\!\!
 \begin{pmatrix} 0\\ 0  \end{pmatrix}\!\!\!\!\!\!&\begin{pmatrix} 0\\ 1  \end{pmatrix}\!\!\!\!\!\!&\begin{pmatrix} 1\\ 1  \end{pmatrix}\!\!\!\!\!\!&\begin{pmatrix} 1\\ 2  \end{pmatrix}\!\!\!\!\!\!&\begin{pmatrix} 1\\ 3  \end{pmatrix}\!\!\!\!\!\!&\begin{pmatrix} 2\\ 3  \end{pmatrix}\!\!\!\!\!\!&\begin{pmatrix} 2\\ 4  \end{pmatrix}\!\!\!\!\!\!&\begin{pmatrix} 2\\ 5  \end{pmatrix}\!\!\!\!\!\!&\begin{pmatrix} 3\\ 5  \end{pmatrix}\!\!\!\!\!\!&\begin{pmatrix} 3\\ 6  \end{pmatrix}\!\!\!\!\!\!&\begin{pmatrix} 0\\ -1  \end{pmatrix}\!\!\!\!\!\!&\begin{pmatrix} -1\\ -1  \end{pmatrix}\!\!\!\!\!\!&\begin{pmatrix} -1\\ -2  \end{pmatrix}\!\!\!\!\!\!&\begin{pmatrix} -1\\ -3  \end{pmatrix}\!\!\!\!\!\!&\begin{pmatrix} -2\\ -3  \end{pmatrix}\!\!\!\!\!\!&\begin{pmatrix} -2\\ -4  \end{pmatrix}
\\
\hline\\
a\!\!\!&\!\!\!0 \!\!\!\!\!\!& 1 \!\!\!\!\!\!& 2 \!\!\!\!\!\!& 3 \!\!\!\!\!\!& 4 \!\!\!\!\!& 5 \!\!\!\!\!\!& 6 \!\!\!\!\!\!& 7 \!\!\!\!\!\!& 8 \!\!\!\!\!\!& 9 \!\!\!\!\!\!& A \!\!\!\!\!\!& B \!\!\!\!\!\!& C \!\!\!\!\!\!& D \!\!\!\!\!\!& E \!\!\!\!\!\!& F 
\end{array}
\end{equation}
On the fig.~\ref{Fig12} the encounter regions of the orbits $\Aa_\Gamma, \Aa_{\bar{\Gamma}}$ are colored in green. 
One can easily verify that the symbolic sequences inside these regions coincide with each other. Note that the shape of the green areas slightly differs on its borders from the rectangular shape (colored in yellow) of the original encounter-patches. As it is clearly seen, the deformation on the border of legitimate fields causes only local transformation in the symbolic representation.

\begin{figure}
\includegraphics[height=7.0cm]{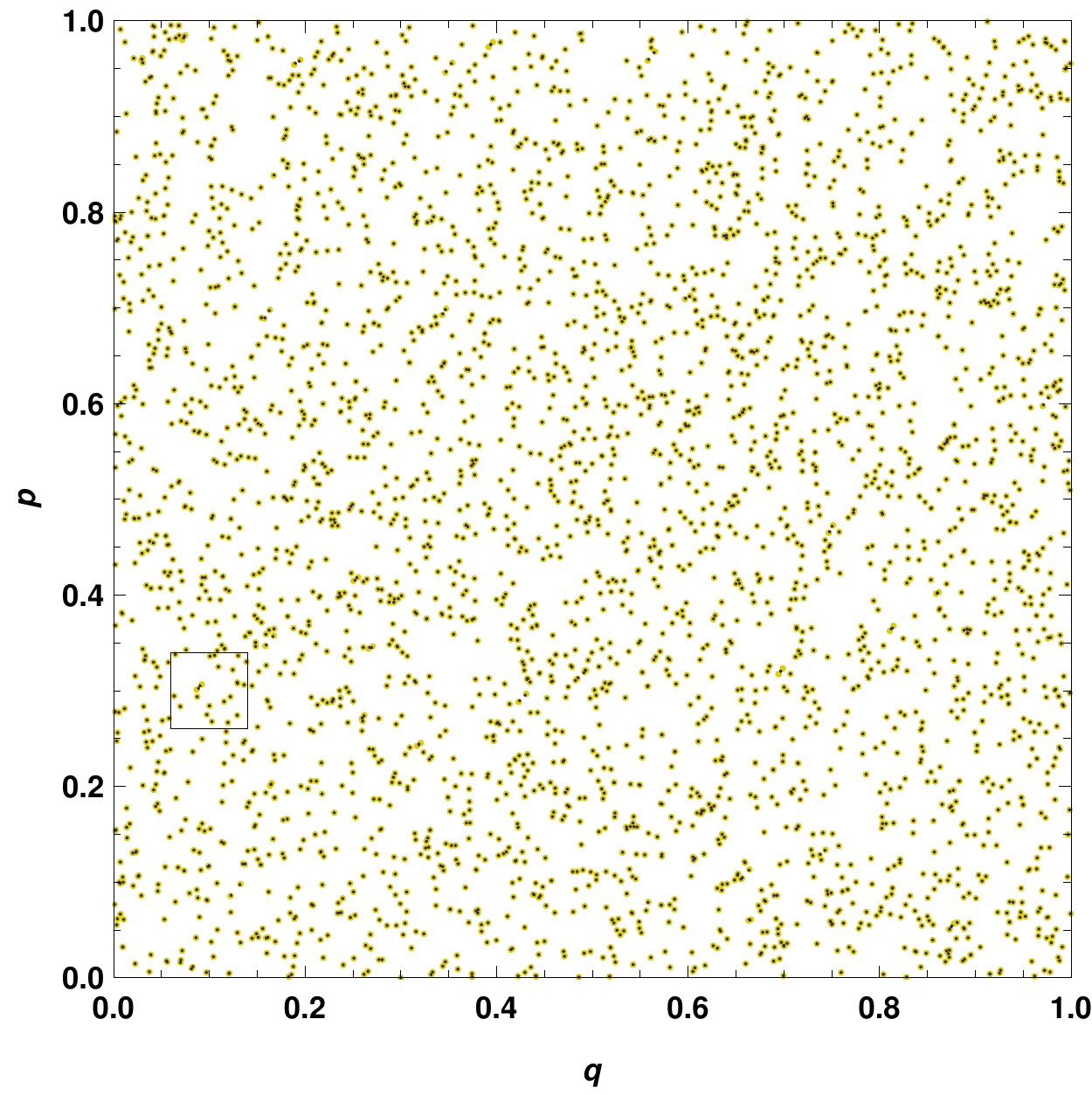}\includegraphics[height=7.0cm]{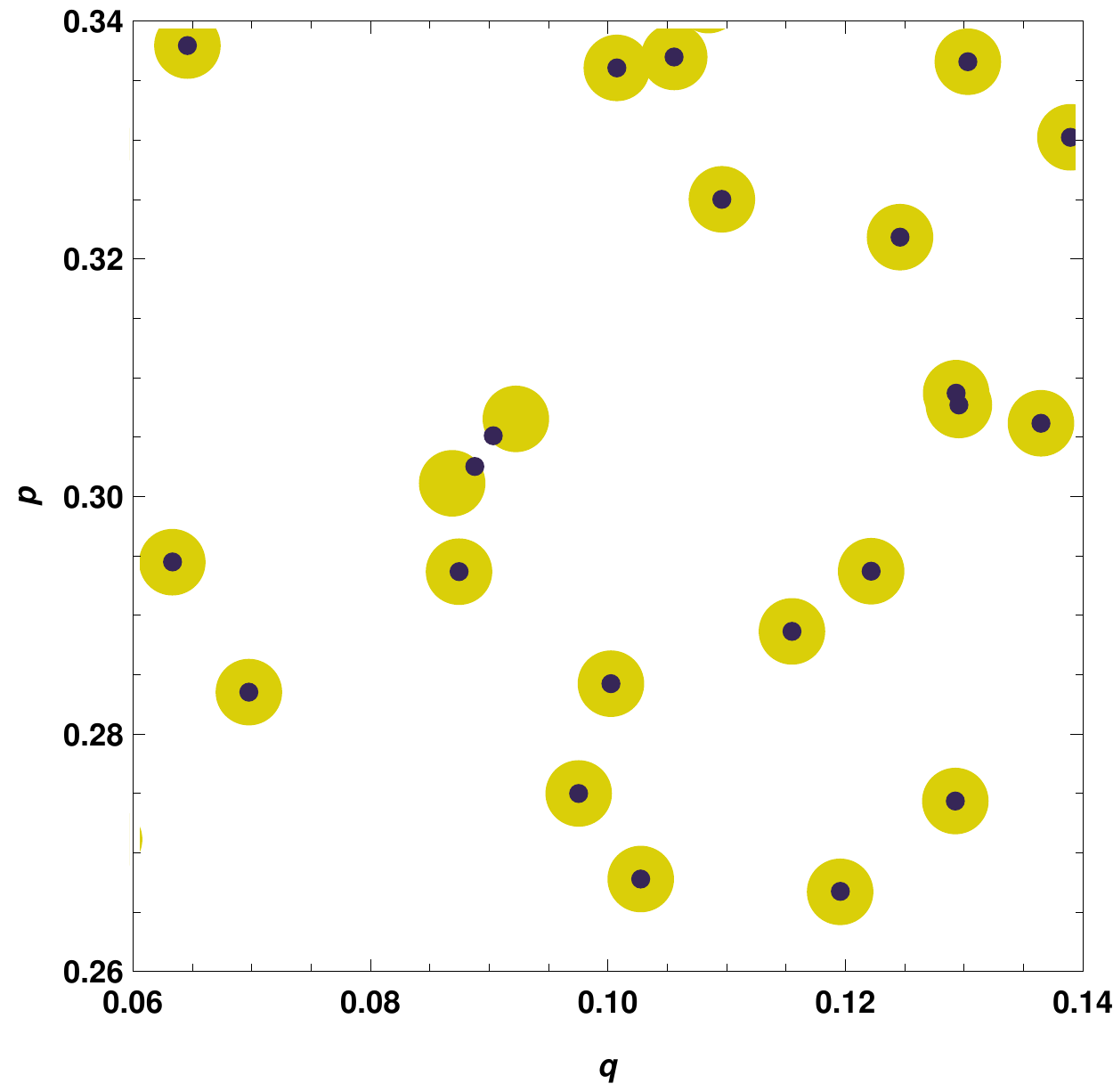}
	\caption{\small \textit{Left:} The phase-space coordinates of the pair of MPOs ${\Gamma}$, $\bar{\Gamma}$ encoded by the symbolic 2D sequences $\Aa_{\Gamma}$, $\Aa_{\bar{\Gamma}}$ on the fig.~\ref{Fig12}. The positions of particles belonging to ${\Gamma}$ are depicted by the green circles, while the positions of particles of the orbit $\bar{\Gamma}$ are the blue circles of smaller sizes. All green and blue points come in pairs and there are no unpaired points. 
\textit{Right:} The enlarged region of the phase space (the selected rectangle on the left picture). The geometrical centers of the circles visually coincide for almost all points except the four points in the middle of the diagram.
}\label{Fig13}
\end{figure}

To evaluate the  actual distances between the partner orbits in the phase space we have restored their coordinates and momenta $\Gamma=\{(q_{n,t}, p_{n,t})| (n,t)\in \Zz\}$, $\bar{\Gamma}=\{(\bar{q}_{n,t}, \bar{p}_{n,t})| (n,t)\in \Zz\}$  from the corresponding  matrices  of winding numbers $\Mm_\Gamma, \Mm_{\bar{\Gamma}}$. The resulting  sets  of the points  in the phase space $V$ are shown on fig.~\ref{Fig13} by the black points for  $\Gamma$ and by  the green circles for its partner $\bar{\Gamma}$.  As one can immediately see, all the points of two orbits are coupled into pairs separated by very small distances. For some points the distances  are somewhat  larger and the pairs form quadruples, see the enlarged region of the phase space on the fig.~\ref{Fig13}. These special points correspond to the encounter region of the orbits.

\begin{figure}
\includegraphics[height=8.0cm]{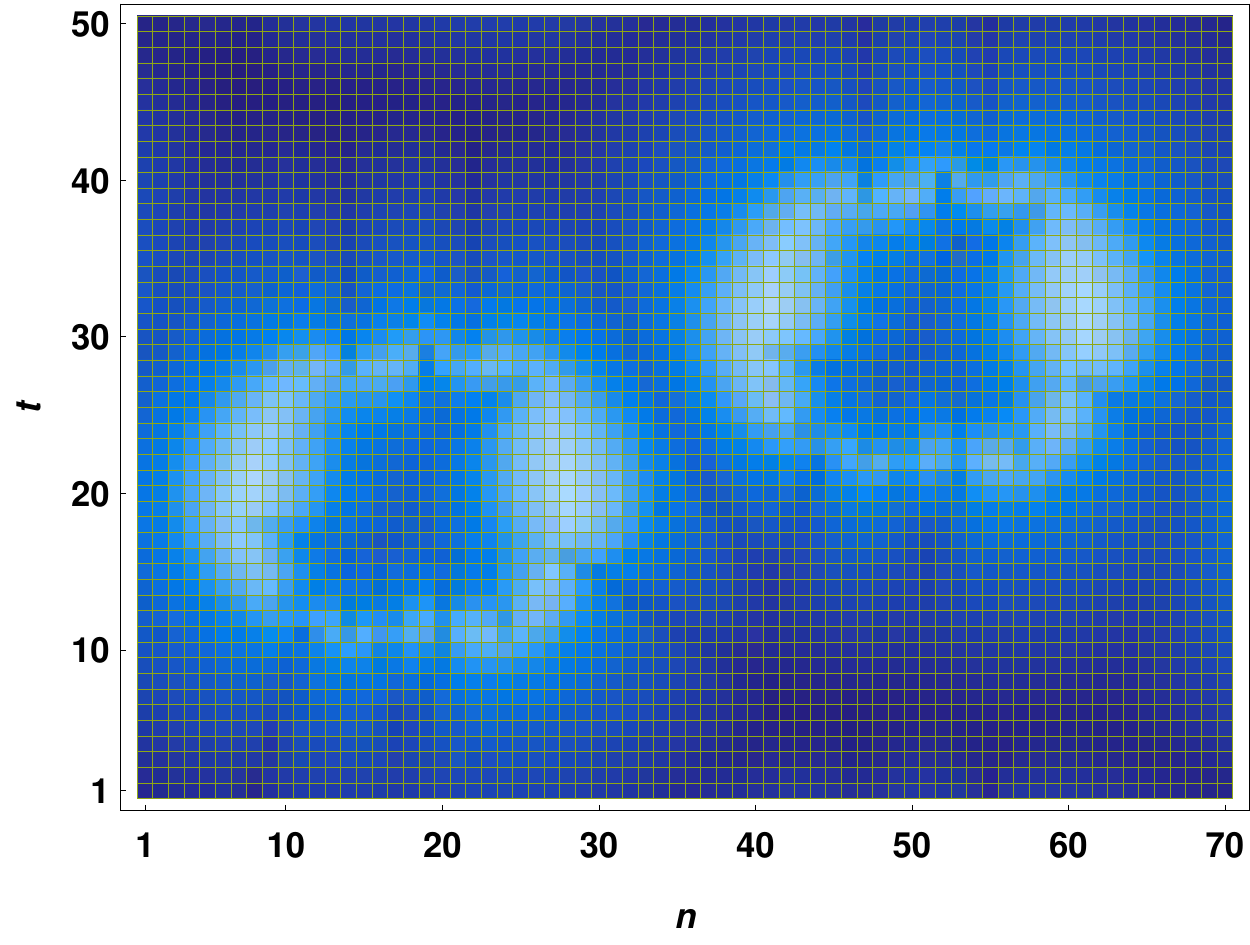} \includegraphics[height=7.0cm]{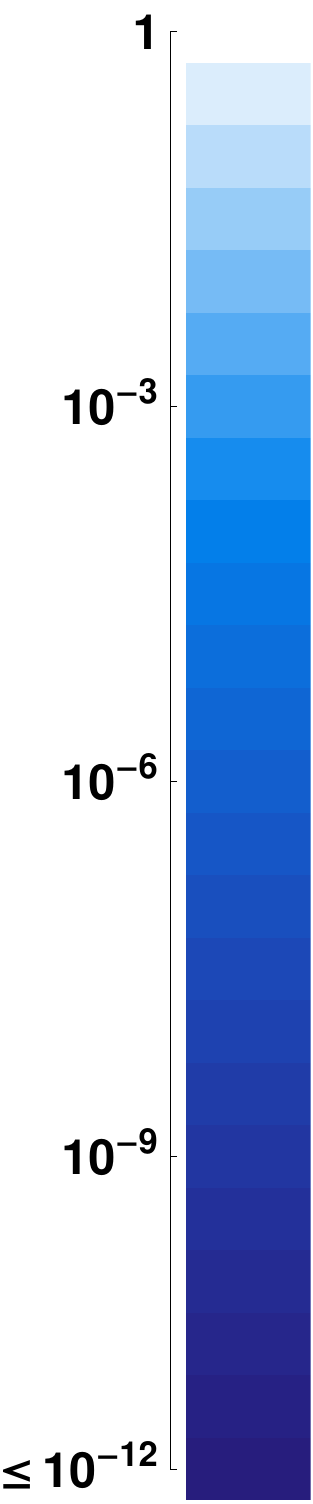}
\caption{Diagram of metric distances between MPOs $\Gamma$ and $\bar{\Gamma}$ from the fig.~\ref{Fig12}. The distances are calculated according to the formula~(\ref{dist}). The maximal distance between points is $\simeq 1.7889\times  10^{-3}$.  The numeric error is of the order $\simeq 10^{-14}$.}\label{Fig15}
\end{figure}

 To quantify  the metric distances between partner orbits we have calculated the mean square displacement  in the phase space for each couple of the paired points. The value of the displacement $d_{n,t}$ for the particle number $n$ at time $t$ is given by 
\begin{equation}\label{dist}
	d_{n,t}= \sqrt{(q_{n,t}- \bar{q}_{n',t'} )^2+(p_{n,t}- \bar{p}_{n',t'} )^2},
\end{equation}
where $(\bar{q}_{n',t'}, \bar{p}_{n',t'})\in \bar{\Gamma}$ is the   point  paired to  the one $(q_{n,t}, p_{n,t})\in\Gamma$.
The  diagram of the metric distances is  plotted on the figure~\ref{Fig15}. As expected, the largest distances  $d_{n,t}\sim 10^{-3}$ are observed 
in the encounter regions (appearing as white areas on the figure~\ref{Fig12}). Outside of the encounters the separations between partner orbits are extremely small with $d_{n,t}\lesssim 10^{-12}$.

\section*{Appendix B} 
\begin{figure}
\centering
\includegraphics[height=7.8cm]{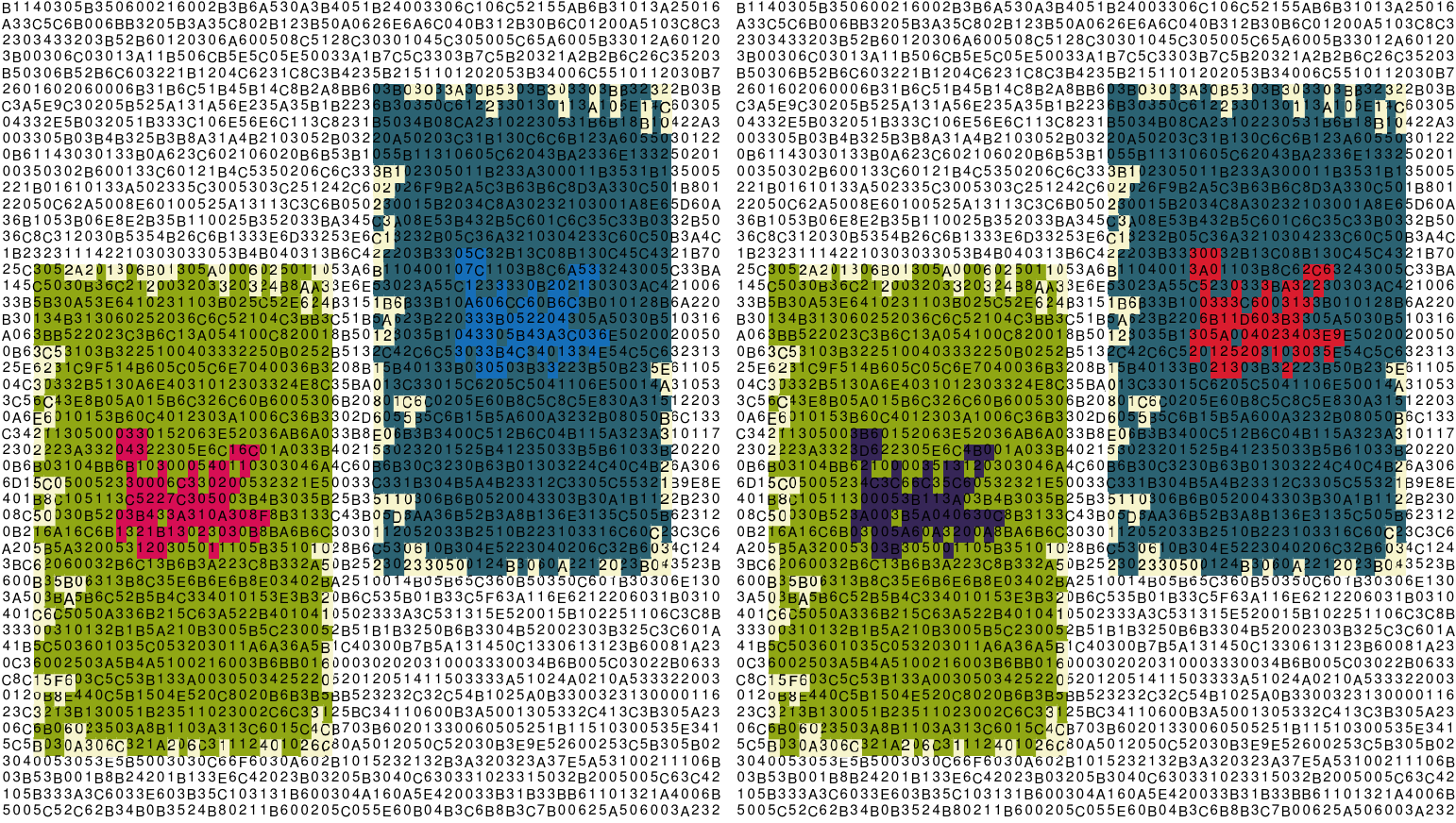}\ \includegraphics[height=7.8cm]{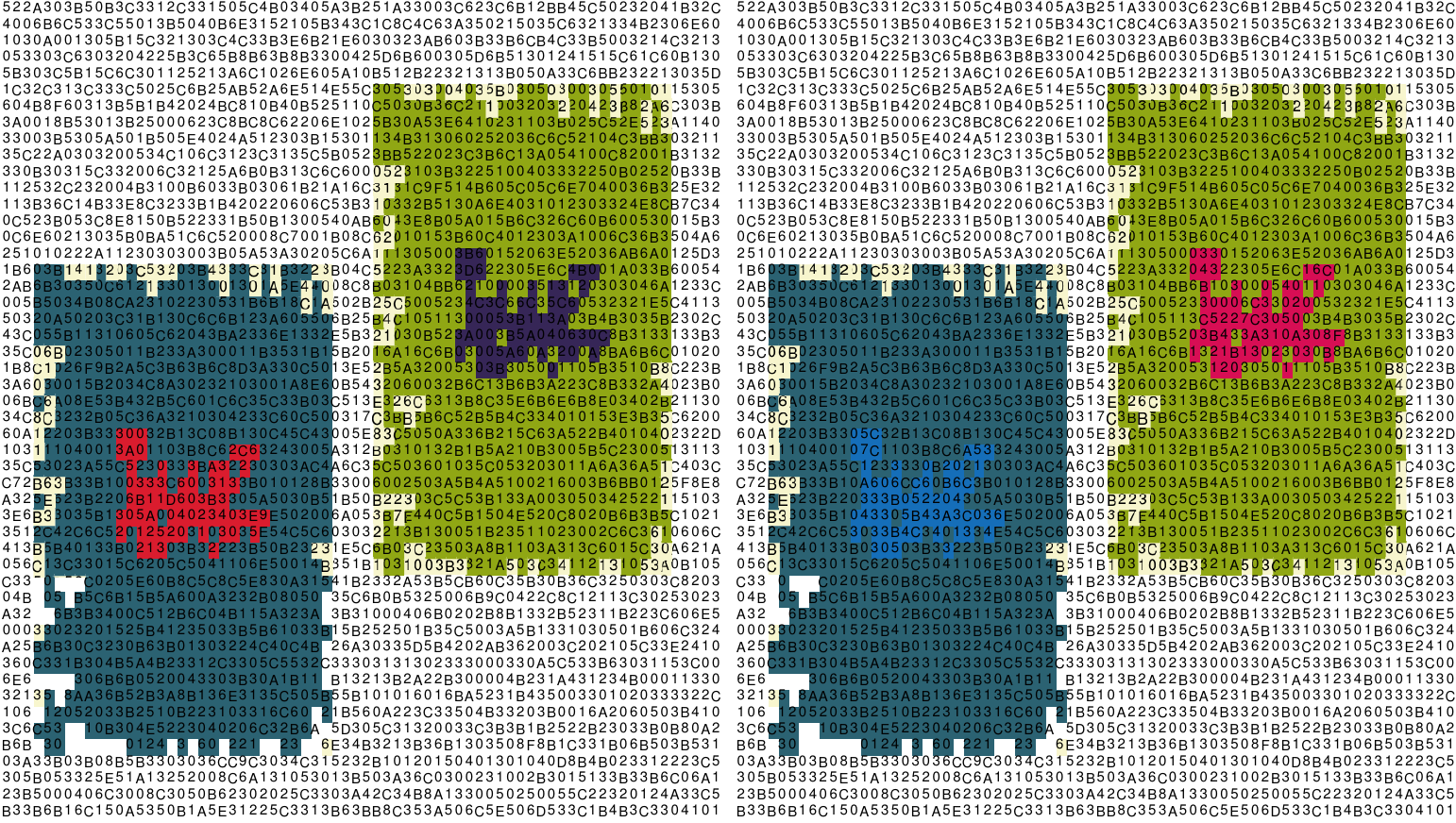}
	\caption{Symbolic representations $\Aa_{\Gamma}$ (upper left) $\Aa_{\bar{\Gamma}}$ (upper right) $\Aa_{\Gamma^*}$ (down left) $\Aa_{\bar{\Gamma}^*}$ (down right) of two partner  periodic orbits (${\Gamma}$, $\bar{\Gamma}$) and their symmetrically conjugate counterparts  (${\Gamma}^*=\Sym_\C\cdot\Gamma$, $\bar{\Gamma}^*=\Sym_\C\cdot\bar{\Gamma}$) for $T=50$, $N=70$. 
 The identical regions of symbols  are highlighted by the same colors (color on line). The two  encounter regions  (green and dark-blue) are pairwise conjugated by the symmetry operation $\Sym_\C$.}\label{Fig17}
\end{figure}

\begin{figure}
\includegraphics[height=7.0cm]{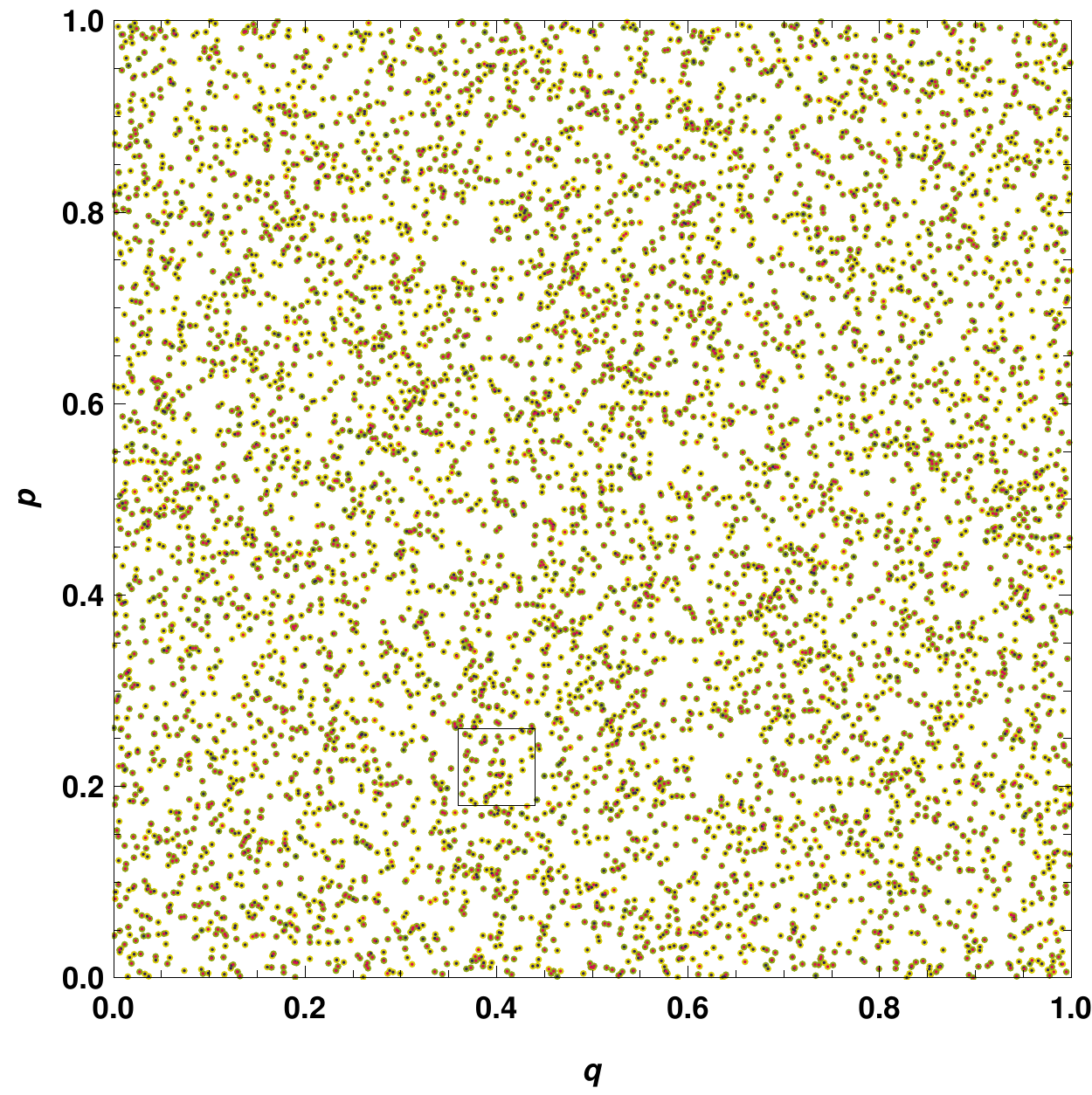}\includegraphics[height=7.0cm]{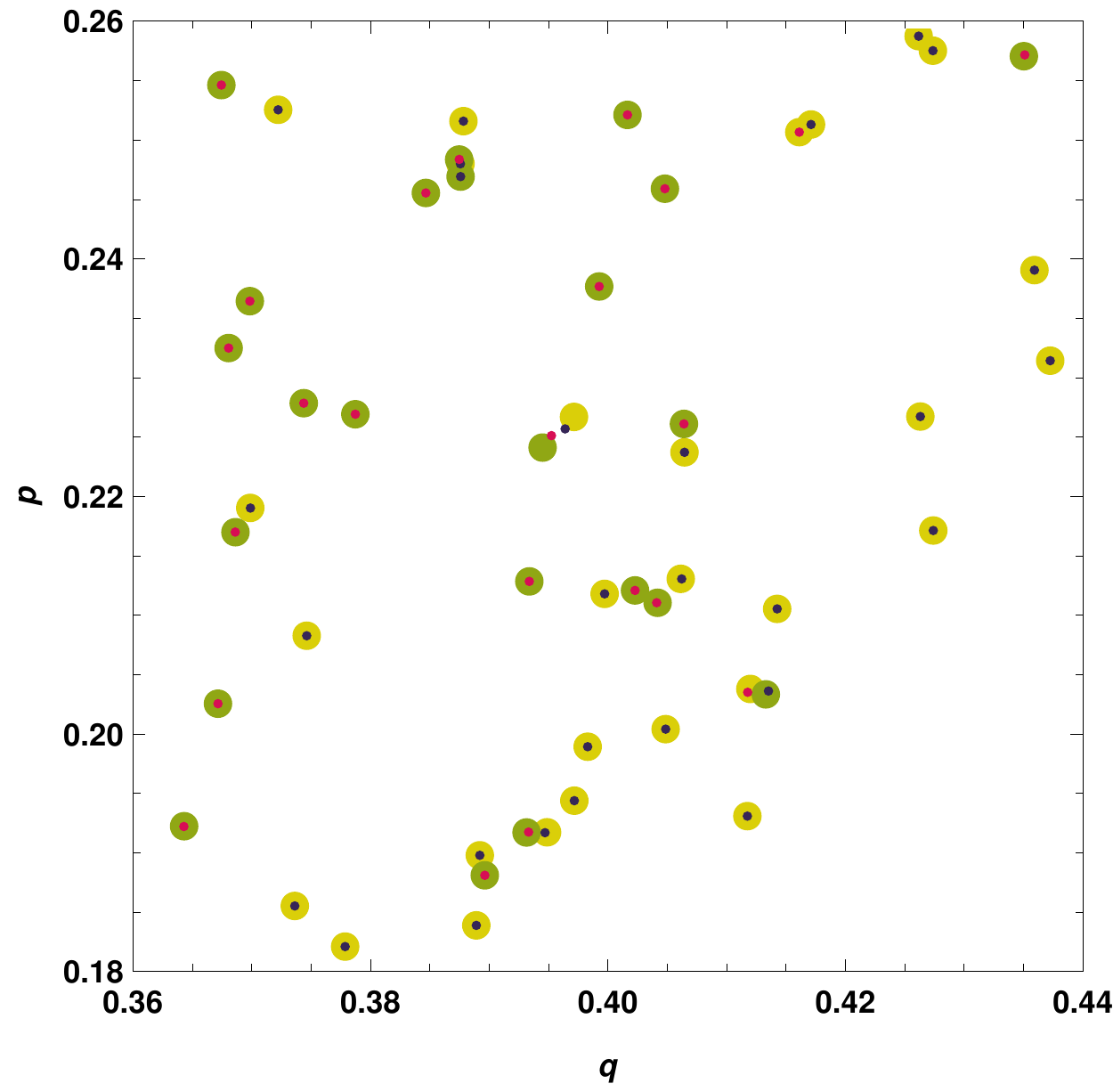}
	\caption{ \small \textit{Left:} The phase-space coordinates of the quadruple of orbits $\Gamma$,  $\Gamma^*$,  $\bar{\Gamma}$, $\bar{\Gamma}^*$. The positions of the particles of the trajectories $\Gamma$ ($\Gamma^*$) are depicted by the large dark (light) green circles. The phase space coordinates of the partner MPOs $\bar{\Gamma}$ ($\bar{\Gamma}^*$) are depicted by blue (red) circles respectively. All points come in pairs and there are no unpaired points. \textit{Right:} The enlarged region of the phase space (the selected rectangle on the left picture). The geometrical centers of the circles visually coincide for almost all points except the four points in the middle of the plot. These points belong to the encounter region, where the distances between partner orbits are maximal.}\label{Fig19} 
\end{figure}

In this appendix we explicitly construct the quadruple of orbits $\Gamma$,  $\Gamma^*$,  $\bar{\Gamma}$, $\bar{\Gamma}^*$   of the period $T=50$ for the same map $\Phi_{\Nn}$  as in Appendix~A. As has been explained in Section~\ref{symmetries} this map posses the symmetry $\Sym_\C$:  $\q\to 1-\q\!\!\!\mod 1$, $\p\to 1-\p\!\!\!\mod 1$.  To construct partner orbits which traverse different points of the phase space $V$ we utilize 4-step protocol similar to one used in Appendix~A. First, we generate a random MPO $\Gamma_0$, its symmetric counterpart $\Gamma_0^*$ and the corresponding  matrices of winding numbers $\Mm_{\Gamma_0}$, $\Mm_{\Gamma^*_0}$. We then take a frame-like region $E_1\subset\Zz$ and superimpose the local winding numbers   of   $\Mm_{\Gamma^*_0}$ at $E_1$ upon another (shifted) region  $E_2\subset\Zz$ of $\Mm_{\Gamma_0}$. The resulting  matrix $\widetilde{\Mm}_{\Gamma_0}$ has two regions $E_1$, $E_2$ satisfying (\ref{encountersym}). Note, however, that  $\widetilde{\Mm}_{
\Gamma_0}$ does not correspond to 
any real 
MPO. In order to obtain matrix of winding numbers for  a valid periodic orbit we then apply eq.~(\ref{validmonodromy}) (see step 3 in Section~\ref{protocol}). This yields to  the matrix  $\Mm_{\Gamma}$  of winding numbers for a valid trajectory $\Gamma$ with the correct encounter structure. Its symmetric counterpart   is obtained by application of the symmetry operation $\Sym_\C$. Finally the partner orbits  $\bar{\Gamma}$, $\bar{\Gamma}^*$ are obtained  by rearrangement of symbols from $\Mm_{\Gamma}$, $\Mm_{\Gamma^*}$ as shown in fig.~\ref{encounter_sym}. The resulting symbolic representations of  orbits $\Gamma$,  $\Gamma^*$,  $\bar{\Gamma}$, $\bar{\Gamma}^*$ are shown in fig.~\ref{Fig17}.  The  elements of the  matrices $\Mm_{\Gamma}$ (resp.) $\Mm_{\bar{\Gamma}}$ are encoded here by the hexadecimal numbers in accordance with the table~(\ref{table}).
In order to see that $\Gamma$,  $\Gamma^*$,  $\bar{\Gamma}$, $\bar{\Gamma}^*$ are indeed partner orbits we plotted on the figure (\ref{Fig19}) the corresponding sets of momenta and coordinates. It is clearly visible on this plot that any  point from  $\bar{\Gamma}$ (resp. $\bar{\Gamma}^*$) closely approaches  one of the points from either $\Gamma$ or $\Gamma^*$. In other words all the points of $\bar{\Gamma}$ and $\Gamma$ are paired up to the symmetry operation $\Sym_\C$. As expected, the largest distances between the pairs of the points are observed in the encounters. 

\section*{Appendix C}

In this appendix we present explicit calculations of  partner  MPOs for the perturbed map. The periodic potential $\V(q)$ is chosen in the form
$$
\V(q)=-\frac{\kappa}{4\pi^2}\cos(2\pi q),
$$
with  the  other parameters of the map  set to the same values as in  Appendix~A. 
For the sake of convenience we require that the alphabet of the symbolic representation (e.g., the number of symbols)  would not change  under the perturbation. This  can be achieved by a proper choice of the perturbation strength, which in our calculations was set to $\kappa=1$. As we found in our numerical study for  this choice of $\kappa$ the dynamics of the system are chaotic and  the symbolic alphabet is still preserved.

\begin{figure}
\includegraphics[height=6.0cm]{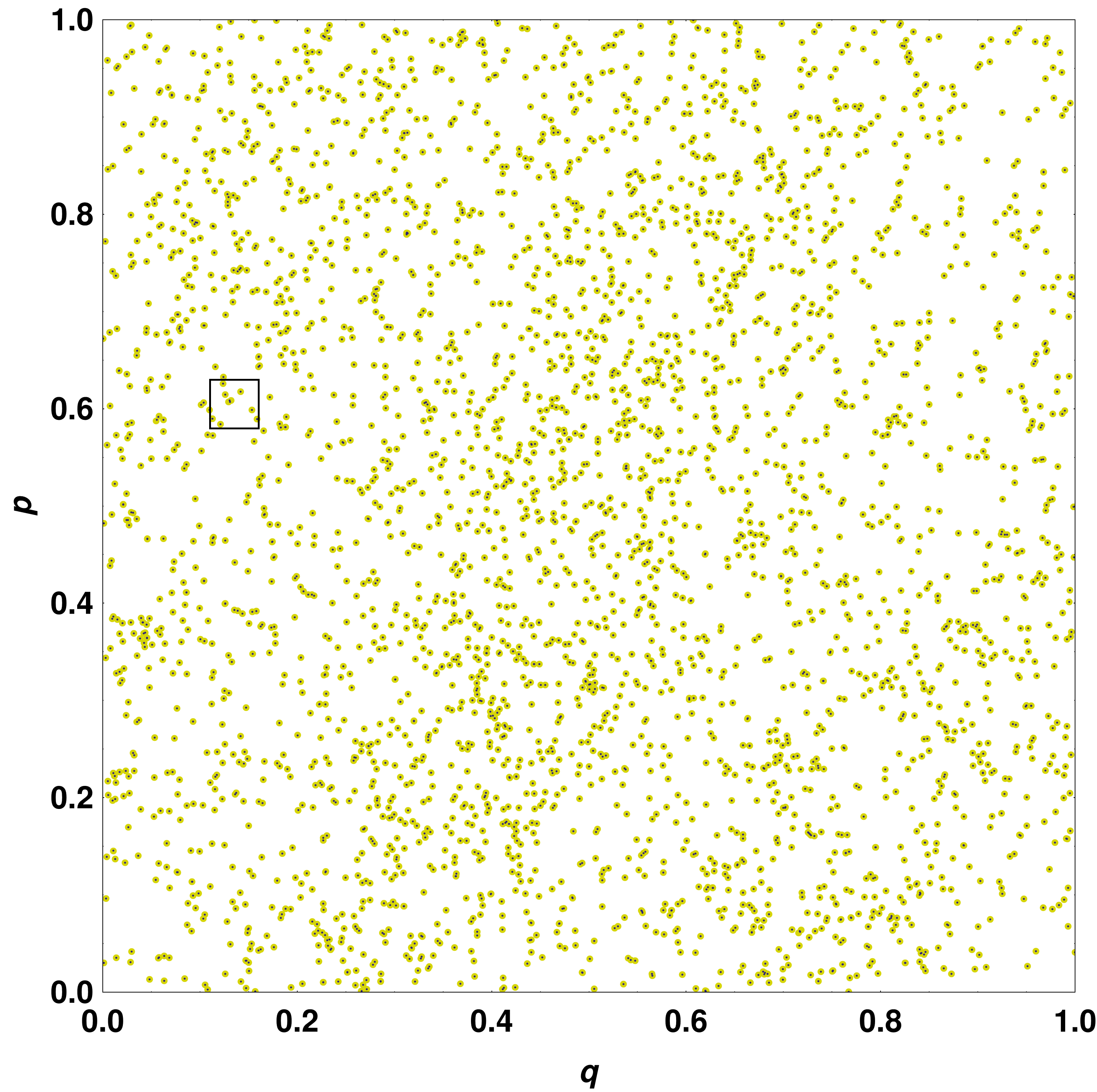}\includegraphics[height=6.0cm]{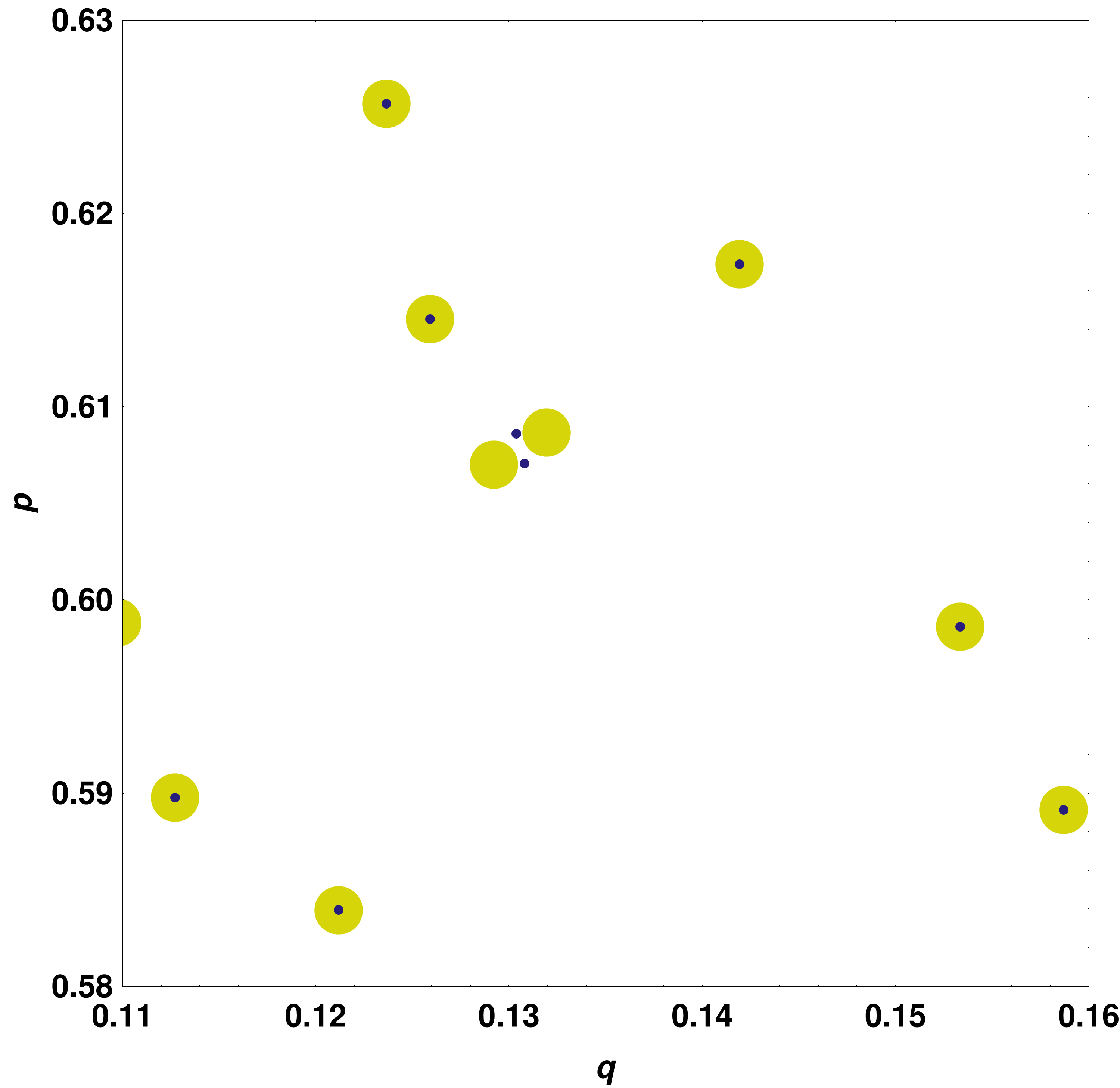}
	\caption{\small  \textit{Left:} The phase-space diagram of the pair of MPOs in the case of perturbed map. The coordinates of the  orbit and its pair are depicted by the green circles and  by the blue circles of smaller size, respectively. All green and blue points come in pairs and there are no unpaired points. 
\textit{Right:} The enlarged region of the phase space (rectangle on the left picture). The geometrical centers of the  circles visually coincide for almost all points except for the four points in the middle of the diagram belonging to the encounter.}\label{Fig21} 
\end{figure}
\begin{figure}
\includegraphics[height=5.5cm]{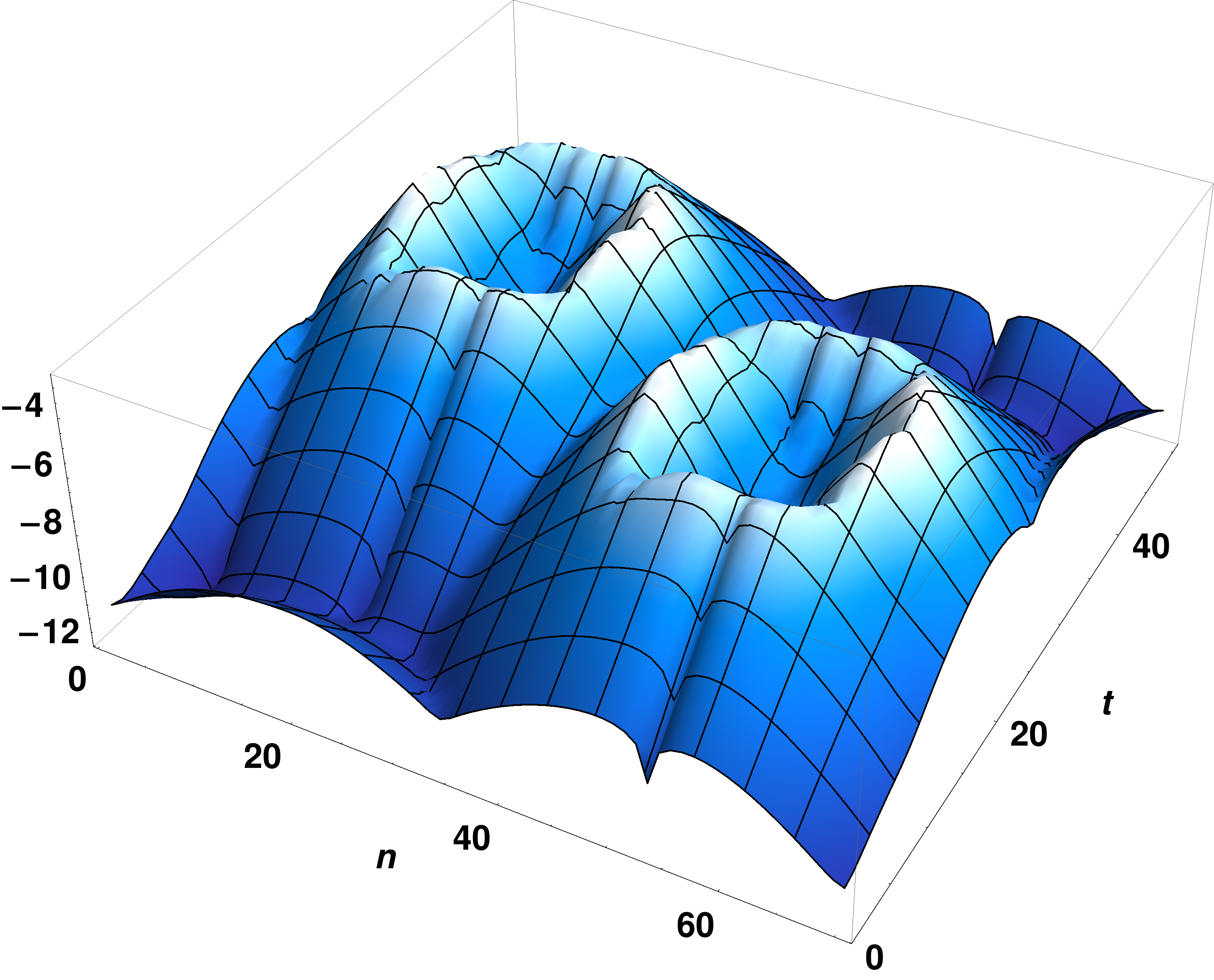}\includegraphics[height=5.5cm]{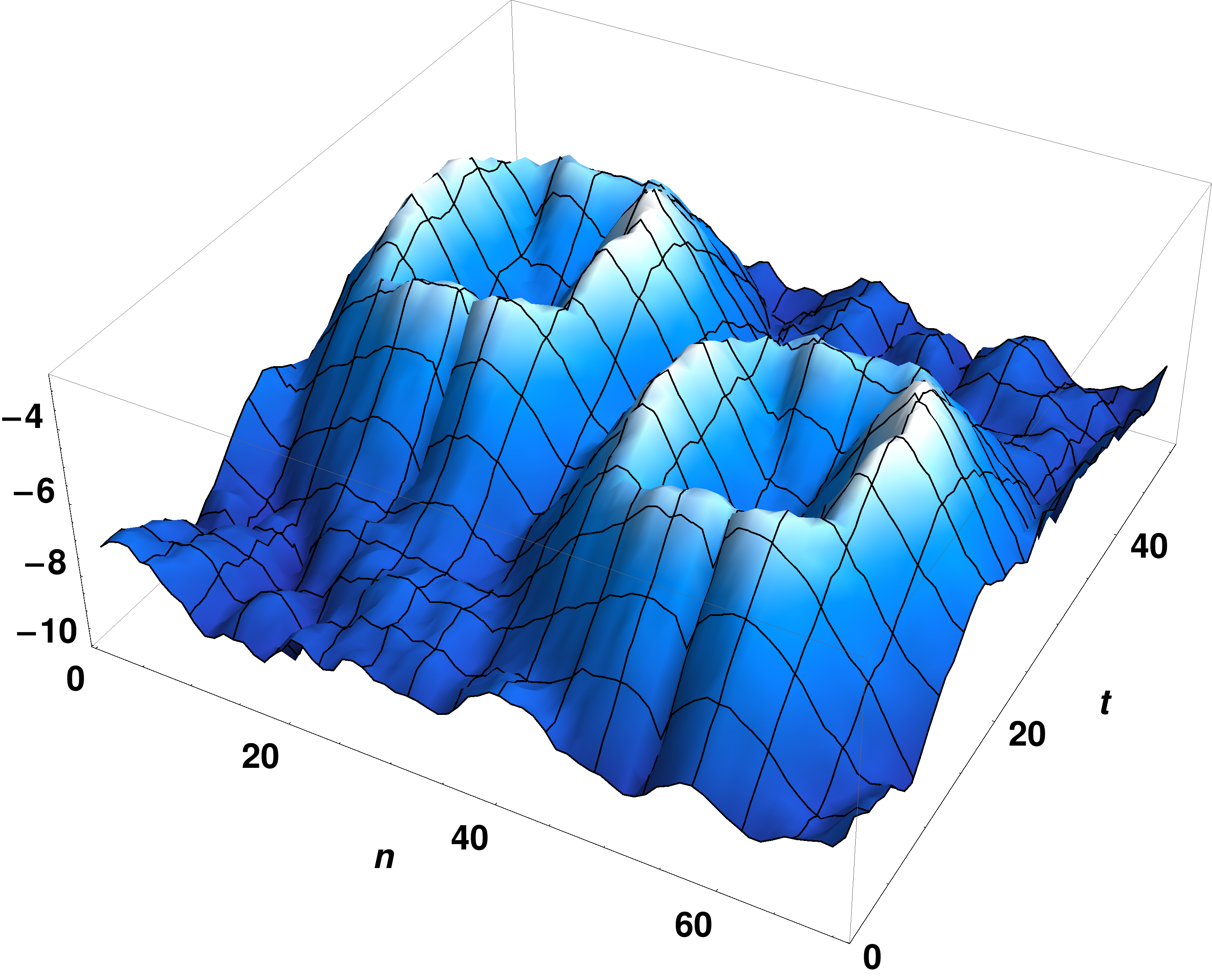}
	\caption{\small 3D diagrams of metric distances for  non-perturbed map (left) (2D version of this figure is shown on the fig.~\ref{Fig15}) and  for the perturbed map (right). The maximal distance in the case of the perturbed map is $\simeq 1.90578\times  10^{-3}$, the numeric  precision is of the order $\simeq 10^{-10}$.}\label{Fig20} 
\end{figure}
The initial partner MPOs $\Gamma_0$, $\bar{\Gamma}_0$ of the non-perturbed map were taken from our calculations in  Appendix~A.   The partner orbits $\Gamma$, $\bar{\Gamma}$ for the perturbed map  were then  constructed following the recipe in Section~\ref{Sec7}. The two obtained sets of points are depicted on the fig.~\ref{Fig21}.  As one can see, all the points of partner orbits are coupled into pairs separated by  very small distances. In the  encounters the distances become larger and the points form quadrapole, see the enlarged region of the phase space on the fig.~\ref{Fig21}. The metric distances between $\Gamma$, $\bar{\Gamma}$  were evaluated using the same method as in Appendix A. The result  is presented on the fig.~\ref{Fig20} in 3D form. For a comparison, the result for the original non-perturbed MPO's $\Gamma_0$, $\bar{\Gamma}_0$ is shown on the same figure. Although the points of the  initial $\Gamma_0$, $\bar{\Gamma}_0$ and perturbed orbits $\Gamma$, $\bar{\Gamma}$ are positioned  far away 
from each other, the generic picture of encounters has no structural changes. One can see only slight changes in their pairwise differences on the fig.~\ref{Fig20}.

\end{document}